\PassOptionsToPackage{unicode}{hyperref}
\PassOptionsToPackage{hyphens}{url}
\PassOptionsToPackage{dvipsnames,svgnames,x11names}{xcolor}
\documentclass[
  12pt]{article}

\usepackage{amsmath,amssymb}
\usepackage{iftex}
\ifPDFTeX
  \usepackage[T1]{fontenc}
  \usepackage[utf8]{inputenc}
  \usepackage{textcomp} 
\else 
  \usepackage{unicode-math}
  \defaultfontfeatures{Scale=MatchLowercase}
  \defaultfontfeatures[\rmfamily]{Ligatures=TeX,Scale=1}
\fi
\usepackage{lmodern}
\ifPDFTeX\else  
\fi
\IfFileExists{upquote.sty}{\usepackage{upquote}}{}
\IfFileExists{microtype.sty}{
  \usepackage[]{microtype}
  \UseMicrotypeSet[protrusion]{basicmath} 
}{}
\makeatletter
\@ifundefined{KOMAClassName}{
  \IfFileExists{parskip.sty}{%
    \usepackage{parskip}
  }{
    \setlength{\parindent}{0pt}
    \setlength{\parskip}{6pt plus 2pt minus 1pt}}
}{
  \KOMAoptions{parskip=half}}
\makeatother
\usepackage{xcolor}
\makeatletter
\ifx\paragraph\undefined\else
  \let\oldparagraph\paragraph
  \renewcommand{\paragraph}{
    \@ifstar
      \xxxParagraphStar
      \xxxParagraphNoStar
  }
  \newcommand{\xxxParagraphStar}[1]{\oldparagraph*{#1}\mbox{}}
  \newcommand{\xxxParagraphNoStar}[1]{\oldparagraph{#1}\mbox{}}
\fi
\ifx\subparagraph\undefined\else
  \let\oldsubparagraph\subparagraph
  \renewcommand{\subparagraph}{
    \@ifstar
      \xxxSubParagraphStar
      \xxxSubParagraphNoStar
  }
  \newcommand{\xxxSubParagraphStar}[1]{\oldsubparagraph*{#1}\mbox{}}
  \newcommand{\xxxSubParagraphNoStar}[1]{\oldsubparagraph{#1}\mbox{}}
\fi
\makeatother

\usepackage{longtable,booktabs,array}
\usepackage{calc} 
\usepackage{etoolbox}
\makeatletter
\patchcmd\longtable{\par}{\if@noskipsec\mbox{}\fi\par}{}{}
\makeatother
\IfFileExists{footnotehyper.sty}{\usepackage{footnotehyper}}{\usepackage{footnote}}
\makesavenoteenv{longtable}
\usepackage{graphicx}
\makeatletter
\def\maxwidth{\ifdim\Gin@nat@width>\linewidth\linewidth\else\Gin@nat@width\fi}
\def\maxheight{\ifdim\Gin@nat@height>\textheight\textheight\else\Gin@nat@height\fi}
\makeatother
\setkeys{Gin}{width=\maxwidth,height=\maxheight,keepaspectratio}
\makeatletter
\def\fps@figure{htbp}
\makeatother

\makeatletter
\@ifpackageloaded{caption}{}{\usepackage{caption}}
\AtBeginDocument{%
\ifdefined\contentsname
  \renewcommand*\contentsname{Table of contents}
\else
  \newcommand\contentsname{Table of contents}
\fi
\ifdefined\listfigurename
  \renewcommand*\listfigurename{List of Figures}
\else
  \newcommand\listfigurename{List of Figures}
\fi
\ifdefined\listtablename
  \renewcommand*\listtablename{List of Tables}
\else
  \newcommand\listtablename{List of Tables}
\fi
\ifdefined\figurename
  \renewcommand*\figurename{Figure}
\else
  \newcommand\figurename{Figure}
\fi
\ifdefined\tablename
  \renewcommand*\tablename{Table}
\else
  \newcommand\tablename{Table}
\fi
}
\@ifpackageloaded{float}{}{\usepackage{float}}
\floatstyle{ruled}
\@ifundefined{c@chapter}{\newfloat{codelisting}{h}{lop}}{\newfloat{codelisting}{h}{lop}[chapter]}
\floatname{codelisting}{Listing}

\makeatother
\makeatletter
\@ifpackageloaded{caption}{}{\usepackage{caption}}
\@ifpackageloaded{subcaption}{}{\usepackage{subcaption}}
\makeatother

\ifLuaTeX
  \usepackage{selnolig}  
\fi
\usepackage[]{natbib}
\usepackage{bookmark}

\IfFileExists{xurl.sty}{\usepackage{xurl}}{} 
\usepackage{bm}
\usepackage[ruled]{algorithm2e}
\usepackage{multirow}
\usepackage{amsthm}
\newtheorem{theorem}{Theorem}[section]
\newtheorem{lemma}[theorem]{Lemma}
\newtheorem{corollary}[theorem]{Corollary}
\newtheorem{remark}[theorem]{Remark}
\theoremstyle{definition}
\newtheorem{definition}[theorem]{Definition}
\usepackage{silence}
\newcommand{\anon}{1}

\begin{document}

\def\spacingset#1{\renewcommand{\baselinestretch}%
{#1}\small\normalsize} \spacingset{1}


\if1\anon
{
  \title{\bf Differentially Private Tests of Fisher's Sharp Null Hypothesis for Binary Outcomes}
  \author{Qingyang Sun\thanks{The authors gratefully acknowledge funding from grant NSF-SES-2217456.}\hspace{.2cm}\\
    Department of Statistical Science, Duke University\\
    and \\
    Jerome P. Reiter \\
    Department of Statistical Science, Duke University}
  \maketitle
} \fi

\if0\anon
{
  \bigskip
  \bigskip
  \bigskip
  \begin{center}
    {\LARGE\bf Differentially Private Tests of Fisher's Sharp Null Hypothesis for Binary Outcomes}
\end{center}
  \medskip
} \fi

\bigskip
\begin{abstract}
Across many disciplines, causal inference often relies on randomized experiments with binary outcomes. In such experiments, often analysts are interested in testing Fisher's sharp null hypothesis, i.e., the treatment has zero effect on the outcome for every study subject.
Sometimes the outcomes are sensitive and must be kept confidential, for example, when they comprise physical or mental health measurements. Releasing test statistics or $p$-values computed with the confidential outcomes can leak information about the individuals in the study. Those responsible for sharing the analysis results may wish to bound this information leakage, which they can do by ensuring the released outputs satisfy differential privacy. In this article, we develop several differentially private tests of Fisher's sharp null for binary outcomes. Specifically, we consider direct perturbation approaches that inject calibrated noise into test statistics or $p$-values, as well as a Bayesian denoising framework that explicitly models the privacy mechanism. We further develop decision-making procedures under privacy constraints, including a Bayes risk-optimal rule and a frequentist-calibrated significance test. Through theoretical results, simulation studies, and an application to the ADAPTABLE clinical trial, we demonstrate that our methods can achieve valid and interpretable causal inference while ensuring the differential privacy guarantee.
\end{abstract}

\vfill

\newpage
\spacingset{1.8} 

\section{Introduction}
\label{sec1}

Randomization-based inference is a cornerstone of causal analysis, offering inferences that derive solely from the randomization of treatment assignments to study subjects. Among such methods, the Fisher randomization test (FRT) is one of the most widely used approaches \citep{Fisher1935}. In the FRT, one presumes Fisher's sharp null hypothesis: each individual's outcome is the same regardless of treatment assignment. Under this sharp null, one can use the observed outcomes to compute the value of the chosen test statistic for any possible randomization and thereby construct a reference distribution for the test statistic. The analyst compares the observed value of the test statistic to this reference distribution, resulting in a $p$-value under the null hypothesis. In this article, we consider tests of Fisher's sharp null for binary outcomes in completely randomized experiments.

In many causal studies, the binary outcomes are inherently sensitive, such as indicators of a disease or a risky behavior, and therefore should be kept confidential.
The literature on data privacy has shown that releasing results of any statistical analysis leaks information about the underlying study subjects. Even releasing summary statistics can introduce disclosure risks \citep{DN2003, DSS2017, AAC2022}.
Thus, those responsible for sharing the results of an FRT with confidential outcomes may want to limit the amount of information leakage.

One way to bound this leakage is to ensure the released results satisfy differential privacy (DP) \citep{Dwork2006, DMNS2006}, which offers a rigorous privacy guarantee.
While researchers have developed DP methods for causal inference \citep[e.g., ][]{OHK2015, LGPM2019, NNQCNK2022, MMS2024, GR2025} and for hypothesis testing \citep[e.g., ][]{GLRV2016, CKSBG2019, AS2018, KSGB2023, KKR2024, PB2025}, to our knowledge, there do not exist DP algorithms for testing Fisher's sharp null.
Arguably, the closest work is \cite{KS2025}, which develops DP permutation tests for kernel-based statistics. The framework in \cite{KS2025} targets two-sample and independence testing in general metric spaces. It does not exploit the combinatorial structure of the $2\times 2$ table that arises under Fisher's sharp null for binary outcomes.

We aim to close this gap by developing differentially private tests of Fisher's sharp null for binary outcomes, which we abbreviate as DP-FRTs. 
We develop methodologies targeted at two distinct inferential goals. For the first set of methods, the goal is 
to recover output from the confidential FRT from a set of differentially private statistics. In particular, we seek inference about the confidential $p$-value from the FRT itself. The analyst can interpret the resulting private inference as a measure of evidence against Fisher's sharp null. With this goal in mind, we first explore direct perturbation mechanisms that add calibrated noise to the confidential $p$-value or test statistic (Section \ref{sec:3.1}). We find that these direct approaches may not be sufficiently accurate for practical use. We therefore propose our preferred approach to learning about the confidential $p$-value: a mechanism-aware, Bayesian denoising approach that explicitly models the DP noise to recover a posterior distribution for the confidential $p$-value (Section \ref{sec:3.2}).
Based on this posterior distribution, we present a decision-theoretic approach that seeks to match the decision about the sharp null from the confidential FRT output (Section \ref{sec:4.1}). This approach includes an option for analysts to abstain from making a decision when the evidence is inconclusive, then refine or update their conclusions by spending additional privacy budget. 

In lieu of a decision-theoretic approach, some users may prefer to make decisions about Fisher's sharp null using a significance testing framework.
We therefore show how to construct a frequentist-calibrated significance test from the posterior distribution of the confidential $p$-value (Section \ref{sec:4.2}). Here, the inferential goal is not to emulate the results from the confidential data; rather, it is to control Type I error rates.
The test ensures that the privacy protection does not compromise the nominal significance level.
We note that one could develop frequentist-calibrated tests with other quantities as well, as we discuss and illustrate in the supplementary material. Indeed, our test is not necessarily the most powerful.
Nonetheless, basing the test on the posterior distribution of the confidential $p$-value has the benefit of using the same underlying quantity regardless of the framework for making the decision. Thus, for example, the analyst could report a summary of the posterior distribution to give a sense of likely values for the confidential $p$-value and the result of the significance test as the formal decision about the sharp null corresponding directly to that posterior distribution. 

The remainder of this article is organized as follows. Section \ref{sec2} reviews the FRT for binary outcomes and key concepts of DP.
Section \ref{sec3} presents the methods that perturb outputs from the confidential FRT, including the Bayesian denoising methods. 
Section \ref{sec4} describes the Bayesian decision-theoretic and significance testing frameworks.
Section \ref{sec5} presents simulation studies and a genuine data analysis that assess the performance of the proposed methods and offer guidance for their implementation. Section \ref{sec6} concludes with a discussion and potential extensions. Proofs are provided in the supplementary material.

\section{Background}
\label{sec2}

In Section \ref{sec:2.1}, we review the FRT in the context of completely randomized experiments with binary outcomes.
In Section \ref{sec:2.2}, we review key concepts and properties of DP.

\subsection{Fisher Randomization Test for Binary Outcomes}
\label{sec:2.1}

We develop DP-FRTs under the potential outcomes framework for causal inference \citep{Rubin1974}.
For each unit $i = 1, \dots, n$ in the study, let $Z_i=1$ when unit $i$ is assigned to treatment and $Z_i=0$ when the unit is assigned to control. For $i = 1, \dots, n$, let $Y_i(1)$ and $Y_i(0)$ be the outcomes for unit $i$ under treatment and control, respectively. Under the stable unit treatment value assumption, the observed outcome is $Y_i^{\rm obs} = Z_iY_i(1)+(1-Z_i)Y_i(0).$

The full collection $\{(Y_i(1),Y_i(0))\}_{i=1}^n$ is often referred to as the ``Science,'' which represents the underlying object of interest in causal inference.
The Science is never fully observed, since only one of $(Y_i(1),Y_i(0))$ is realized for each unit $i$.
Let $n_{11}$ and $n_{10}$ be the numbers of treated units with $Y_i^{\rm obs} =1$ and $Y_i^{\rm obs} =0$, respectively. Let $n_{01}$ and $n_{00}$ be the corresponding counts in the control group. Let $n_1$ and $n_0$ be the sizes of the treatment and control groups, and $n_{+1}$ and $n_{+0}$ be the marginal counts of each outcome.

Fisher \cite{Fisher1935} proposed to test the null hypothesis of no individual causal effects, that is, $H_0^{\mathrm{F}}: Y_i(1) = Y_i(0)$ for $i=1, \dots, n$.
For completely randomized experiments (CRE), the FRT under $H_0^{\mathrm{F}}$ proceeds as follows. Let $\bm{Z}=(Z_1,\dots,Z_n)$ be the vector of treatment assignments, with $\bm Z^{\rm obs}$ the realized assignment. Let $\bm{Y}^{\rm obs}=(Y_1^{\rm obs},\dots,Y_n^{\rm obs})$ be the vector of observed outcomes.
First, the analyst selects a test statistic $T(\bm{Z};\bm{Y}^{\rm obs})$ that is sensitive to deviations from $H_0^{\mathrm F}$.
The value of the observed statistic is $T^{\rm obs}=T(\bm Z^{\rm obs}; \bm{Y}^{\rm obs}).$
Under the known assignment mechanism of CRE, the analyst evaluates $T(\bm Z; \bm{Y}^{\rm obs})$ for each possible randomization $\bm Z$ that adheres to the design to obtain the randomization distribution under $H_0^{\mathrm F}$.
The (one-sided) FRT $p$-value is
\begin{equation}
p_{\mathrm{FRT}} = \Pr\left(T(\bm Z; \bm{Y}^{\rm obs}) \ge T^{\rm obs}\right) = \frac{1}{|\mathcal Z|}\sum_{\bm Z\in\mathcal Z}\mathbf{1}(T(\bm Z; \bm{Y}^{\rm obs})\ge T^{\rm obs}),
\end{equation}
where $|\mathcal Z|$ is the total number of possible random assignments. When $|\mathcal Z|$ is large, one can approximate $p_{\mathrm{FRT}}$ using Monte Carlo methods.

For binary outcomes, a natural choice of the test statistic is the difference in sample proportions, $\hat{\tau}=n_{11} / n_1 - n_{01} / n_0$. In this case, the FRT coincides numerically with Fisher's exact test for $2 \times 2$ tables. That is, under $H_0^{\mathrm F}$, $n_{11}$ follows a hypergeometric distribution so that, for $a = \max\{0,~n_1-n_{+0}\}, \dots, \min\{n_1,~n_{+1}\}$, we have
\begin{equation}\label{eq:hypergeom}
n_{11} \sim \mathrm{Hypergeometric}(n,~n_{+1},~n_1), \quad
\Pr(n_{11}=a) =\frac{\binom{n_{+1}}{a}\binom{n_{+0}}{n_1-a}}{\binom{n}{n_1}}.
\end{equation}

\subsection{Differential Privacy}
\label{sec:2.2}

In DP, we view a data-release procedure as a randomized algorithm $\mathcal{M}$ that takes a dataset $D$ as input and produces a randomized output. The privacy guarantee is defined with respect to neighboring datasets, i.e., datasets that differ in the data of a single individual, such as by adding, removing, or modifying one record.
DP requires that the output distributions produced by $\mathcal{M}$ on any pair of neighboring datasets be nearly indistinguishable. Consequently, it becomes difficult for an adversary to determine whether a specific individual is present in the dataset or to infer sensitive attributes with high confidence.

\begin{definition} \label{def:dp}
A randomized algorithm $\mathcal{M}$ satisfies $\epsilon$-differential privacy ($\epsilon$-DP) if, for any pair of neighboring datasets $D$ and $D'$ and for all measurable subsets $S \subseteq \mathcal{R}(\mathcal{M})$,
\begin{equation}
\Pr (\mathcal{M}(D) \in S) \le e^{\epsilon} \Pr (\mathcal{M}(D') \in S).
\end{equation}
\end{definition}

The privacy parameter $\epsilon > 0$ is referred to as the privacy budget.
Smaller values of $\epsilon$ provide stronger privacy guarantees; however, they also typically result in greater perturbations to the results computed with the confidential data. The literature on DP suggests values of $\epsilon \leq 1$ to provide robust privacy guarantees, although larger values often are used in practical applications \citep{KR2024prior}.

We make use of three properties of DP when designing the private algorithms for testing Fisher's sharp null. The first is post-processing invariance.
If $\mathcal{M}$ is an $\epsilon$-DP mechanism and $g$ is any (possibly randomized) function that does not depend on the data, then $g \circ \mathcal{M}$ also satisfies $\epsilon$-DP.
The second is sequential composition. Suppose $\mathcal{M}_1$ satisfies $\epsilon_1$-DP and $\mathcal{M}_2$ satisfies $\epsilon_2$-DP. Then, the joint release of $(\mathcal{M}_1(D), \mathcal{M}_2(D))$ satisfies $(\epsilon_1 + \epsilon_2)$-DP.
The third is parallel composition.
If $\mathcal{M}_1$ and $\mathcal{M}_2$ are applied to disjoint subsets of data, then the overall mechanism satisfies $\max(\epsilon_1, \epsilon_2)$-DP.

To ensure DP, a typical approach is to add random noise calibrated to the sensitivity of the released quantity.
Let $f : \mathcal{D} \rightarrow \mathbb{R}^d$ be a function defined on datasets. The $\ell_1$-sensitivity of $f$ is defined as $ \Delta f = \max_{D, D'} \|f(D) - f(D')\|_1$, where the maximum is taken over all pairs of neighboring datasets $D$ and $D'$. For example, when $f(D)$ produces a count, $\Delta f = 1$.

A common $\epsilon$-DP mechanism used in continuous domains is the Laplace mechanism \citep{Dwork2006}.
The Laplace mechanism releases $\tilde{f}(D) = f(D) + \eta$, where $\eta \sim \text{Lap}\left(0, \Delta f/\epsilon\right)$ with $p_\eta(h) = \left(\epsilon/(2\Delta f)\right) \exp\left( -\epsilon |h|/\Delta f \right).$
Another common mechanism, particularly for count data, is the Geometric mechanism \citep{GRS2012}.
Let $\rho = \exp(-\epsilon / {\Delta f})$. The Geometric mechanism releases $\tilde{f}(D) = f(D) + \eta$ where $\eta \sim \text{Geom}(\rho)$ with $\Pr (\eta = h) = ((1 - \rho)/(1 + \rho))\rho^{|h|}$, where $h \in \mathbb{Z}.$ We use both the Laplace and Geometric mechanisms in the DP-FRT algorithms.

\section{DP Estimation of the FRT \texorpdfstring{$p$-value}{p-value}}
\label{sec3}

This section presents $\epsilon$-DP approaches for privatizing the FRT $p$-value, including the direct perturbation methods (Section \ref{sec:3.1}) and the Bayesian denoising framework (Section \ref{sec:3.2}).
For both approaches, to define the DP guarantee we consider neighboring databases as two study populations that differ in one individual but receive the same treatment assignments. This presumes $(n_1,n_0)$ and $\bm Z^{\rm obs}$ are fixed by the design of CRE and are public knowledge.

\subsection{Direct Perturbation Approaches}
\label{sec:3.1}

One approach is to directly perturb the exact $p$-value using a Laplace mechanism.
Lemma \ref{lemma:sen_p} provides the sensitivity of $p_{\rm FRT}$.

\begin{lemma}
\label{lemma:sen_p}
Under the design of CRE with binary outcomes and the test statistic $\hat{\tau}$, the $\ell_1$-sensitivity of $p_{\rm FRT}$ is $\Delta_p = \max \left\{n_1/n, n_0/n\right\}$.
\end{lemma}

Based on Lemma \ref{lemma:sen_p}, we can release privatized FRT $p$-values by adding Laplace noise calibrated to the sensitivity and clipping to the feasible range.
Define $[L,U]=\left[  |\mathcal Z|^{-1},  1  \right]$. We release
$\tilde p = \Pi_{[L,U]} \left(p_{\mathrm{FRT}}+\eta\right)$, where
$\eta\sim{\rm Lap} \left(0, \Delta_p/\epsilon\right)$
and $\Pi_{[L,U]}(x) = \min\{U, \max\{L, x\} \}$ is the clipping operator. We call this method DP-FRT-p. Theorem \ref{thm:lap_exact_p} states the privacy guarantee for DP-FRT-p, which follows from the Laplace mechanism and the post-processing invariance of DP. The latter ensures that clipping does not degrade privacy.

\begin{theorem}
\label{thm:lap_exact_p}
For fixed $\epsilon$, $\tilde p$ produced by DP-FRT-p satisfies $\epsilon$-DP.
\end{theorem}

Another alternative is to perturb the test statistic and pair it with a privatized randomization distribution.
Lemma \ref{lem:sen_T} provides the $\ell_1$-sensitivity of $\hat{\tau} = n_{11} / n_1 - n_{01} / n_0$.

\begin{lemma}
\label{lem:sen_T}
Under the design of CRE with binary outcomes, the $\ell_1$-sensitivity of $\hat{\tau}$ is $\Delta_{\hat{\tau}} = \max\left\{1/n_1, 1/n_0\right\}.$
\end{lemma}

To implement this approach, we separately perturb the observed test statistic and its randomization distribution. Privatizing the randomization distribution is necessary since it depends on the total number of observed successes $n_{+1}$, which is computed from the confidential data.
Based on Lemma \ref{lem:sen_T}, we first release $\tilde T^{\mathrm{obs}}=\Pi_{[-1,1]}(\hat{\tau}+\eta_{\mathrm{obs}})$ with $\eta_{\mathrm{obs}}\sim{\rm Lap}(0,\Delta_{\hat{\tau}}/\epsilon_{\mathrm{obs}})$. We then release $\tilde n_{+1}=\Pi_{\{0,1,\dots,n\}}(n_{+1}+\eta_{\mathrm{ref}})$ with $\eta_{\mathrm{ref}}\sim{\rm Geom}(\exp(-\epsilon_{\mathrm{ref}}))$. As a default, we set $\epsilon_{\mathrm{obs}} = \epsilon_{\mathrm{ref}} = \epsilon/2$, where $\epsilon$ is the total privacy budget. Given $(\tilde T^{\mathrm{obs}},\tilde n_{+1})$, we compute the privatized Fisher tail probability $\tilde p=\Pr \big(\tilde n_{11}/n_1-(\tilde n_{+1}-\tilde n_{11})/n_0\ge \tilde T^{\mathrm{obs}}\big)$ for $\tilde n_{11}\sim{\rm Hypergeometric}(n,\tilde n_{+1},n_1)$. We refer to this procedure as DP-FRT-t. The privacy guarantee is stated in Theorem \ref{thm:sep_perturb}. It follows from the properties of the Laplace and Geometric mechanisms, sequential composition, and the post-processing invariance of DP.

\begin{theorem}
\label{thm:sep_perturb}
For fixed $\epsilon=\epsilon_{\mathrm{obs}}+\epsilon_{\mathrm{ref}}$, $\tilde p$ produced by DP-FRT-t satisfies $\epsilon$-DP.
\end{theorem}

The direct perturbation approaches have some limitations.
For DP-FRT-p, the addition of DP noise can strongly distort the FRT $p$-value. Per Lemma \ref{lemma:sen_p}, the sensitivity of $p_{\textrm{FRT}}$ is at least $1/2$, which is substantial relative to $[0,1]$. Moreover, clipping outputs can introduce bias. 
Additionally, DP-FRT-p and DP-FRT-t do not provide uncertainty quantification for the privatized outputs.
The absence of uncertainty measures limits the interpretability and transparency of the results.
Simulation studies illustrating the unsatisfactory performance of DP-FRT-p and DP-FRT-t are provided in the supplementary material.

It may be possible to reduce the distortions in DP-FRT-p and DP-FRT-t by using other privacy mechanisms, such as smooth sensitivity \citep{NRS2007} or propose-test-release \citep{DL2009}.
However, even with DP mechanisms that offer lower variances in the noise distributions, the analyst still could confront biases from clipping. Further, it can be difficult to quantify uncertainty appropriately when the scale of the noise depends on the confidential values, e.g., as in \cite{NRS2007}.

\subsection{DP Inference for the FRT $p$-value via Bayesian Denoising}
\label{sec:3.2}

The shortcomings of the methods in Section \ref{sec:3.1} motivate us to develop a Bayesian approach for obtaining evidence about $H_0^{\rm F}$ that explicitly accounts for the DP mechanism. The overarching strategy is to privatize the sufficient statistics, update a Bayesian model for the underlying true counts, and map the posterior distribution onto the space of $p$-values.
The resulting posterior distribution for $p_{\mathrm{FRT}}$ is the DP release. We call this approach DP-FRT-Bayes. We note that Bayesian denoising strategies have been applied for other inferential tasks as well \citep[e.g., ][]{BS2019, JAGR2022, LR2022}.

The algorithm begins by adding noise to $(n_{11}, n_{10}, n_{01}, n_{00})$. Since $(n_1,n_0)$ are fixed by design, it suffices to privatize $n_{11}$ and $n_{01}$.
Specifically, given $\epsilon$, we perturb $(n_{11},n_{01})$ by
\begin{equation}
\tilde{\bm{n}} = (\tilde n_{11}, \tilde n_{01}) = (n_{11}+\eta_{11},~n_{01}+\eta_{01}),
\qquad \eta_{11},\eta_{01}\stackrel{\text{i.i.d.}}{\sim}\mathrm{Geom}\left(\exp(-\epsilon)\right).
\label{eq:geom_mech_counts}
\end{equation}
Modifying the outcome of a single individual only changes the success count within that individual's assigned group. Consequently, at most one of $n_{11}$ or $n_{01}$ can change by $\pm1$. Therefore, the $\ell_1$-sensitivity of the pair $(n_{11}, n_{01})$ is one.

To describe the post-processing Bayesian inference, we introduce $\bm{m} = (m_{11}, m_{01})$ as random variables for
$(n_{11}, n_{01})$, which are unknown to the analyst who has the DP noisy counts $\tilde{\bm{n}}$.
We specify a data-independent prior $\pi$ on $(m_{11},m_{01})$ with support $\{0,\dots,n_1\}\times\{0,\dots,n_0\}$.
The posterior distribution is then $\Pr(m_{11}=a,m_{01}=b \mid \tilde{\bm{n}}) \propto w(a,b)$, where $w(a,b)=\pi(a,b) \kappa_\rho(\tilde n_{11}-a) \kappa_\rho(\tilde n_{01}-b).$
Here, $\kappa_\rho(h)=((1-\rho)/(1+\rho))\rho^{|h|}$ is the two-sided geometric kernel with $\rho=\exp(-\epsilon)$.

For each candidate pair of true counts $(a,b)$, let $K=a+b$ denote the corresponding total number of observed successes.
Under the sharp null $H_0^{\rm F}$ and using \eqref{eq:hypergeom}, the one-sided FRT $p$-value corresponding to $(a,b)$ is given by:
\begin{equation}
p_{ab}=\sum_{t=\max\{a, K-n_0\}}^{\min\{n_1,K\}}\frac{\binom{K}{t}\binom{n-K}{ n_1-t }}{\binom{n}{n_1}}.
\label{eq:pvalue_ab}
\end{equation}
Denote $\gamma(a,b)=\Pr(m_{11}=a,m_{01}=b\mid \tilde{\bm n})$. The deterministic mapping $(a,b)\mapsto p_{ab}$ induces the following posterior distribution of $p_{\mathrm{FRT}}$ given $(\tilde n_{11},\tilde n_{01})$:
\begin{equation}
\Pr(p_{\mathrm{FRT}}\in B\mid \tilde{\bm n})
=\sum_{a=0}^{n_1}\sum_{b=0}^{n_0}\gamma(a,b) \bm{1}(p_{ab}\in B),\qquad B\subseteq[0,1].
\label{eq:posterior_p}
\end{equation}
Note that using DP-FRT-Bayes allows $\tilde{n}_{11}$ and $\tilde{n}_{01}$ to fall outside $[0, n_1]$ or $[0, n_0]$, respectively. Clipping the DP counts to the feasible range is not necessary, as the posterior distribution enforces the constraints automatically.
Further, as shown in the supplementary material, under the Geometric mechanism posterior inferences are identical regardless of whether or not the DP counts are clipped.

By the Geometric mechanism, the release of $(\tilde n_{11},\tilde n_{01})$ satisfies $\epsilon$-DP.
Since all later steps use post-processing, DP-FRT-Bayes satisfies $\epsilon$-DP, as formally stated in Theorem \ref{thm:dp_bayes}.
\begin{theorem}
\label{thm:dp_bayes}
$\Pr(p_{\mathrm{FRT}} \mid \tilde{\bm n})$ released by DP-FRT-Bayes satisfies $\epsilon$-DP.
\end{theorem}

\begin{figure}[t]
    \centering
    \includegraphics[width=0.9\linewidth]{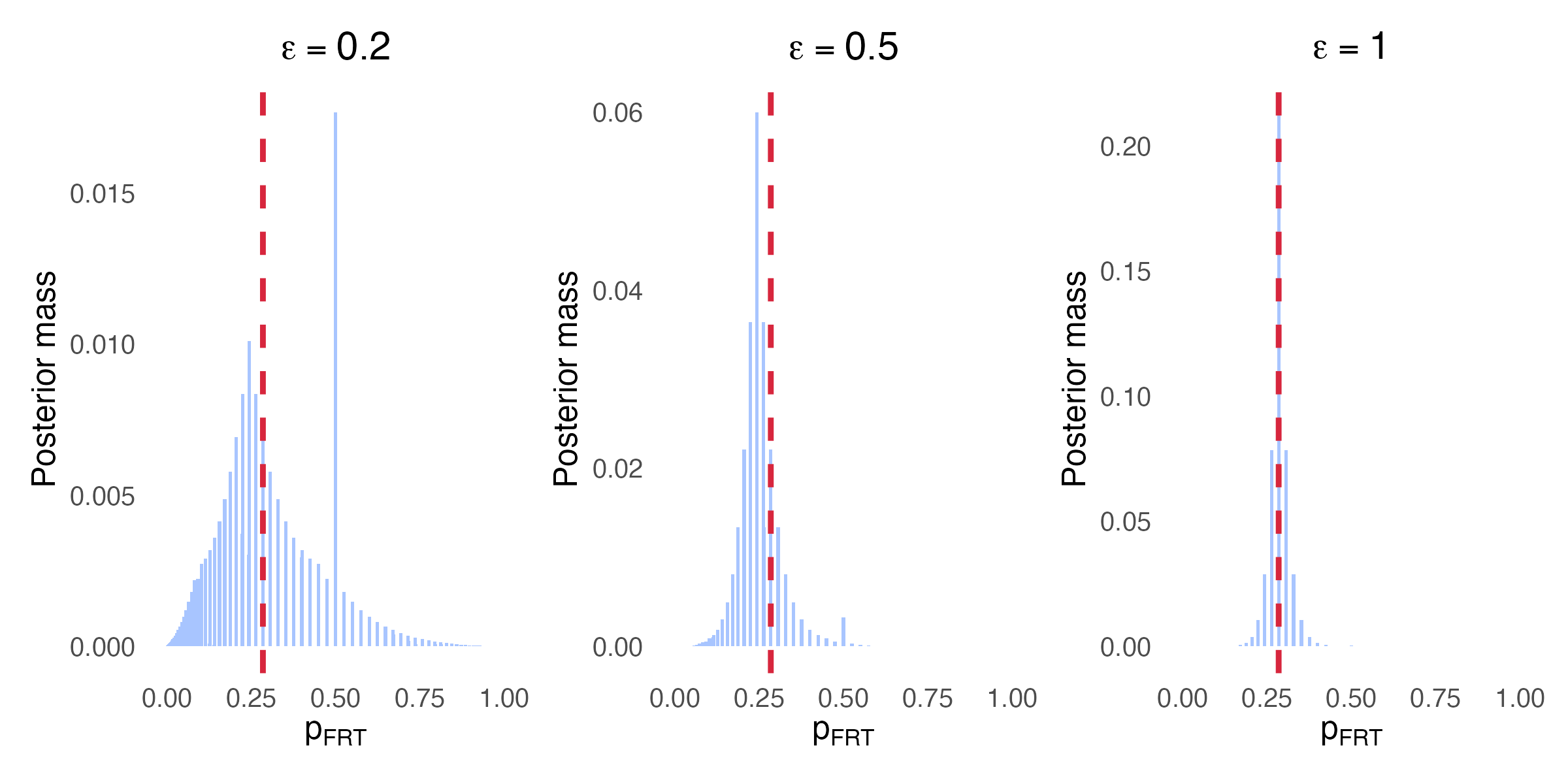}
    \caption{Posterior distributions from DP-FRT-Bayes under $\epsilon \in \{0.2, 0.5, 1\}$ for an illustrative dataset with $n_1=n_0=500$, $n_{11}=260$, $n_{01}=250$. The red dashed line indicates the non-private $p$-value, which equals 0.2846.}

    \label{fig:dpfrt_posteriors}
\end{figure}

To provide an illustrative example of DP-FRT-Bayes, we consider data with $n_1 = n_0 = 500$, $n_{11} = 260$, and $n_{01} = 250$.
We use a two-dimensional discrete uniform prior on $(m_{11}, m_{01})$.
Figure \ref{fig:dpfrt_posteriors} shows the posterior probability mass functions when $\epsilon \in \{0.2, 0.5, 1\}$.
When $\epsilon$ is small, the posterior is highly diffuse and exhibits a mode at $p_{\rm FRT} = 0.5$, reflecting greater uncertainty due to the DP noise.
As $\epsilon$ increases, the posterior concentrates more sharply around the non-private value.
In fact, one can show theoretically that the posterior probability that $p_{\mathrm{FRT}}$ falls far from the non-private value decays exponentially in $\epsilon$; see the supplementary material for a relevant theorem. We note that the simulations of Section \ref{sec:dp_sim} also reflect this concentration behavior. Indeed, for any particular setting, the analyst can use simulations like those in Section \ref{sec:dp_sim} to assess the concentration behavior for a specific design without expending privacy budget. Specifically, the analyst generates repeated privatized draws over a grid of plausible tables and compares the resulting DP-FRT-Bayes posterior with the FRT $p$-value for each table. 
We provide additional discussion of interpreting the posterior of $p_{\rm FRT}$ as a measure of evidence against Fisher's sharp null in Section \ref{sec:interpretBayes}.

Analysts could use other DP mechanisms in \eqref{eq:geom_mech_counts} in lieu of the Geometric mechanism. Provided that the noise distribution does not depend on the values of $(n_{11}, n_{01})$, i.e., it is a data-independent mechanism, it should be straightforward to compute the posterior distribution $p(m_{11}, m_{01} \mid \tilde{\bm{n}})$ for use in the DP-FRT-Bayes post-processing steps. To streamline the presentation, in this article we consider only the Geometric mechanism; see Section \ref{sec6} for additional discussion.

\subsubsection{Prior Specification for Counts}
\label{sec:3.2.1}

While conceptually analysts can posit any prior distribution for $(m_{11}, m_{01})$ for the Bayesian denoising, it is well known in DP that informative prior distributions can have a strong influence on the posterior \citep{KR2026}. Therefore, we expect many analysts will prefer to use default prior distributions with DP-FRT-Bayes.
A natural choice is the discrete uniform prior,
\begin{equation}
\pi_{\mathrm{unif}}(a,b)=\frac{1}{(n_1+1)(n_0+1)},\qquad
a\in\{0,\dots,n_1\},~b\in\{0,\dots,n_0\}.
\label{eq:prior_uniform}
\end{equation}
This choice facilitates computations and avoids privileging any specific configuration between the treatment and the control groups.

Another option arises from the independent binomial formulation commonly used in clinical practice. Specifically, we independently posit $m_{11}\sim\mathrm{Binom}(n_1,p_1)$ and $m_{01}\sim\mathrm{Binom}(n_0,p_0)$, together with independent priors $p_1\sim\mathrm{Beta}(\alpha_1,\beta_1)$ and $p_0\sim\mathrm{Beta}(\alpha_0,\beta_0)$. Integrating out $(p_1,p_0)$ yields the independent Beta-binomial prior for the counts,
\begin{equation}
\pi_{\mathrm{BB}}(a,b;\alpha_1,\beta_1,\alpha_0,\beta_0)
=\frac{\binom{n_1}{a} B(a+\alpha_1, n_1-a+\beta_1)}{B(\alpha_1,\beta_1)}
 \frac{\binom{n_0}{b} B(b+\alpha_0, n_0-b+\beta_0)}{B(\alpha_0,\beta_0)},
\label{eq:prior_bb}
\end{equation}
where the Beta function $B(x,y) = \Gamma(x)\Gamma(y)/\Gamma(x+y)$. This specification assumes that each group has its own baseline success rate and that the two groups are a priori independent. 
In particular, when $(\alpha_1,\beta_1)=(\alpha_0,\beta_0)=(1,1)$, $\pi_{\mathrm{BB}}$ reduces to the discrete uniform prior \eqref{eq:prior_uniform}. Thus, the uniform prior can be viewed as a special case of this independent Beta-binomial construction.
However, as \cite{DD2016} caution, the independent binomial specification is a convenient practice rather than an implication of Fisher's sharp null and the CRE design.

A third alternative uses independent binomial distributions with a common success rate across groups by setting $p_1=p_0=\theta$ with $\theta \sim \mathrm{Beta}(\alpha_c,\beta_c)$.
Marginalizing over $\theta$, we have
\begin{equation}
\pi_{\mathrm{CR}}(a,b;\alpha_c,\beta_c)
=\binom{n_1}{a}\binom{n_0}{b}
  \frac{B\big(a+b+\alpha_c, n_1+n_0-(a+b)+\beta_c\big)}{B(\alpha_c,\beta_c)}.
\label{eq:prior_common}
\end{equation}
When $(\alpha_c,\beta_c)=(1,1)$, the induced distribution of the FRT $p$-value is close to the uniform distribution on $(0,1)$ up to discreteness.

We recommend the discrete uniform prior in \eqref{eq:prior_uniform} as the default, since it requires no hyperparameter selection and treats all prior configurations as {\em a priori} equally likely.
Analysts with reliable domain knowledge about baseline event rates can use the Beta-binomial or common-rate priors with informative hyperparameters to incorporate such information. Our approach also accommodates hyperpriors on the Beta parameters as in \cite{JAGR2022}; the MCMC sampler in the supplementary material can be extended to sample from the resulting posterior distribution. 

In the supplementary material, we include sensitivity analyses using the uniform, Beta-binomial, and common-rate prior distributions. These analyses suggest that the posterior mean estimator of $p_{\mathrm{FRT}}$ is largely insensitive to the choice among these priors once $\epsilon$ and the sample sizes are large enough to support accurate inferences with the noisy counts.

\subsubsection{Monte Carlo Sampling and Aggregation}
\label{sec:3.2.2}

When both $n_1$ and $n_0$ are moderate, one can compute the posterior distribution for $p_{\mathrm{FRT}}$ by enumerating all $(a, b) \in \{0, \dots, n_1\} \times \{0, \dots, n_0\}$. This requires $O(n_1n_0)$ operations to compute and normalize the weights $w(a, b)$. For each pair, evaluating the tail sum in \eqref{eq:pvalue_ab} takes $O(\min\{n_1, n_0\})$ time, resulting in an overall complexity of $O(n_1n_0\times \min\{n_1, n_0\})$.

For large $(n_1, n_0)$, it can be preferable to sample from the posterior distribution $\gamma(a,b)$ rather than exhaustively evaluate the entire grid. A Metropolis-Hastings sampler for drawing from $\gamma(a,b)$ is provided in the supplementary material. When the prior distribution can be factorized as $\pi(a,b) = \pi_1(a)\pi_0(b)$, as in the case of \eqref{eq:prior_uniform} and \eqref{eq:prior_bb}, $\gamma(a,b)$ factorizes as $\gamma(a,b) = \gamma_{11}(a)\gamma_{01}(b)$, where
$\gamma_{11}(a) ~\propto~ \pi_1(a)  \kappa_\rho(\tilde n_{11} - a)$ and $\gamma_{01}(b) ~\propto~ \pi_0(b)  \kappa_\rho(\tilde n_{01} - b)$.
As a result, we can sample $a^{(r)} \sim \gamma_{11}$ and $b^{(r)} \sim \gamma_{01}$ independently at each iteration $r = 1, \dots, R$ of the sampler.
The cost of normalizing the two distributions is $O(n_1 + n_0)$; sampling $R$ pairs is $O(R)$; and, evaluating the $p$-value for each draw via \eqref{eq:pvalue_ab} costs $O(\min\{n_1, n_0\})$ per sample. The resulting total computational cost is $O(n_1 + n_0 + R \min\{n_1, n_0\})$.

Given Monte Carlo samples $\{(a^{(r)}, b^{(r)})\}_{r=1}^R$, posterior summaries are obtained by mapping each pair to $u^{(r)} = p_{a^{(r)} b^{(r)}}$ and aggregating over $\{u^{(r)}\}$. For example, the posterior mean can be approximated by $\tilde p_{\mathrm{mean}} = \sum_{r=1}^R u^{(r)}/R$. The posterior distribution can be approximated by the empirical distribution of $\{u^{(r)}\}$, from which the posterior median and credible sets can be extracted.
The MAP estimate can be obtained by tabulating the frequencies of the distinct values in $\{u^{(r)}\}$ and selecting the mode.

The Monte Carlo draws also enable posterior predictive generation of synthetic data and effect estimation. Specifically, each draw $(a^{(r)},b^{(r)})$ determines
$(n_{11}^{(r)}, n_{01}^{(r)}, n_{10}^{(r)}, n_{00}^{(r)}) =(a^{(r)}, b^{(r)}, n_1-a^{(r)}, n_0-b^{(r)}).$ These counts can be released as multiply-imputed, synthetic data \citep{Reiter2003}.
Based on these tables, analysts also can compute estimands such as
\begin{equation}\label{eq:others}
\tau^{(r)} = \frac{a^{(r)}}{n_1} - \frac{b^{(r)}}{n_0}, \qquad
\mathrm{RR}^{(r)} = \frac{a^{(r)} / n_1}{b^{(r)} / n_0}, \qquad
\mathrm{OR}^{(r)} = \frac{a^{(r)} (n_0 - b^{(r)})}{(n_1 - a^{(r)}) b^{(r)}}.
\end{equation}
Here, $\tau^{(r)}$ is the risk difference, $\mathrm{RR}^{(r)}$ is the risk ratio, and $\mathrm{OR}^{(r)}$ is the odds ratio. Their empirical distributions provide point estimates and credible intervals of treatment effects.

\subsection{Interpreting the Output of DP-FRT-Bayes}
\label{sec:interpretBayes}

The posterior distribution from DP-FRT-Bayes reflects uncertainty arising from DP noise as well as the analyst's prior beliefs about the distribution of the confidential counts. Because of the latter, DP-FRT-Bayes does not offer results of a pure randomization test. In a sense, this departure from pure randomization inference is a price the analyst pays for not having access to the confidential counts.

Despite this departure, the output of DP-FRT-Bayes can be used as a measure of evidence against Fisher's sharp null. In particular, its posterior distribution represents the analyst's beliefs about the value of the unobserved, confidential $p$-value, given the noisy counts and using the analyst's prior distribution over possible true counts. The analyst can interpret plausible ranges for $p_{\textrm{FRT}}$ to weigh the evidence against Fisher's sharp null. For example, consider two 95\% posterior intervals for $p_{\rm FRT}$, one with $B_1= (0.01, 0.25)$ and another with $B_2 = (0.001, 0.025)$. The analyst may find the evidence in $B_1$ to be too diffuse to discredit Fisher's sharp null, as the $p$-value from the confidential data plausibly could be small or large; the DP noise obfuscates its magnitude. On the other hand, the analyst may be convinced by $B_2$ that Fisher's sharp null is implausible, even accounting for the noise from ensuring DP. Of course, it is more demanding for the analyst to consider posterior uncertainty about the unobserved $p$-value rather than simply a single summary. Nonetheless, the posterior distribution offers the analyst means to fully consider the uncertainty introduced by the privacy protection mechanism in assessing the evidence against the sharp null.

\section{Decision Making Using DP-FRT-Bayes}
\label{sec4}

As originally proposed in \cite{Fisher1935}, the FRT does not adhere to strict accept-reject regions for testing Fisher's sharp null. Rather, the analyst interprets $p_{\rm FRT}$ as a measure of evidence against the sharp null; this is the perspective used in Section \ref{sec:interpretBayes}. However, we also can utilize outputs from the Bayesian denoising approach to construct more formal decision-making procedures for tests of Fisher's sharp null, as we show in this section.

Specifically, we utilize $\Pr\left(p_{\mathrm{FRT}} \le \alpha \mid \tilde{\bm n} \right)$, where $\alpha$ is a user-specified level.
This quantity represents the posterior probability that the FRT $p$-value does not exceed $\alpha$. We presume that the analyst decides to reject Fisher's sharp null when $p_{\mathrm{FRT}} \le \alpha$. Thus, $\Pr\left(p_{\mathrm{FRT}} \le \alpha \mid \tilde{\bm n} \right)$ summarizes the strength of evidence for rejection, incorporating the full posterior distribution.

In Section \ref{sec:4.1}, we continue to follow the inferential goal of recovering the decision from the non-private analysis. Here, we develop a Bayes rule that minimizes posterior risk of departing from the non-private decision about the sharp null. In Section \ref{sec:4.2}, we switch inferential goals to develop a significance test based on $\Pr\left(p_{\mathrm{FRT}} \le \alpha \mid \tilde{\bm n} \right)$ that controls Type I error rate.

\subsection{Bayes Risk-optimal Decision Framework}
\label{sec:4.1}

In developing a Bayes rule, our goal is to recover as faithfully as possible the significance label that would be obtained without DP noise.
Let the decision $\delta\in\{1,0\}$ indicate rejection or non-rejection, respectively.
For fixed threshold $\alpha\in(0,1)$, we consider the loss
\begin{equation}
L(\delta,p_{\rm FRT})=
\begin{cases}
\lambda_0, & \delta=1,~p_{\rm FRT}>\alpha,\\
\lambda_1, & \delta=0,~p_{\rm FRT}\le\alpha,\\
0, & (\delta=0,~p_{\rm FRT}>\alpha) \textrm{ or }
(\delta=1,~p_{\rm FRT}\le\alpha).
\end{cases}
\label{eq:bayes_loss_binary}
\end{equation}
Here, $\lambda_0>0$ is the loss of being too aggressive, i.e., we reject in cases where the non-private test would not. Conversely, $\lambda_1$ is the loss of being too conservative, i.e., we fail to reject in cases where the non-private test would.
The corresponding posterior risks given $\tilde{\bm n}$ are
\begin{align}
R(\delta = 1\mid  \tilde{\bm n}) &= \lambda_0 \Pr(p_{\rm FRT}>\alpha\mid  \tilde{\bm n}), \\
R(\delta = 0\mid  \tilde{\bm n}) &= \lambda_1 \Pr(p_{\rm FRT}\le\alpha\mid  \tilde{\bm n}).
\label{eq:posterior_risks_binary2}
\end{align}
The Bayes risk-optimal decision rule $\delta_{\mathrm{Bayes}}(\tilde{\bm n})
= \arg\min_{\delta \in \{0,1\}} \mathbb{E}[L(\delta, p_{\rm FRT}) \mid  \tilde{\bm n}]$ is
\begin{equation}
\begin{aligned}
\delta_{\mathrm{Bayes}}(\tilde{\bm n})
&=
\begin{cases}
1, & R(\delta = 1 \mid  \tilde{\bm n}) < R(\delta = 0 \mid  \tilde{\bm n}) \iff \Pr(p_{\rm FRT} \le \alpha \mid  \tilde{\bm n}) > \dfrac{\lambda_0}{\lambda_0 + \lambda_1}\\
0, & \text{otherwise}.
\end{cases}
\end{aligned}
\label{eq:bayes_rule}
\end{equation}
This rule can be viewed as a decision based on a posterior quantile, where the quantile level is determined by the trade-off between the losses of being overly aggressive and overly conservative relative to the non-private decision. When there is no justification for assigning different losses to the two types of discordance, i.e., $\lambda_0 = \lambda_1$, the decision rule reduces to one based on the posterior median. The release of $\delta_{\mathrm{Bayes}}(\tilde{\bm n})$ should be accompanied by $\Pr(p_{\rm FRT} \le \alpha \mid  \tilde{\bm n})$, which conveys the strength of evidence supporting the decision.

The binary Bayes rule can faithfully recover the decision that the FRT would reach on the confidential counts. As shown in a theorem in the supplementary material, the probability that the private Bayes decision differs from the confidential FRT decision generally is small, so the two agree with high probability even under moderate privacy budgets. An exception arises when the confidential $p$-value lies very close to the significance level $\alpha$, in which case relatively larger privacy budgets are needed for the private and non-private decisions to match.

\subsubsection{Decision Confidence and Abstention under Uncertainty}
\label{sec:4.1.1}

When $\Pr(p_{\rm FRT} \le \alpha \mid  \tilde{\bm n})$ is close to $1$ or $0$, there is strong evidence to reject or not, respectively. However, with a tight privacy budget or small sample size, the posterior for $p_{\mathrm{FRT}}$ can be diffuse due to the added noise, and this probability may lie near the binary Bayes threshold $\lambda_0/(\lambda_0+\lambda_1)$. In such cases, the decision rule might be unreliable.

In the spirit of Chow's rule \citep{Chow1957,Chow1970}, we equip the decision procedure with an explicit abstention option to deal with these situations. Specifically, we now let $\delta\in\{0,1,u\}$, with $u$ representing abstention, and $\lambda_u>0$ be the loss incurred for abstaining, such as the operational cost of deferring the decision.
The Bayes-optimal decision rule becomes
\begin{equation}
\begin{aligned}
\delta^*_{\rm Bayes}(\tilde{\bm n})
&=
\begin{cases}
1, & \Pr(p_{\rm FRT} \le \alpha \mid  \tilde{\bm n}) > \max\left\{\dfrac{\lambda_0}{\lambda_0 + \lambda_1},~1 - \dfrac{\lambda_u}{\lambda_0} \right\},\\
0, & \Pr(p_{\rm FRT} \le \alpha \mid  \tilde{\bm n}) < \min\left\{\dfrac{\lambda_0}{\lambda_0 + \lambda_1},~\dfrac{\lambda_u}{\lambda_1} \right\},\\
u, & \text{otherwise}.
\end{cases}
\end{aligned}
\label{eq:bayes_rule_trinary}
\end{equation}

\begin{figure}[tbp]
    \centering
    \includegraphics[width=0.9\linewidth]{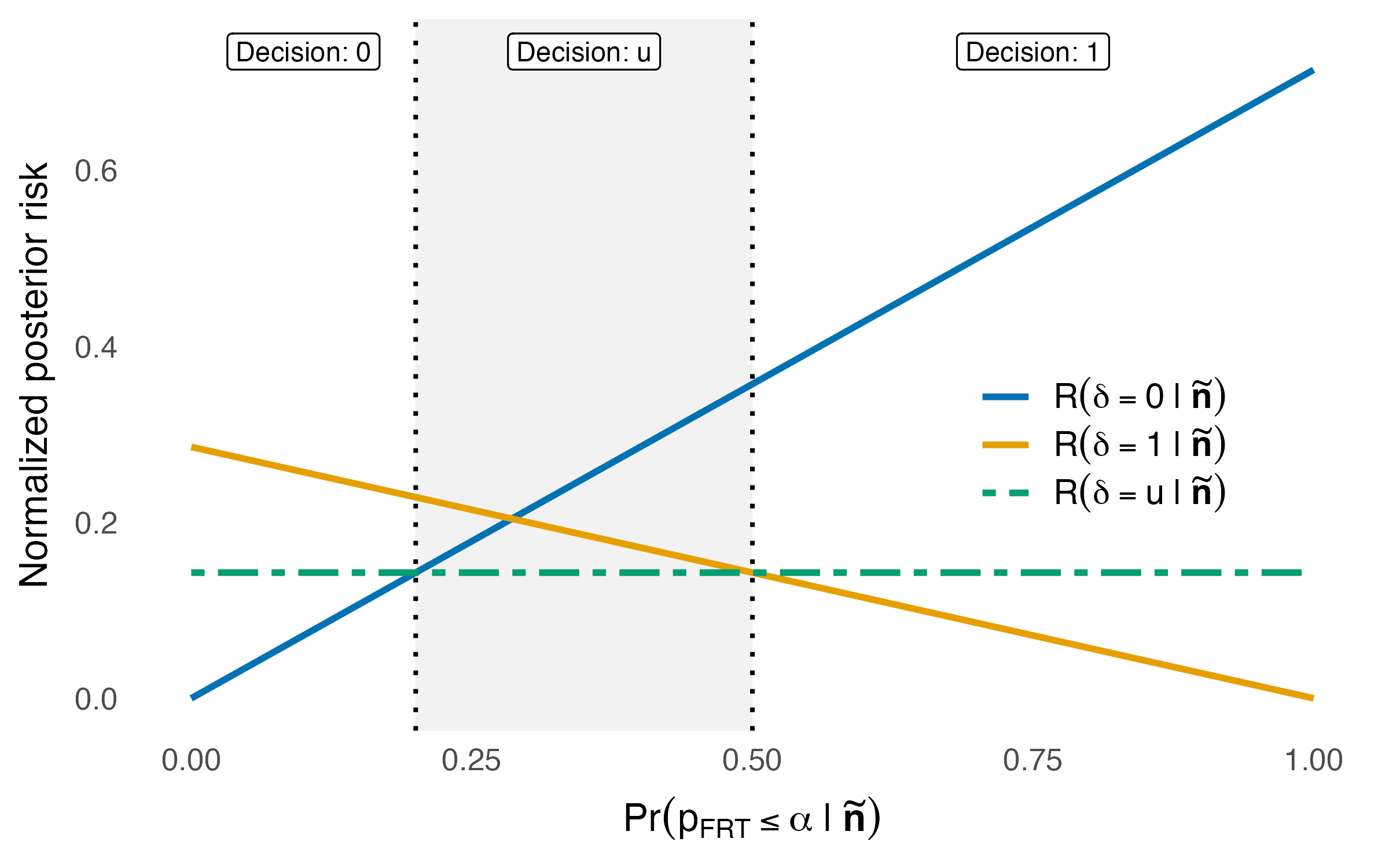}
    \caption{Illustration of the Bayes risk-optimal decision with an abstention option. The losses are specified as $\lambda_0=0.2$, $\lambda_1=0.5$, and $\lambda_u=0.1$. Blue, orange, and green lines correspond to the posterior risks normalized by $(\lambda_0+\lambda_1)$ for decisions $\delta=0$, $\delta=1$, and $\delta=u$, respectively. The shaded gray band marks the abstention region.}
    \label{fig:decision_plot}
\end{figure}

As evident in \eqref{eq:bayes_rule_trinary} and illustrated in Figure \ref{fig:decision_plot}, $\lambda_u$ determines the width of the abstention region.
Increasing $\lambda_u$ reduces the width, thereby decreasing the likelihood of abstention.
When $\lambda_u$ is large relative to $\lambda_0$ and $\lambda_1$, the abstention region degenerates, and \eqref{eq:bayes_rule_trinary} reduces to \eqref{eq:bayes_rule}.
Lemma \ref{lem:degenerate} provides a necessary and sufficient condition.

\begin{lemma}
\label{lem:degenerate}
The abstention option degenerates if and only if $\lambda_u \ge H/2$, where $H = (2\lambda_0\lambda_1)/(\lambda_0 + \lambda_1)$ is the harmonic mean
of $\lambda_0$ and $\lambda_1$.
\end{lemma}

\subsubsection{Sequential Decision under Additional Privacy Budget}
\label{sec:4.1.2}

Given an initial $\tilde{\bm{n}}=(\tilde n_{11},\tilde n_{01})$, analysts who reach $\delta^*_{\rm Bayes} =u$ may wish to improve decision certainty while still ensuring DP. In this section, we describe a sequential procedure to refine initial decisions. The basic idea is to augment the privacy budget and obtain additional noisy counts, which are used to update the posterior distributions and hence the decision.

Let $\epsilon_{\mathrm{plus}}>0$ be the added budget. We generate an independent, second noisy release
\begin{equation}
\tilde {\bm{n}}^{+}=\left(\tilde n^{+}_{11},\tilde n^{+}_{01}\right)
=(n_{11}+\eta^{+}_{11},~n_{01}+\eta^{+}_{01}), \qquad
\eta^{+}_{11},\eta^{+}_{01}\stackrel{\text{i.i.d.}}{\sim}
\mathrm{Geom}  \left(\exp(-\epsilon_{\mathrm{plus}})\right).
\end{equation}
The total privacy budget of the two releases is $\epsilon_{\mathrm{tot}}=\epsilon+\epsilon_{\mathrm{plus}}$
by the sequential composition property. Given $(\tilde {\bm{n}}^{+}, \tilde {\bm{n}})$, the posterior distribution becomes
\begin{align}
\Pr(m_{11}, m_{01} \mid \tilde{\bm{n}}, \tilde{\bm{n}}^{+})
&\propto~
\Pr(\tilde{\bm{n}}^{+} \mid m_{11}, m_{01})~
\Pr(\tilde{\bm{n}} \mid m_{11}, m_{01})~
\pi(m_{11}, m_{01}).
\label{eq:bayes_rule_sequential}
\end{align}
Let $\gamma^{+}(a,b) = \Pr(m_{11}=a,m_{01}=b\mid \tilde{\bm{n}},\tilde{\bm{n}}^{+})$. Let $\rho_{+}=\exp(-\epsilon_{\mathrm{plus}}/\Delta f)$.
By substituting the kernels from the Geometric mechanism defined in Section \ref{sec:2.2} with $\Delta f=1$, we have
\begin{equation}
\gamma^{+}(a,b)
=\frac{\pi(a,b)  \kappa_{\rho}(\tilde n_{11}-a)  \kappa_{\rho}(\tilde n_{01}-b)  \kappa_{\rho_{+}}(\tilde n^{+}_{11}-a)  \kappa_{\rho_{+}}(\tilde n^{+}_{01}-b)}
{\displaystyle \sum_{a'=0}^{n_1}\sum_{b'=0}^{n_0}
 \pi(a',b')  \kappa_{\rho}(\tilde n_{11}-a')  \kappa_{\rho}(\tilde n_{01}-b')
 \kappa_{\rho_{+}}(\tilde n^{+}_{11}-a')  \kappa_{\rho_{+}}(\tilde n^{+}_{01}-b')}
\label{eq:posterior_counts_tworeleases}
\end{equation}
for $(a,b) \in \{0,\dots,n_1\} \times \{0,\dots,n_0\}$.

We now investigate how allocating additional privacy budget can improve decision certainty.
Denote the abstention region as
$$
A = (t_{\mathrm{low}},~t_{\mathrm{high}})
=\left(\min \left\{\dfrac{\lambda_0}{\lambda_0+\lambda_1},~\dfrac{\lambda_u}{\lambda_1}\right\},~\max  \left\{\dfrac{\lambda_0}{\lambda_0+\lambda_1},~1-\dfrac{\lambda_u}{\lambda_0}\right\}\right),
$$
which is determined solely by the loss parameters.
By allocating extra privacy budget, the posterior evidence of rejection given $\alpha$ is updated from
$\Pr(p_{\mathrm{FRT}}\le\alpha\mid\tilde{\bm{n}})$ to $\Pr(p_{\mathrm{FRT}}\le\alpha\mid\tilde{\bm{n}},\tilde{\bm{n}}^{+})$,
which may become more concentrated around $0$ or $1$. Hence, although the width of the abstention region remains fixed, the probability that the updated evidence falls within this abstention region may decrease. In fact, the chance of remaining in the abstention region decays exponentially in $\epsilon_{\mathrm{plus}}$, as shown in the supplementary material.

We can find an upper bound on how likely the refined posterior is to exit the abstention region after the top-up release.
To do so, we adopt an information-theoretic perspective: conditioned on $\tilde {\bm{n}}$, the second release induces a channel whose ability to alter the posterior is governed by its total variation contraction under DP. By bounding this contraction and relating posterior movement to the distance from the abstention boundaries, we obtain an upper bound on the reduction of abstention probability.

To formalize this argument, let $\Psi=\Pr(p_{\mathrm{FRT}}\le\alpha\mid\tilde{\bm{n}})$ and $\Psi^{+}=\Pr(p_{\mathrm{FRT}}\le\alpha\mid\tilde{\bm{n}},\tilde{\bm{n}}^{+}).$
Let $S = \{0,\dots,n_1\}\times\{0,\dots,n_0\}$ be the support of $(m_{11},m_{01})$ and, for $h\in\{0,1\}$, let $S_h = \{(a,b)\in S:  \mathbf{1}(p_{ab} \le \alpha)=h\}$ be the sets corresponding to acceptance ($h=0$) and rejection ($h=1$).
Define the data-adaptive separation between these two classes by
$$
\Delta(\tilde{\bm n}; \epsilon_{\rm plus})
=
\sum_{(a,b)\in S_1}\sum_{(a',b')\in S_0}
\mu_1(a,b)\mu_0(a',b')~
s\left(\epsilon_{\rm plus}~\left(|a-a'|+|b-b'|\right)\right),$$
where $\mu_h(a,b)=\Pr\left(m_{11}=a,m_{01}=b\mid (m_{11},m_{01})\in S_h, \tilde{\bm n}\right)$ and $s(x)=\tanh(x/2)$. This quantity can be computed without accessing any private data.
Theorem \ref{thm:upper_abstention} presents the upper bound.

\begin{theorem}
\label{thm:upper_abstention}
Let $A=(t_{\mathrm{low}},t_{\mathrm{high}})$ be the abstention region with $0<t_{\mathrm{low}}<t_{\mathrm{high}}<1$, and define $r(\Psi) = \min\{\Psi-t_{\mathrm{low}},~t_{\mathrm{high}}-\Psi\}$ whenever $\Psi\in A$.
Then, for every $\epsilon_{\mathrm{plus}}>0$,
\begin{equation}
\label{eq:cond_escape}
\Pr\left(\Psi^{+}\notin A \mid  \Psi\in A\right)
 \le
2~
\mathbb E \left(
\frac{\Psi(1-\Psi)}{r(\Psi)}~
\Delta(\tilde{\bm n}; \epsilon_{\rm plus})
 ~\Bigg|~  \Psi\in A
\right).
\end{equation}
\end{theorem}

The quantity $s(\epsilon)$ is the standard conversion from a privacy budget $\epsilon$ to an upper bound on the adversary's distinguishing advantage between neighboring datasets, measured by the total variation distance between their output distributions under an $\epsilon$-DP mechanism \citep{GI2024}.
Theorem \ref{thm:upper_abstention} shows that, under the additional privacy budget $\epsilon_{\mathrm{plus}}$, the probability of exiting the abstention region is controlled by a data-adaptive weighted average of the total variation bounds $s(\cdot)$ over acceptance and rejection classes.

Corollary \ref{cor:budget_escape} quantifies a necessary lower bound on $\epsilon_{\mathrm{plus}}$ for the refined posterior to escape the abstention region with high probability.
\begin{corollary}
\label{cor:budget_escape}
Let $A=(t_{\mathrm{low}},t_{\mathrm{high}})$ be the abstention region with
$0<t_{\mathrm{low}}<t_{\mathrm{high}}<1$.
Fix a confidence level $1-\xi\in(0,1)$.
For $\Psi\in A$, if the refined posterior $\Psi^+$ satisfies $\Pr\left(\Psi^+\notin A \mid \tilde{\bm n}\right) \ge 1-\xi$, then
\begin{equation}
\label{eq:budget_conditional_eps}
\epsilon_{\mathrm{plus}} \ge \inf\left\{\epsilon>0:
\Delta(\tilde{\bm n};\epsilon)\ \ge\
\frac{(1-\xi)~r(\Psi)}{2 \Psi(1-\Psi)}
\right\},
\end{equation}
where $r(\Psi) = \min\{\Psi-t_{\mathrm{low}},~t_{\mathrm{high}}-\Psi\}$, and $\Delta(\tilde{\bm n}; \epsilon)$ is defined as in Theorem \ref{thm:upper_abstention}.
\end{corollary}

The necessary $\epsilon_{\mathrm{plus}}$ depends on $\xi$, $\Psi$, and $\Delta$. Together, these quantities capture (i) how different the acceptance and rejection classes are under the current release (through $\Delta$), (ii) how confident a decision the analyst requires (through $\xi$), and (iii) how close the current posterior is to the abstention boundary (through $r(\Psi)$). When the acceptance and rejection configurations are less distinguishable under the top-up mechanism (smaller $\Delta$ at a given $\epsilon_{\mathrm{plus}}$), when higher confidence is required (smaller $\xi$), or when the current posterior lies deeper inside the abstention region (larger $r(\Psi)$), a larger additional privacy budget is needed to push the decision outside the abstention region. A detailed algorithm for calculating the lower bound in \eqref{eq:budget_conditional_eps} is presented in the supplementary material.

The supplementary material also establishes a fidelity guarantee for the refined ternary decision after a top-up release, where the refined decision is defined by evaluating \eqref{eq:bayes_rule_trinary} at $\Psi^{+}$.
Specifically, we bound the probability of a committed misclassification after abstention---the event that the first stage abstains and the refined stage then commits to a decision differing from the confidential FRT decision---and show that it remains small and decreases in the total privacy budget. 
The decision rules and sequential top-up procedure are evaluated empirically in Section \ref{sec:causal_sim} and the supplementary material.

In practice, when the decision rule in \eqref{eq:bayes_rule_trinary} abstains, the analyst can select $\epsilon_{\mathrm{plus}}$ larger than the lower bound in \eqref{eq:budget_conditional_eps} for a pre-specified confidence level, e.g., $\xi = 0.05$ for $95\%$ certainty. If the decision still abstains after generating $\tilde{\bm n}^{+}$, the analyst can repeat the procedure (if it is acceptable to increase the total privacy budget), or terminate and report the decision as inconclusive.
This strategy spends privacy budget adaptively and only when necessary, allowing practitioners to decide at each stage whether to continue.

Before committing this budget, the analyst can assess its likely benefit through simulation without expending privacy budget. Conditional on the released $\tilde{\bm n}$, candidate tables $(a,b)$ are drawn from the current posterior. For each such table, a hypothetical top-up release $\tilde{\bm n}^{+}$ is generated; the refined rule $\delta_{\mathrm{Bayes}}^{+}$ is applied; and, the analyst records whether the rule commits to a decision and whether that decision matches the FRT decision $\mathbf{1}(p_{ab}\le\alpha)$ at $(a,b)$. Sweeping over $\epsilon_{\mathrm{plus}}$ shows how often the top-up resolves abstention and how faithful the committed decision is, complementing the lower bound in Corollary \ref{cor:budget_escape} and the fidelity guarantee in the supplementary material.

\subsection{Frequentist-calibrated Decision Framework}
\label{sec:4.2}

We next describe how to use $\Pr(p_{\mathrm{FRT}}\le\alpha\mid\tilde{\bm{n}})$ in a significance test that controls frequentist Type I error.
Let 
\begin{equation}
\delta_{\rm Freq}=
\begin{cases}
1,& \Pr(p_{\rm FRT} \le \alpha \mid  \tilde{\bm n})>t^*,\\
0,& \text{otherwise},
\end{cases}
\label{eq:freq_rule}
\end{equation}
where the threshold $t^*\in[0,1]$ is chosen as the smallest value such that, under $H_0^{\rm F}$, $\Pr(\delta_{\rm Freq} = 1) \le \alpha_{\mathrm{Freq}}$, using $\alpha_{\mathrm{Freq}}$ to denote the target Type I error level. Specifically, let $F_\Psi$ be the distribution of $\Pr(p_{\rm FRT} \le \alpha \mid  \tilde{\bm n})$ under $H_0^{\rm F}$. Its $(1-\alpha_{\mathrm{Freq}})$-quantile,
\begin{equation}
t^*=F_\Psi^{-1}(1-\alpha_{\mathrm{Freq}})=\inf\{t\in[0,1]:F_\Psi(t)>1-\alpha_{\mathrm{Freq}}\},
\label{eq:freq_quantile}
\end{equation}
yields the desired calibrated threshold.
Equivalently, \eqref{eq:freq_rule} can be viewed as a test of Fisher's sharp null based on the test statistic $\Psi=\Pr(p_{\rm FRT} \le \alpha \mid  \tilde{\bm n})$.
This use of $\Psi$ accords with a calibrated Bayes perspective \citep{Little2006}: we use a posterior probability as the test statistic to leverage Bayesian uncertainty quantification and calibrate $t^*$ to control frequentist Type I error. 

To determine the critical value $t^*$ defined in \eqref{eq:freq_quantile}, the key step is to derive $F_\Psi$, the distribution of $\Psi$ under the sharp null hypothesis. However, the construction of $F_\Psi$ depends on the total number of successes, $n_{+1} = n_{11} + n_{01}$, which is also subject to privatization and therefore unknown. Two calibration strategies are proposed below.

\subsubsection{Worst-case Calibration}
\label{sec:4.2.1}

We first construct a threshold that is valid for all possible total numbers of successes.
For each $K\in\{0,\dots,n\}$, let $Q_K$ denote the probability mass function of $\tilde{\bm n} = (\tilde n_{11}, \tilde n_{01})$ under $H_0^{\rm F}$ with total number of successes equal to $K$, given by
\begin{equation}
Q_K(a,b)=\sum_{j=\max\{0,K-n_0\}}^{\min\{n_1,K\}}
\frac{\binom{K}{j}\binom{n-K}{n_1-j}}{\binom{n}{n_1}}
 \kappa_{\rho}(a-j) \kappa_{\rho}(b-(K-j))
\label{eq:QK}
\end{equation}
for $(a,b) \in \mathbb{Z}^2$.
We define $F_\Psi^{(K)}$ as the cumulative distribution function of $\Psi=\Pr(p_{\mathrm{FRT}}\le\alpha\mid \tilde{\bm{n}})$ when $\tilde{\bm n}\sim Q_K$, that is,
\begin{equation}
F_\Psi^{(K)}(t)=\sum_{(a,b)\in\mathbb{Z}^2} Q_K(a,b) \bm 1 (\Psi(a,b)\le t),\qquad t\in[0,1],
\label{eq:FRK}
\end{equation}
and denote $t_K = \inf\{t : F_\Psi^{(K)}(t) > 1 - \alpha_{\mathrm{Freq}}\}$ as the right-continuous $(1 - \alpha_{\mathrm{Freq}})$-quantile of $F_\Psi^{(K)}$.
We set the threshold as $t_{\mathrm{LFC}}^* = \max_{K \in \{0, \dots, n\}} t_K$. This leads to the decision rule
\begin{equation}
\delta_{\mathrm{LFC}}(\tilde{\bm n}) =
\begin{cases}
1, & \Pr(p_{\rm FRT} \le \alpha \mid  \tilde{\bm n}) > t_{\mathrm{LFC}}^*, \\
0, & \text{otherwise}.
\end{cases}
\label{eq:tLFC_Psirule}
\end{equation}
Theorem \ref{thm:uncond} ensures Type I error control under \eqref{eq:tLFC_Psirule}.
\begin{theorem}
\label{thm:uncond}
Under the sharp null $H_0^{\rm F}$, we have $\Pr \left(\delta_{\mathrm{LFC}}(\tilde{\bm n})=1\right)\le \alpha_{\mathrm{Freq}}$, where the probability averages over randomization and the privacy mechanism.
\end{theorem}

In practice, the threshold can be approximated by simulation. For any value of $K \in \{0, \dots, n\}$, we take a random draw of the number of successes in the treated group, $a_{11} \sim \mathrm{Hypergeometric}(n, K, n_1)$, and set $b_{01} = K - a_{11}$.
We apply the Geometric mechanism to $(a_{11}, b_{01})$ to generate $(\tilde a_{11}, \tilde b_{01})$. Using these simulated noisy counts, we compute $\Pr(p_{\rm FRT} \le \alpha \mid  \tilde a_{11}, \tilde b_{01})$. We repeat this process many times and approximate $F_\Psi^{(K)}$ using its empirical distribution. We estimate $t_K$ using the right-continuous empirical quantile of this simulated distribution.
We perform this simulation for each $K=0, \dots, n$ and set $t^*_{\mathrm{LFC}} = \max_K t_K$.
A detailed algorithm is presented in the supplementary material.

\subsubsection{Data-adaptive Calibration with Confidence Sets}
\label{sec:4.2.2}

Instead of the worst case, we can use a data-adaptive confidence set for $n_{+1}$.
For each $K$, order the lattice points $(a,b) \in \mathbb{Z}^2$ by decreasing $Q_K(a,b)$. We take the smallest set $A_K$ whose total mass is at least $1-\zeta$, with $\zeta\in(0, \alpha_{\rm Freq})$.
Define the $(1-\zeta)$-confidence set, $C_{1-\zeta}(\tilde{\bm n})=\{K \in \{0,\dots,n\}: \tilde{\bm n} \in A_K\}$.
For $\alpha'=\alpha_{\rm Freq} - \zeta$, define $t_K'=\inf\{t:F_\Psi^{(K)}(t)>1-\alpha'\}$ as the right-continuous $(1-\alpha')$ quantile of $F_\Psi^{(K)}$ from \eqref{eq:FRK}. Letting $t^*_{\mathrm{Neyman}}=\max_{K\in C_{1-\zeta}(\tilde{\bm n})} t_K'$, with the convention $t^*_{\mathrm{Neyman}}=t^*_{\mathrm{LFC}}$ when $C_{1-\zeta}(\tilde{\bm n})=\emptyset$, the decision rule is
\begin{equation}
\delta_{\mathrm{Neyman}}(\tilde{\bm n}) =
\begin{cases}
1, & \Pr(p_{\rm FRT} \le \alpha \mid  \tilde{\bm n}) > t^*_{\mathrm{Neyman}}, \\
0, & \text{otherwise}.
\end{cases}
\label{eq:neyman_rule_case}
\end{equation}
Intuitively, \eqref{eq:neyman_rule_case} spends at most $\zeta$ on coverage of $n_{+1}$ and uses the least favorable threshold within the confidence set, which yields Type I error control while typically reducing conservativeness.
For example, one can choose $\zeta = 0.05$ to make $C_{1-\zeta}(\tilde{\bm n})$ a $95\%$ confidence set.
Lemma \ref{lem:neyman} and Theorem \ref{thm:neyman} formally state the frequentist properties of the data-adaptive procedure.

\begin{lemma}
\label{lem:neyman}
For each $K$, we have $\Pr_K(K\in C_{1-\zeta}(\tilde{\bm n}))=\Pr_K(\tilde{\bm n} \in A_K)\ge 1-\zeta,$ where $\Pr_K$ denotes probability under $Q_K$.
\end{lemma}

\begin{theorem}
\label{thm:neyman}
Fix $\zeta \in (0,\alpha_{\mathrm{Freq}})$ and set $\alpha'=\alpha_{\mathrm{Freq}}-\zeta$. Under the sharp null $H_0^{\rm F}$, we have $\Pr \left(\delta_{\mathrm{Neyman}}(\tilde{\bm n})=1\right)\le \alpha_{\mathrm{Freq}}$.
\end{theorem}

We can approximate $t^*_{\mathrm{Neyman}}$ using simulation, proceeding as in the simulation in Section \ref{sec:4.2.1}. However, after obtaining $(\tilde a_{11}, \tilde b_{01})$, we use their empirical distribution to approximate $Q_K$.
We construct $A_K$ as the smallest empirical set with probability mass at least $1 - \zeta$. We form the set $C_{1 - \zeta}(\tilde{\bm{n}})$, estimate $F_\Psi^{(K)}$, compute $t_K'$, and finally set $t^*_{\mathrm{Neyman}}$ to determine the decision rule. A detailed algorithm is presented in the supplementary material.

As noted in Section \ref{sec1}, analysts who wish to use significance testing to make decisions about Fisher's sharp null can use other test statistics besides $\Psi$ that are functions of $\tilde{\bm n}$; the same calibration pipeline applies. In the supplementary material, we present one such test statistic based on the noisy counts themselves and compare its power properties to those of the test based on $\Psi$ via simulation. The two tests have similar power profiles. 
The supplementary material also includes some additional heuristic interpretations of $\Psi$ as a test statistic.

\section{Simulation Studies and Data Analysis}
\label{sec5}

This section presents simulation studies that illustrate the performance of DP-FRT-Bayes.
We also apply DP-FRT-Bayes to the ADAPTABLE trial data \citep{JMW2021}.
Code for reproducing the results is available at \url{https://github.com/qy-sun/dp_frt}.

\subsection{Assessing DP-FRT-Bayes Estimates of \texorpdfstring{$p$-value}{p-values}}
\label{sec:dp_sim}

We first evaluate the ability of DP-FRT-Bayes to estimate the FRT $p$-value. To do so, we fix twelve sets of $(n_{11}, n_{10}, n_{01}, n_{00})$ at the values displayed in Table \ref{tab:setting1}.
For each dataset, we generate 1000 independent draws of privatized $\tilde {\bm{n}}$ and estimate the posterior distribution of $p_{\textrm{FRT}}$ using the methods from Section \ref{sec:3.2}. Results are based on the uniform prior distribution for the true counts.

\begin{table}[t]
\spacingset{1.4}
\centering
\caption{Settings for simulations in Section \ref{sec:dp_sim} of how accurately DP-FRT-Bayes estimates the non-private $p$-value, $p_{\textrm{FRT}}$.}
\begin{tabular*}{\linewidth}{@{\extracolsep{\fill}}lccccc}
\toprule
\textbf{~~~~~~~~~~Case}  & $n_{11}$ & $n_{10}$ & $n_{01}$ & $n_{00}$ & $p_{\rm FRT}$ \\
\midrule

\multirow{3}{*}{~~~Case 1: No effect}
   & 25  & 25  & 25  & 25  & 0.579 \\
  & 125 & 125 & 125 & 125 & 0.536 \\
  & 250 & 250 & 250 & 250 & 0.525 \\
\midrule

\multirow{3}{*}{~~~Case 2: Small effect}
  & 28  & 22  & 25  & 25  & 0.344 \\
  & 138 & 112 & 125 & 125 & 0.141 \\
  & 275 & 225 & 250 & 250 & $6.43\times10^{-2}$ \\
\midrule

\multirow{3}{*}{~~~Case 3: Medium effect}
 & 30  & 20  & 25  & 25  & 0.211 \\
  & 150 & 100 & 125 & 125 & $1.54\times10^{-2}$ \\
  & 300 & 200 & 250 & 250 & $9.14\times10^{-4}$ \\
\midrule

\multirow{3}{*}{~~~Case 4: Large effect}
  & 32 & 18 & 25 & 25 & 0.113 \\
   & 162 & 88 & 125 & 125 & $5.54\times10^{-4}$ \\
  & 325 & 175 & 250 & 250 & $1.05\times10^{-6}$ \\
\bottomrule
\end{tabular*}
\label{tab:setting1}
\end{table}

\begin{table}[t]
\spacingset{1.4}
\centering
\caption{Simulated biases of posterior means along with coverage rates (\%) and average widths of 95\% credible sets in Cases 1--4 when $\epsilon \in \{0.2, 0.5, 1\}$.}
\begin{tabular*}{\linewidth}{@{\extracolsep{\fill}}l ccc ccc ccc}
\toprule
 & \multicolumn{3}{c}{$\epsilon=0.2$}
 & \multicolumn{3}{c}{$\epsilon=0.5$}
 & \multicolumn{3}{c}{$\epsilon=1$} \\
\textbf{~~~~~Case} & Bias & Cov & Width & Bias & Cov & Width & Bias & Cov & Width \\
\midrule

\multicolumn{10}{l}{\textbf{(a) $n=100$}} \\
~1: No effect   & -0.059 & 93.5 & 0.938 & -0.022 & 95.0 & 0.797 & -0.007 & 95.0 & 0.538 \\
~2: Small       & 0.093 & 94.2 & 0.935 & 0.033 & 94.8 & 0.780 & 0.025 & 95.6 & 0.520 \\
~3: Medium      & 0.174 & 94.2 & 0.920 & 0.089 & 96.0 & 0.721 & 0.024 & 95.1 & 0.431 \\
~4: Large       & 0.211 & 94.8 & 0.902 & 0.095 & 95.1 & 0.621 & 0.030 & 96.1 & 0.324 \\

\midrule
\multicolumn{10}{l}{\textbf{(b) $n=500$}} \\
~1: No effect   & -0.024 & 93.6 & 0.835 & -0.005 & 94.0 & 0.500 & 0.002 & 95.1 & 0.275 \\
~2: Small       & 0.099 & 94.3 & 0.704 & 0.027 & 93.7 & 0.323 & 0.008 & 95.6 & 0.162 \\
~3: Medium      & 0.077 & 93.4 & 0.406 & 0.012 & 94.9 & 0.085 & 0.002 & 95.3 & 0.032 \\
~4: Large       & 0.020 & 94.1 & 0.132 & 0.001 & 94.0 & 0.009 & 0.000 & 95.7 & 0.002 \\

\midrule
\multicolumn{10}{l}{\textbf{(c) $n=1000$}} \\
~1: No effect   & -0.004 & 92.8 & 0.732 & 0.003 & 93.6 & 0.376 & -0.001 & 96.1 & 0.198 \\
~2: Small       & 0.066 & 93.6 & 0.419 & 0.011 & 96.1 & 0.137 & 0.003 & 96.7 & 0.065 \\
~3: Medium      & 0.009 & 93.4 & 0.058 & 0.001 & 95.0 & 0.006 & 0.000 & 96.5 & 0.002 \\
~4: Large       & 0.000 & 91.5 & 0.003 & 0.000 & 94.7 & 0.000 & 0.000 & 94.3 & 0.000 \\

\bottomrule
\end{tabular*}
\label{tab:DP-FRT-Bayes}
\end{table}

Table \ref{tab:DP-FRT-Bayes} reports the simulated biases of the posterior mean $\tilde p_{\textrm{mean}}$, as well as the percentages of the 1000 intervals based on the $95\%$ credible sets that cover the non-private $p_{\textrm{FRT}}$ and the average width of those intervals.
The posterior means offer reasonable estimates of $p_{\textrm{FRT}}$, particularly when $n\ge 500$ and $\epsilon \ge 0.5$. When $\epsilon=0.2$, the posterior mean tends to be inaccurate unless $n=1000$. The coverage of the intervals is slightly below 95\% when $\epsilon = 0.2$ and near 95\% when $\epsilon \ge 0.5$.
When $n=100$, the intervals tend to be wide, suggesting that DP-FRT-Bayes at this small sample size is not particularly informative. Indeed, when $\epsilon = 0.2$ and $n=100$, the intervals typically cover nearly the full support of $p_{\textrm{FRT}}$. This is not surprising, as the noise from DP often can swamp signals when $\epsilon$ and $n$ are small. When $n=1000$, the interval widths decline, so that even with this modest sample size DP-FRT-Bayes offers reasonably precise inferences. We note that across all $n$, the interval widths are largest for the no effect scenarios. In these situations, the noisy counts can be consistent with a relatively wide range of plausible true counts, which in turn correspond to a relatively wide range of plausible $p$-values.

\begin{figure}[t]
    \centering
    \includegraphics[width=0.9\linewidth]{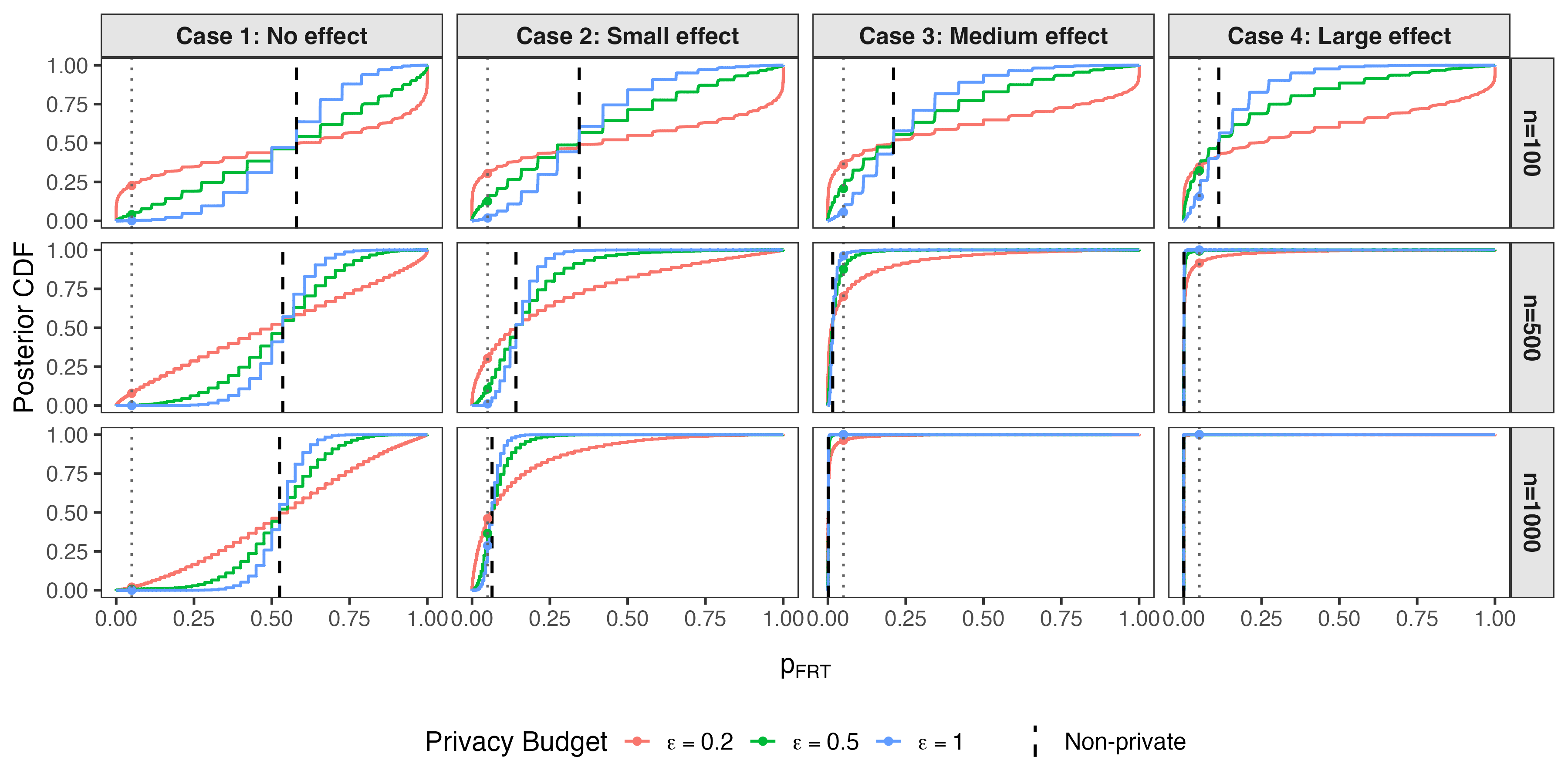}
    \caption{Posterior CDF of the DP-FRT-Bayes $p$-value from 100 runs in Cases 1--4 when $n\in\{100,500,1000\}$ for $\epsilon\in\{0.2,0.5,1\}$.
    Dotted gray vertical line indicates the nominal significance level $\alpha=0.05$.
    Dashed black vertical line marks the non-private FRT $p$-value.
    }

    \label{fig3}
\end{figure}

As a further illustration of the behavior of DP-FRT-Bayes, Figure \ref{fig3} displays the cumulative distribution function (CDF) of all posterior draws of the DP-FRT-Bayes $p$-value from 100 runs of each scenario.
The CDF shifts upward when the treatment effect size increases, leading to higher values of $\Psi$ and stronger evidence against the null.

The supplementary material includes additional simulation results. They show that the posterior mean and median perform similarly and are slightly preferable to the MAP as a point estimator of $p_{\textrm{FRT}}$. They also show that results are similar under the independent Beta-binomial prior of \eqref{eq:prior_bb} and the common success rate Beta prior of \eqref{eq:prior_common}.

\begin{table}[t]
\spacingset{1.4}
\centering
\caption{Simulation settings for repeated sampling studies of causal inferences in Section \ref{sec:causal_sim}.
$N_{11} = \sum_{i=1}^n Y_i(1)$,
$N_{10} = \sum_{i=1}^n (1-Y_i(1))$,
$N_{01} = \sum_{i=1}^n Y_i(0)$,
$N_{00} = \sum_{i=1}^n (1-Y_i(0))$.}
\begin{tabular*}{\linewidth}{@{\extracolsep{\fill}}lcccc}
\toprule
\textbf{~~~~~~~~~~~~Case} & $N_{11}$ & $N_{10}$ & $N_{01}$ & $N_{00}$ \\
\midrule

\multirow{3}{*}{~~~Case 5: No effect ($\tau = 0$)}
   & 50  & 50  & 50  & 50  \\
   & 250 & 250 & 250 & 250 \\
   & 500 & 500 & 500 & 500 \\
\midrule

\multirow{3}{*}{~~~Case 6: Small effect ($\tau = 0.05$)}
  & 55  & 45  & 50  & 50 \\
  & 275 & 225 & 250 & 250 \\
  & 550 & 450 & 500 & 500 \\
\midrule

\multirow{3}{*}{~~~Case 7: Medium effect ($\tau = 0.1$)}
   & 60  & 40  & 50  & 50  \\
   & 300 & 200  & 250 & 250 \\
   & 600 & 400 & 500 & 500 \\
\midrule

\multirow{3}{*}{~~~Case 8: Large effect ($\tau = 0.15$)}
  & 65  & 35  & 50  & 50  \\
  & 325 & 175 & 250 & 250 \\
  & 650 & 350 & 500 & 500 \\
\bottomrule
\end{tabular*}
\label{tab:setting2}
\end{table}

\subsection{Evaluating Decision Rules under DP-FRT-Bayes}
\label{sec:causal_sim}

We next consider the decision rules described in Section \ref{sec4}.
We fix twelve sets of potential outcomes $\{Y_i(1), Y_i(0)\}_{i=1}^n$ representing different study populations and treatment effects, summarized in Table \ref{tab:setting2}. In each study population, we randomly assign $n_1$ out of $n$ units to the treatment group. This results in a realized outcome table, $(n_{11}, n_{10}, n_{01}, n_{00})$. We then apply DP-FRT-Bayes. We compute the Bayes-optimal decision using loss parameters $(\lambda_0, \lambda_1, \lambda_u) = (1, 1, 0.025)$ and $\alpha = 0.05$.
We repeat this process 1000 times for each scenario. The supplementary material includes analogous results for the frequentist decision rule of Section \ref{sec:4.2}, as well as simulations illustrating the use of DP-FRT-Bayes for sequential decision-making described in Section \ref{sec:4.1.2}.

\begin{figure}[t]
    \centering
    \includegraphics[width=0.9\linewidth]{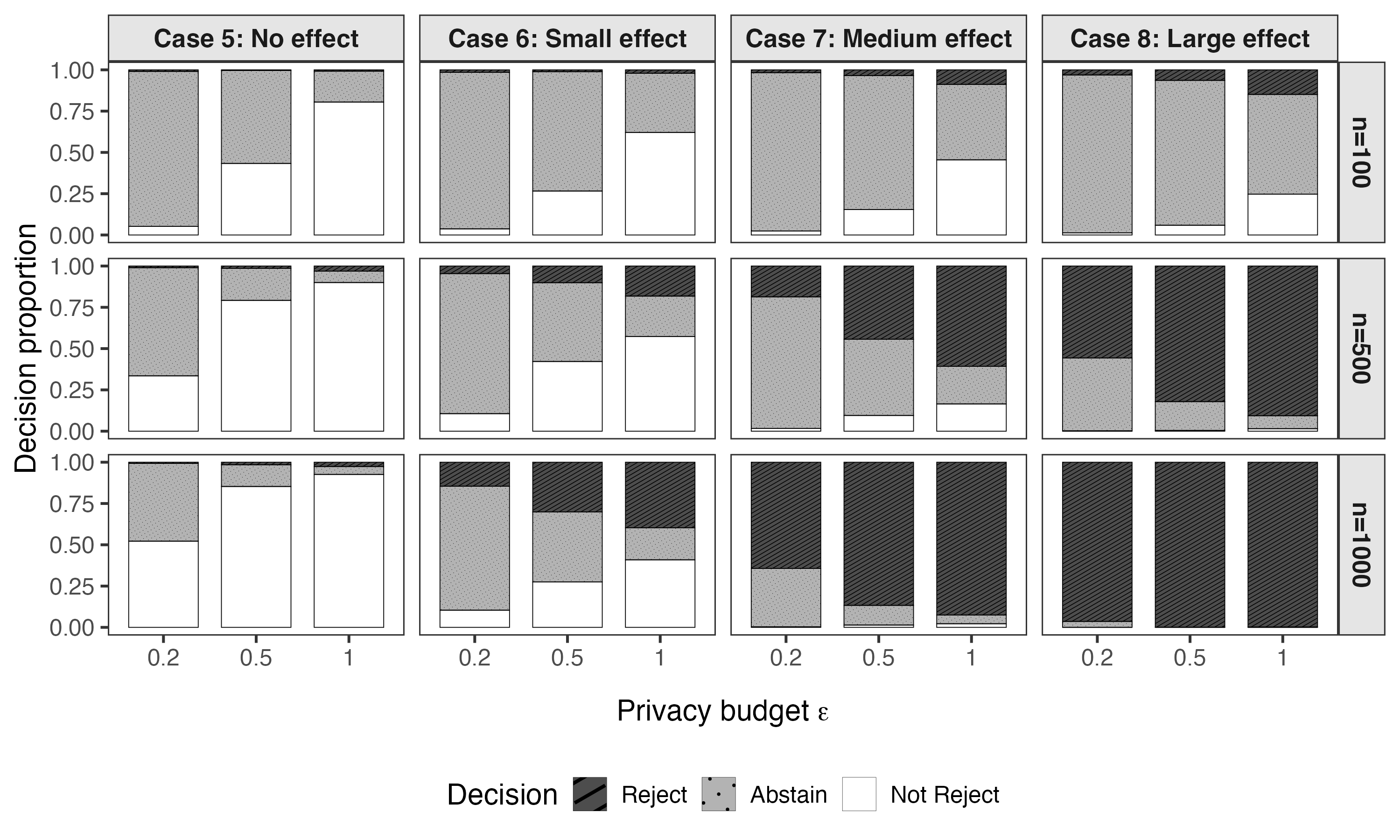}

    \caption{Relative frequencies of each decision from the Bayes risk-optimal rule with abstention for $\epsilon\in\{0.2,0.5,1\}$ in Cases 5--8 for $n\in\{100,500,1000\}$ and $(\lambda_0,\lambda_1,\lambda_u)=(1,1,0.025)$.}

    \label{fig:Bayes_stack}
\end{figure}

\begin{figure}[t]
    \centering
    \includegraphics[width=0.9\linewidth]{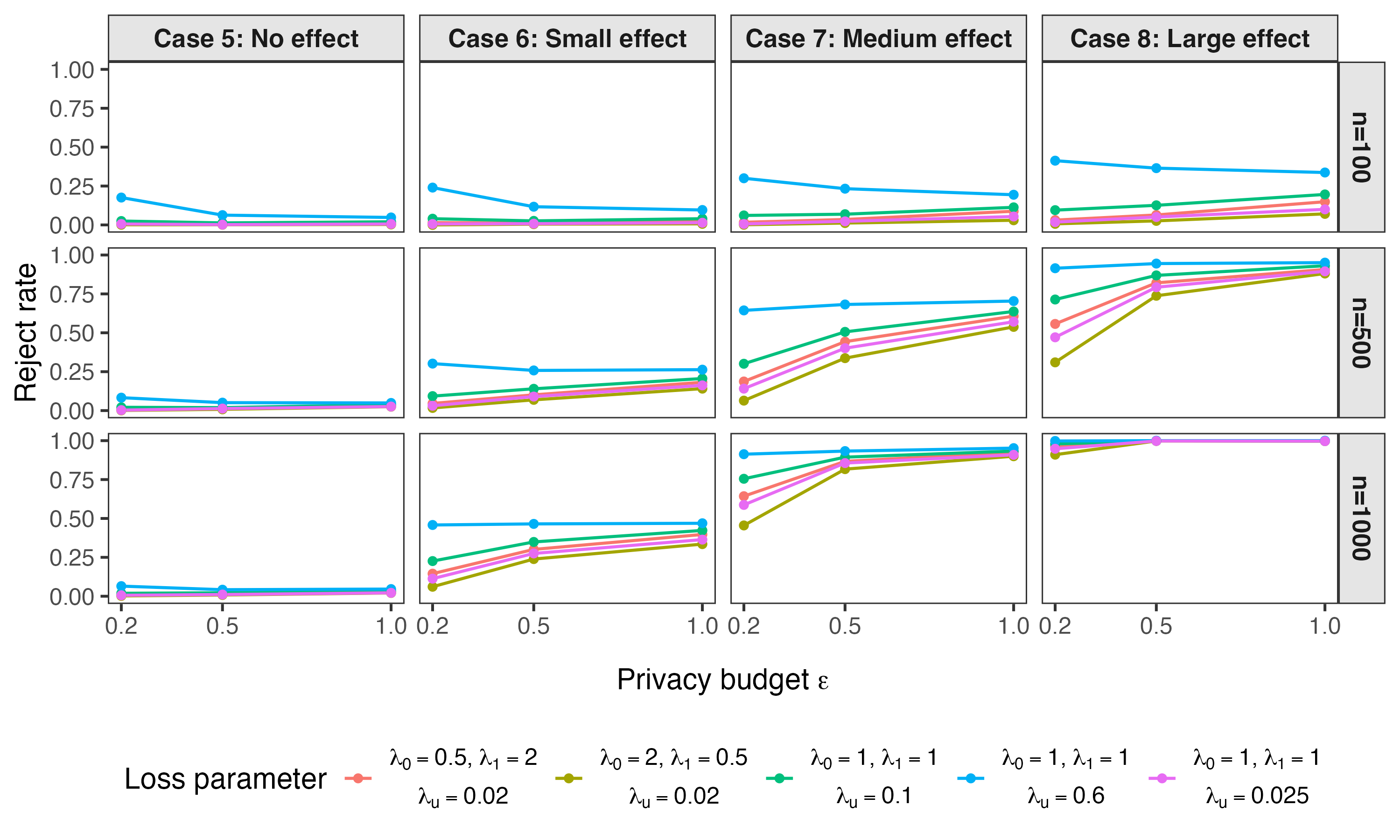}
    \caption{Rejection rates of the Bayes risk-optimal rule with abstention under varying loss parameters $(\lambda_0, \lambda_1, \lambda_u)$ for $\epsilon \in \{0.2, 0.5, 1\}$ in Cases 5--8 under $n \in \{100, 500, 1000\}$.}
    \label{fig:Bayes_line}
\end{figure}

Figure \ref{fig:Bayes_stack} summarizes the relative frequencies of each decision in the Bayes risk-optimal rule with abstention under the different study populations when $\epsilon \in \{0.2, 0.5, 1\}$.
When $\epsilon = 0.2$, the proportion of abstentions can be substantial, particularly when $n=100$, reflecting greater decision uncertainty induced by the relatively large impacts of adding DP noise. As $\epsilon$ or $n$ increases, abstentions gradually diminish, and the proportions of correct rejections rise. For large effects, nearly all decisions favor rejection when either $\epsilon$ or $n$ is sufficiently large, indicating that the Bayes risk-optimal rule can have power to detect treatment effects while managing privacy-induced uncertainty through the abstention option.

Figure \ref{fig:Bayes_line} shows the rejection rates of the Bayes risk-optimal rule with abstention under varying loss parameters $(\lambda_0, \lambda_1, \lambda_u)$ across privacy budgets and sample sizes. The results highlight the sensitivity of decision outcomes to the relative weighting of rejection, non-rejection, and abstention losses. When $\lambda_u$ is small, the rule tends to abstain more and reject less frequently, leading to lower rejection rates across all settings. As $\lambda_u$ increases, abstention becomes less likely. Under the sharp null (Case 5), rejection rates remain near zero across all loss parameters. For cases with genuine effects (Cases 6--8), rejection rates increase with both $\epsilon$ and $n$ as expected.

\subsection{Analysis of ADAPTABLE trial}
\label{sec:real}

In this section, we illustrate the DP-FRT-Bayes framework using data from the ADAPTABLE trial \citep{JMW2021}. This is a randomized comparison of two daily aspirin dosing strategies for secondary prevention in patients with established atherosclerotic cardiovascular disease.
We analyze the data under intention-to-treat assignment of taking 81 mg ($Z=0; n_0 = 7540$) versus 325 mg ($Z=1; n_1 = 7536$) aspirin.
The binary outcome $Y$ is a composite of death from any cause, hospitalization for myocardial infarction, or hospitalization for stroke. At median follow-up, we have $(n_{11}, n_{10}, n_{01}, n_{00}) = (569, 6967, 590, 6950)$. The non-private FRT $p$-value is 0.746, offering essentially no evidence against the sharp null hypothesis of no difference between the 325 mg and 81 mg aspirin groups.

\begin{figure}[t]
    \centering
    \includegraphics[width=0.9\linewidth]{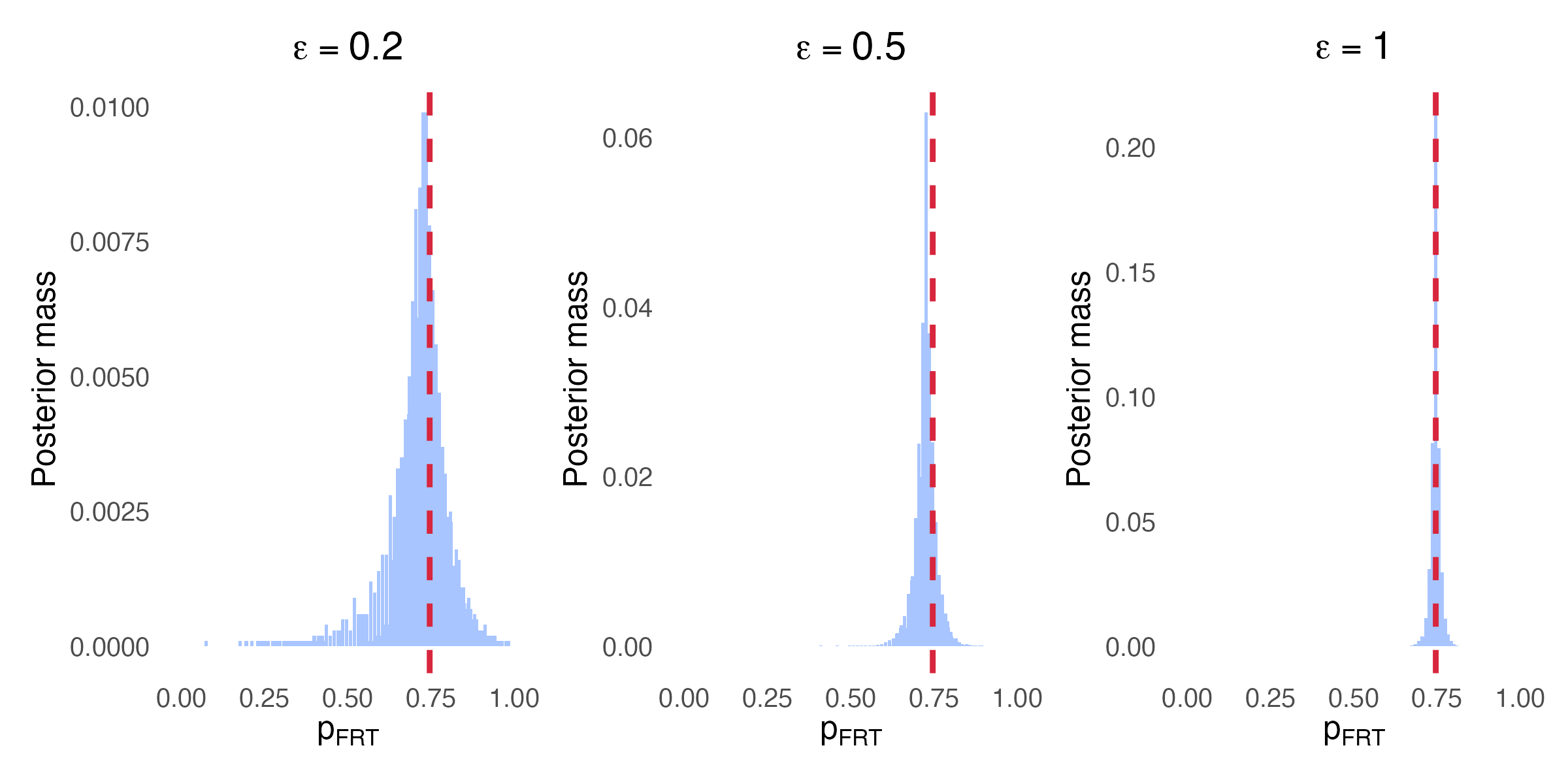}
    \caption{Posterior distributions of the DP-FRT-Bayes $p$-value under $\epsilon \in \{0.2, 0.5, 1\}$ in the ADAPTABLE trial, with $n_1=7536$ (325~mg), $n_0=7540$ (81~mg), and observed counts $(n_{11},n_{10},n_{01},n_{00})=(569,6967,590,6950)$. Red dashed lines indicate the non-private $p$-value of 0.746.}

    \label{fig:real}
  \end{figure}

For illustrative purposes, we apply DP-FRT-Bayes using $\epsilon \in \{0.2, 0.5, 1\}$, where $\alpha = 0.05$ and loss parameters $\lambda_0 = \lambda_1 = 1$ and $\lambda_u = 0.025$. As shown in Figure \ref{fig:real}, the posterior distributions of the differentially private FRT $p$-values increasingly concentrate around the non-private counterpart as the privacy budget grows, with most of the posterior mass remaining well away from the rejection region.
The posterior rejection probabilities $\Psi$ are nearly zero across all privacy levels, yielding a Bayes-optimal decision of ``do not reject.''

\begin{figure}[t]
    \centering
    \includegraphics[width=0.9\linewidth]{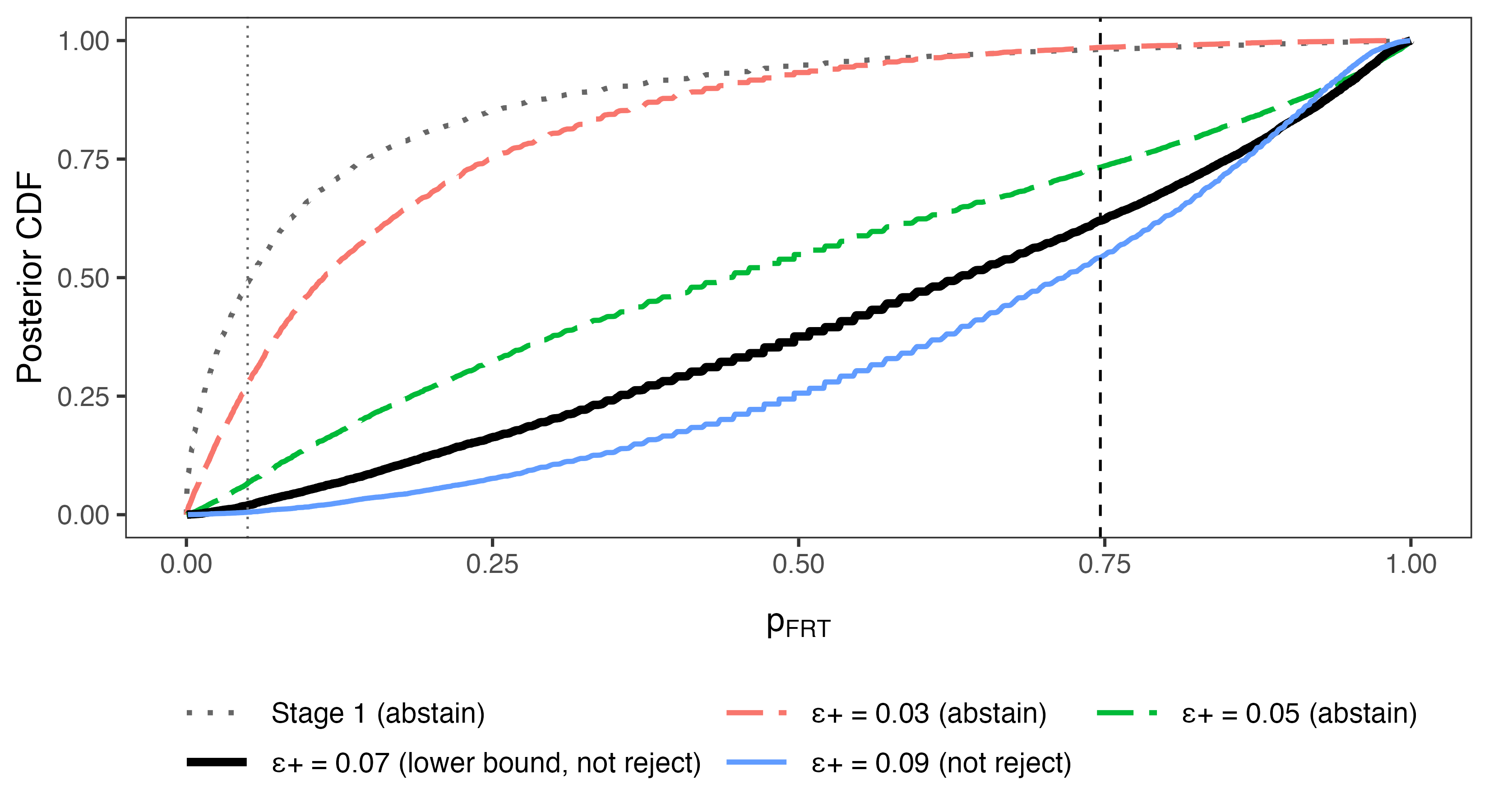}
    \caption{Posterior CDFs of $p_{\mathrm{FRT}}$ under the sequential DP-FRT-Bayes procedure applied to the ADAPTABLE trial. The dotted gray curve is the first-stage posterior at initial budget $\epsilon=0.06$. The four colored curves are the updated posteriors after a top-up release at additional $\epsilon_{\mathrm{plus}}\in\{0.03,0.05,0.07,0.09\}$; the solid black curve at $\epsilon_{\mathrm{plus}}=0.07$ is the analyst-side lower bound corresponding to target second-stage abstention probability $\xi=0.05$. Dashed line style indicates that the stage-2 decision is ``abstain''; solid line style indicates ``do not reject''. The dotted vertical line marks $\alpha=0.05$ and the dashed vertical line marks the non-private $p$-value of $0.746$.}
    \label{fig:real_seq}
\end{figure}

To illustrate the sequential procedure of Section \ref{sec:4.1.2}, we use a small initial budget $\epsilon=0.06$, under which the first-stage posterior is highly diffuse.
The posterior rejection probability $\Psi \approx 0.49$ lies near the middle of the abstention region, so the stage-1 decision is ``abstain.''
At a target second-stage abstention probability of $\xi=0.05$, the analyst-side lower bound for the additional budget is $\epsilon_{\mathrm{plus}}\approx 0.07$, which can be interpreted as the minimum extra privacy cost needed for at most a $5\%$ chance of abstaining again at stage 2.
Figure \ref{fig:real_seq} shows the updated posterior CDF of $p_{\mathrm{FRT}}$ under four top-up budgets $\epsilon_{\mathrm{plus}}\in\{0.03,0.05,0.07,0.09\}$ that straddle this bound. When $\epsilon_{\mathrm{plus}}$ falls below the bound, the combined posterior remains spread out across $[0,1]$ and the stage-2 decision is again ``abstain.'' At or above the bound, the posterior shifts towards the non-private $p$-value of $0.746$ and the stage-2 decision becomes ``do not reject,'' in agreement with the Bayes-optimal analysis at larger initial budgets.
Because the bound is necessary rather than tight, the stage-2 decision need not change exactly at $\epsilon_{\mathrm{plus}}\approx 0.07$ in any particular run. When an analyst finds even this modest additional cost too high, accepting the stage-1 abstention remains a sensible option.

Finally, we apply the two frequentist calibrations of Section \ref{sec:4.2} for controlling Type I error at $\alpha_{\mathrm{Freq}}=0.05$. 
For $\epsilon \in \{0.2, 0.5, 1\}$, the worst-case thresholds that protect against the least favorable configuration are $t^*_{\mathrm{LFC}} = (0.986, 1.000, 1.000)$, respectively, and the data-adaptive thresholds, computed with $\zeta = 0.025$, are $t^*_{\mathrm{Neyman}} = (0.952, 0.996, 1.000)$.
Because the posterior rejection probability $\Psi \approx 0$ under all three privacy budgets, both calibrations return the same ``do not reject'' decision, matching the Bayes-optimal analysis.

\section{Discussion}
\label{sec6}

We develop a framework for differentially private tests of Fisher's sharp null for binary outcomes, including methods for differentially private estimation of the FRT $p$-value and formal decision rules for hypothesis testing. The methods exhibit reasonable finite-sample performance while also supporting uncertainty quantification and adaptive use of privacy budgets.

The DP-FRT methods can be extended directly to bounded discrete responses, such as ordinal or count data with finite support.
For bounded continuous responses, direct application of the binary FRT requires discretization or binning. A principled approach is to treat the binning as a latent process and marginalize over the latent contingency tables rather than conditioning on a fixed discretization. This can be embedded within the Bayesian denoising framework, thereby propagating both binning and DP-induced uncertainties.

The DP-FRT methods can be generalized beyond CREs to stratified or block-randomized designs, paired or cluster randomization, and rerandomization procedures. In each case, the structure of the assignment mechanism changes the randomization distribution but does not affect the privacy mechanism applied to the observed outcomes. Combining DP-FRT with covariate-adjusted or rerandomized FRTs may improve efficiency, though the resulting randomization distributions will generally depend on more complex sufficient statistics requiring tailored privacy analysis.

Another promising direction is to extend the DP-FRT framework to the testing of weak null hypotheses \citep{DD2018, WD2021}.
Under weak nulls, exact randomization inference is generally infeasible without additional modeling or asymptotic justification because unit-level treatment effects are not fully specified. A natural bridge arises through the posterior predictive checking framework \citep{Rubin1984, Meng1994}, where one computes posterior predictive $p$-values by averaging the FRT $p$-values over the posterior distribution of unobserved potential outcomes.
Incorporating this averaging into the DP-FRT could yield a unified framework for privacy-preserving inference under both sharp and weak nulls.

Certain designs naturally induce non-sensitive randomization distributions. For example, under matched-pair randomization, the paired sign test depends only on the number of discordant pairs and not on individual outcomes. Consequently, the randomization distribution itself requires no privacy noise, which instead can be applied only to the observed test statistic. Further exploration of design-adaptive privacy mechanisms could result in algorithms with increased accuracy for a given privacy cost.

The Bayesian denoising framework is not tied to the Geometric mechanism. Any data-independent privacy mechanism with a known noise distribution can be used, since the posterior distribution of the true counts given the noisy release depends only on the prior, the observed noisy counts, and the mechanism's likelihood. For example, one could apply the Laplace mechanism to the cell counts and use the corresponding Laplace kernel in place of the geometric kernel. Future research could systematically compare the power and accuracy of different noise distributions within this framework, as well as develop guidance on how practitioners can choose among them without leaking information about the confidential data.

Finally, as suggested by a reviewer, there may be other effective Bayesian denoising strategies for testing Fisher's sharp null that utilize posterior predictive $p$-values.
One possible algorithm proceeds as follows.
First, the analyst draws a plausible value of the true counts given the noisy counts using Bayesian denoising. Second, the analyst uses Fisher's sharp null to generate all possible randomizations for those true counts. Third, the analyst adds DP noise to the counts in each possible randomization. Fourth, the analyst computes some test statistic of interest in each noisy randomization. Fifth, the analyst repeats these four steps many times. Sixth, the analyst determines the fraction of the noisy replicated test statistics that exceed the noisy test statistic in the released data. Future research could evaluate the performance of such methods. Additionally, research could assess whether practitioners find such posterior predictive $p$-values more or less interpretable as measures of evidence against Fisher's sharp null than the posterior distribution of the confidential $p_{\rm FRT}$.

\section*{Disclosure statement}
\label{disclosure-statement}

On behalf of all authors, the corresponding author states that there is no conflict of interest.

\section*{Data Availability Statement}\label{data-availability-statement}

The data analyzed in this study are derived from the summary reported in the published ADAPTABLE trial \citep{JMW2021}. No individual-level or proprietary data were accessed. All quantities required to reproduce the analyses are fully contained within the manuscript; therefore, no additional data are publicly archived.

\bibliography{Refer}

@article{GR2025,
  title={Differentially private estimation of weighted average treatment effects for binary outcomes},
  author={Guha, Sharmistha and Reiter, Jerome P},
  journal={Computational Statistics \& Data Analysis},
  volume={207},
  pages={108145},
  year={2025},
  publisher={Elsevier}
}

@article{DSS2017,
  title={Exposed! A survey of attacks on private data},
  author={Dwork, Cynthia and Smith, Adam and Steinke, Thomas and Ullman, Jonathan},
  journal={Annual Review of Statistics and Its Application},
  volume={4},
  number={1},
  pages={61--84},
  year={2017},
  publisher={Annual Reviews}
}

@article{Reiter2003,
  title={Inference for partially synthetic, public use microdata sets},
  author={Reiter, Jerome P},
  journal={Survey Methodology},
  volume={29},
  number={2},
  pages={181--188},
  year={2003}
}

@article{KR2026,
  title={Prior distributions for Gaussian models under differential privacy with bounds on data values},
  author={Kazan, Zeki and Reiter, Jerome P},
  journal={Journal of Privacy and Confidentiality},
  year={forthcoming}
}

@article{LR2022,
  author={Li, Linlin and Reiter, Jerome P.},
  title={{Bayesian} inference for estimating subset proportions using differentially private counts}, 
  journal={Journal of Survey Statistics and Methodology}, 
  volume={10},
  number={3},
  pages={785--803},
  year={2022}
}

@article{Rubin1974,
  title={Estimating causal effects of treatments in randomized and nonrandomized studies},
  author={Rubin, Donald B},
  journal={Journal of Educational Psychology},
  volume={66},
  number={5},
  pages={688--701},
  year={1974},
  publisher={American Psychological Association}
}

@book{Fisher1935,
  author={Fisher, Ronald A.},
  title={The Design of Experiments},
  edition={1st},
  year={1935},
  publisher={Oliver and Boyd},
  address={Edinburgh}
}

@inproceedings{DN2003,
  title={Revealing information while preserving privacy},
  author={Dinur, Irit and Nissim, Kobbi},
  booktitle={Proceedings of the Twenty-second ACM SIGMOD-SIGACT-SIGART Symposium on Principles of Database Systems},
  pages={202--210},
  publisher={ACM},
  year={2003}
}

@article{WD2021,
  title={Randomization tests for weak null hypotheses in randomized experiments},
  author={Wu, Jason and Ding, Peng},
  journal={Journal of the American Statistical Association},
  volume={116},
  number={536},
  pages={1898--1913},
  year={2021},
  publisher={Taylor \& Francis}
}

@article{DD2016,
  title={A potential tale of two-by-two tables from completely randomized experiments},
  author={Ding, Peng and Dasgupta, Tirthankar},
  journal={Journal of the American Statistical Association},
  volume={111},
  number={513},
  pages={157--168},
  year={2016},
  publisher={Taylor \& Francis}
}

@inproceedings{Dwork2006,
  title={Differential privacy},
  author={Dwork, Cynthia},
  booktitle={International Colloquium on Automata, Languages, and Programming},
  volume={4052},
  pages={1--12},
  year={2006},
  publisher={Springer}
}

@article{GRS2012,
  title={Universally utility-maximizing privacy mechanisms},
  author={Ghosh, Arpita and Roughgarden, Tim and Sundararajan, Mukund},
  journal={SIAM Journal on Computing},
  volume={41},
  number={6},
  pages={1673--1693},
  year={2012},
  publisher={SIAM}
}

@inproceedings{DMNS2006,
  title={Calibrating noise to sensitivity in private data analysis},
  author={Dwork, Cynthia and McSherry, Frank and Nissim, Kobbi and Smith, Adam},
  booktitle={Theory of Cryptography Conference},
  volume={3876},
  pages={265--284},
  year={2006},
  publisher={Springer}
}

@inproceedings{BS2019,
author = {Garrett Bernstein and Daniel R. Sheldon},
title = {Differentially Private {B}ayesian Linear Regression},
booktitle = {Advances in Neural Information Processing Systems},
volume = {32}, 
year = {2019}, 
pages = {523--533}}

@inproceedings{JAGR2022,
author = {Ju, Nianqiao Phyllis and Jordan Awan and Ruobin Gong and Vinayak Rao}, 
title = {Data Augmentation {MCMC} for {B}ayesian Inference from Privatized Data},
booktitle = {Advances in Neural Information Processing Systems}, 
editor = {S. Koyejo and S. Mohamed and A. Agarwal and D. Belgrave and K. Cho and A. Oh},
pages = {12732--12743},
 publisher = {Curran Associates, Inc.},
 volume = {35},
 year = {2022}}

@article{DD2018,
  title={A randomization-based perspective on analysis of variance: a test statistic robust to treatment effect heterogeneity},
  author={Ding, Peng and Dasgupta, Tirthankar},
  journal={Biometrika},
  volume={105},
  number={1},
  pages={45--56},
  year={2018},
  publisher={Oxford University Press}
}

@article{Rubin1984,
  title={Bayesianly justifiable and relevant frequency calculations for the applied statistician},
  author={Rubin, Donald B},
  journal={The Annals of Statistics},
  volume={12},
  number={4},
  pages={1151--1172},
  year={1984},
  publisher={Institute of Mathematical Statistics}
}

@article{Meng1994,
  title={Posterior predictive $p$-values},
  author={Meng, Xiao Li},
  journal={The Annals of Statistics},
  volume={22},
  number={3},
  pages={1142--1160},
  year={1994},
  publisher={Institute of Mathematical Statistics}
}

@misc{OHK2015,
  author={D'Orazio, Vito and Honaker, James and King, Gary},
  title={Differential privacy for social science inference},
  year={2015},
  howpublished={Sloan Foundation Economics Research Paper No. 2676160},
  note={Available at SSRN: \url{https://ssrn.com/abstract=2676160}}
}

@article{LGPM2019,
  title={Privacy-preserving causal inference via inverse probability weighting},
  author={Lee, Si Kai and Gresele, Luigi and Park, Mijung and Muandet, Krikamol},
  journal={arXiv preprint arXiv:1905.12592},
  year={2019}
}

@inproceedings{NNQCNK2022,
  title={Differentially private estimation of heterogeneous causal effects},
  author={Niu, Fengshi and Nori, Harsha and Quistorff, Brian and Caruana, Rich and Ngwe, Donald and Kannan, Aadharsh},
  booktitle={Conference on Causal Learning and Reasoning},
  volume={177},
  pages={618--633},
  year={2022},
  publisher={PMLR}
}

@incollection{MMS2024,
  author={Mukherjee, Soumya and Mustafi, Aratrika and Slavkovi\'c, Aleksandra and Vilhuber, Lars},
  title={Improving privacy for respondents in randomized controlled trials: A differential privacy approach},
  booktitle={Data Privacy Protection and the Conduct of Applied Research: Methods, Approaches and their Consequences},
  year={2024},
  publisher={University of Chicago Press},
  editor={Gong, Ruobin and Hotz, V. Joseph and Schmutte, Ian M.}
}

@article{AAC2022,
  author={Abowd, John M. and Ashmead, Robert and Cumings-Menon, Ryan and Garfinkel, Simson and Heineck, Micah and Heiss, Christine and others},
  journal={Harvard Data Science Review},
  issue={Special Issue 2},
  year={2022},
  publisher={The MIT Press},
  title={The 2020 {Census} {Disclosure} {Avoidance} {System} {TopDown} {Algorithm}},
}

@inproceedings{GLRV2016,
  title={Differentially private chi-squared hypothesis testing: Goodness of fit and independence testing},
  author={Gaboardi, Marco and Lim, Hyun and Rogers, Ryan and Vadhan, Salil},
  booktitle={International Conference on Machine Learning},
  volume={48},
  pages={2111--2120},
  year={2016},
  publisher={PMLR}
}

@inproceedings{KR2024prior,
 author={Kazan, Zeki and Reiter, Jerome P.},
 booktitle={Advances in Neural Information Processing Systems},
 editor={A. Globerson and L. Mackey and D. Belgrave and A. Fan and U. Paquet and J. Tomczak and C. Zhang},
 pages={90384--90430},
 publisher={Curran Associates, Inc.},
 title={Prior-itizing privacy: A {Bayesian} approach to setting the privacy budget in differential privacy},
 volume={37},
 year={2024}
}

@inproceedings{CKSBG2019,
  title={Differentially private nonparametric hypothesis testing},
  author={Couch, Simon and Kazan, Zeki and Shi, Kaiyan and Bray, Andrew and Groce, Adam},
  booktitle={Proceedings of the 2019 ACM SIGSAC Conference on Computer and Communications Security},
  year={2019},
  pages={737--751},
  publisher={ACM}
}

@inproceedings{AS2018,
  title={Differentially private uniformly most powerful tests for binomial data},
  author={Awan, Jordan and Slavkovi{\'c}, Aleksandra},
  booktitle={Proceedings of the 32nd International Conference on Neural Information Processing Systems},
  pages={4212--4222},
  year={2018},
  publisher={Curran Associates Inc.}
}

@inproceedings{KSGB2023,
  title={The test of tests: A framework for differentially private hypothesis testing},
  author={Kazan, Zeki and Shi, Kaiyan and Groce, Adam and Bray, Andrew P},
  booktitle={International Conference on Machine Learning},
  volume={202},
  pages={16131--16151},
  year={2023},
  publisher={PMLR}
}

@article{PB2025,
  title={Differentially private hypothesis testing with the subsampled and aggregated randomized response mechanism},
  author={Pe{\~n}a, V{\'\i}ctor and Barrientos, Andr{\'e}s F},
  journal={Statistica Sinica},
  volume={35},
  pages={671--691},
  year={2025}
}

@article{KS2025,
author = {Ilmun Kim and Antonin Schrab},
title = {Differentially Private Permutation Tests},
journal = {Journal of the American Statistical Association},
volume = {},
number = {},
pages = {},
year = {2025},
publisher = {Taylor \& Francis},
note = {\url{https://doi.org/10.1080/01621459.2025.2610033} (\textbf{in press})}
}

@article{Chow1957,
  title={An optimum character recognition system using decision functions},
  author={Chow, CK},
  journal={IRE Transactions on Electronic Computers},
  volume={EC--6},
  number={4},
  pages={247--254},
  year={1957},
  publisher={IEEE}
}

@article{Chow1970,
  title={On optimum recognition error and reject tradeoff},
  author={Chow, C},
  journal={IEEE Transactions on Information Theory},
  volume={16},
  number={1},
  pages={41--46},
  year={1970},
  publisher={IEEE}
}

@article{GI2024,
  title={Total variation meets differential privacy},
  author={Ghazi, Elena and Issa, Ibrahim},
  journal={IEEE Journal on Selected Areas in Information Theory},
  volume={5},
  pages={207--220},
  year={2024},
  publisher={IEEE}
}

@article{JMW2021,
  title={Comparative effectiveness of aspirin dosing in cardiovascular disease},
  author={Jones, W Schuyler and Mulder, Hillary and Wruck, Lisa M and Pencina, Michael J and Kripalani, Sunil and Mu{\~n}oz, Daniel and others},
  journal={New England Journal of Medicine},
  volume={384},
  number={21},
  pages={1981--1990},
  year={2021},
  publisher={Mass Medical Soc}
}

@inproceedings{KKR2024,
  title={Differentially private conditional independence testing},
  author={Kalemaj, Iden and Kasiviswanathan, Shiva and Ramdas, Aaditya},
  booktitle={International Conference on Artificial Intelligence and Statistics},
  volume={238},
  pages={3700--3708},
  year={2024},
  publisher={PMLR}
}

@inproceedings{NRS2007,
  title={Smooth sensitivity and sampling in private data analysis},
  author={Nissim, Kobbi and Raskhodnikova, Sofya and Smith, Adam},
  booktitle={Proceedings of the Thirty-Ninth Annual ACM Symposium on Theory of Computing},
  pages={75--84},
  year={2007},
  publisher={ACM}
}

@inproceedings{DL2009,
  title={Differential privacy and robust statistics},
  author={Dwork, Cynthia and Lei, Jing},
  booktitle={Proceedings of the Forty-First Annual ACM Symposium on Theory of Computing},
  pages={371--380},
  year={2009},
  publisher={ACM}
}

@article{Little2006,
  title={Calibrated {B}ayes: A {B}ayes/frequentist roadmap},
  author={Little, Roderick J.},
  journal={The American Statistician},
  volume={60},
  number={3},
  pages={213--223},
  year={2006},
  publisher={Taylor \& Francis}
}

\newpage
\appendix
\begin{center}

{\LARGE\bf Supplemental Material for ``Differentially Private Tests of Fisher's Sharp Null Hypothesis for Binary Outcomes''}

\end{center}
\bigskip

This supplementary material contains additional algorithmic details, simulation and theoretical results, and proofs of lemmas and theorems in the main text. 
Section \ref{sec:algo} provides detailed algorithms.
Section \ref{sec:addsims} shows results of additional simulation studies referenced in the main text.
Section \ref{sec:addtheorys} presents several results. First, we show that clipping does not influence the posterior inferences of DP-FRT-Bayes. Second, we establish the rate at which the DP-FRT-Bayes posterior concentrates around the confidential FRT $p$-value. Third, we bound the probability that the binary Bayes decision differs from the confidential FRT decision. Fourth, we show that the probability of remaining in the abstention region when spending additional privacy budget decays exponentially. Fifth, we extend the decision-fidelity bound to the sequential procedure. Sixth, we provide two heuristic interpretations of the test statistic $\Psi$ from Section 4.2 in the main text. Finally, we compare the power of the test based on $\Psi$ to the power of a differentially private significance test based on an alternative statistic.
Section \ref{sec:proof} includes proofs and derivations of lemmas and theorems stated in the main text.

\renewcommand{\thesection}{S.\arabic{section}}
\renewcommand{\thesubsection}{S.\arabic{section}.\arabic{subsection}}
\renewcommand{\theequation}{S.\arabic{section}.\arabic{equation}}
\renewcommand{\thefigure}{S.\arabic{figure}}
\renewcommand{\thetable}{S.\arabic{table}}
\renewcommand{\thealgocf}{S.\arabic{algocf}}
\renewcommand{\thetheorem}{S.\arabic{section}.\arabic{theorem}}
\setcounter{section}{0}
\setcounter{equation}{0}
\setcounter{figure}{0}
\setcounter{table}{0}
\setcounter{algocf}{0}
\setcounter{theorem}{0}

\section{Algorithms and Implementation Details}
\label{sec:algo}

\subsection{MCMC Samplers for Posterior Computation}

The following two algorithms provide implementation details for posterior computation in Section 3.2 of the main text.
Algorithm \ref{algo:mh_counts} describes a generic Metropolis-Hastings sampler and Algorithm \ref{algo:mh_counts_commonrate} illustrates a sampler tailored for the common success rate Beta prior.

\begin{algorithm}[H]
\spacingset{1.3}
\setlength{\baselineskip}{0.75\baselineskip}
\caption{Metropolis-Hastings Sampler for $\gamma(a,b)$}
\label{algo:mh_counts}

\KwIn{group sizes $(n_1,n_0)$; privatized counts $\tilde{\bm n} = (\tilde n_{11},\tilde n_{01})$;
prior $\pi(a,b)$; privacy budget $\epsilon$; Monte Carlo size $R$; Burn-in size $B$.}

\KwOut{posterior samples $\{(a^{(r)},b^{(r)})\}_{r=B+1}^{B+R}$ and the corresponding
$\{p^{(r)}\}_{r=B+1}^{B+R}$.}

Initialize $(a^{(0)},b^{(0)})$ on $\{0,\dots,n_1\}\times\{0,\dots,n_0\}$\;

\For{$r = 0,\dots,B+R-1$}{
    Draw an independent proposal $(a^*,b^*) \sim \pi(\cdot)$ for $(a,b)\in\{0,\dots,n_1\}\times\{0,\dots,n_0\}$\;

    Compute
    $
      \alpha_{\mathrm{MH}}
      = \min\left\{1,~\dfrac{\kappa_\rho(\tilde n_{11} - a^*)\kappa_\rho(\tilde n_{01} - b^*)}{\kappa_\rho(\tilde n_{11} - a^{(r)})\kappa_\rho(\tilde n_{01} - b^{(r)})}
      \right\}
    $\;

    Draw $U \sim \mathrm{Unif}(0,1)$ and set $(a^{(r+1)}, b^{(r+1)}) \gets 
    \begin{cases}
    (a^*, b^*), & U \le \alpha_{\mathrm{MH}}, \\
    (a^{(r)}, b^{(r)}), & \text{otherwise}.
    \end{cases}$
    
    Compute $p^{(r+1)} = p_{a^{(r+1)}b^{(r+1)}}$ via (5) in the main text\;
}

Use $\{(a^{(r)},b^{(r)})\}_{r=B+1}^{B+R}$ as draws from $\gamma(a,b)$ and record the corresponding $\{p^{(r)}\}_{r=B+1}^{B+R}$.

\end{algorithm}

\begin{algorithm}[H]
\spacingset{1.3}
\setlength{\baselineskip}{0.75\baselineskip}
\caption{Metropolis-Hastings sampler for $\gamma(a,b)$ with common-rate prior}
\label{algo:mh_counts_commonrate}

\KwIn{group sizes $(n_1,n_0)$; privatized counts $\tilde{\bm n} = (\tilde n_{11},\tilde n_{01})$;
common success rate Beta prior $\pi_{\mathrm{CR}}(a,b)$;
privacy budget $\epsilon$; Monte Carlo size $R$; Burn-in size $B$.}

\KwOut{samples $\{(a^{(r)},b^{(r)})\}_{r=B+1}^{B+R}$ from $\gamma(a,b)$ and the corresponding
$\{p^{(r)}\}_{r=B+1}^{B+R}$.}

Define $q_A(a) \propto \kappa_\rho(\tilde n_{11} - a)$ for $a = 0,\dots,n_1$ and $q_B(b) \propto \kappa_\rho(\tilde n_{01} - b)$ for $b = 0,\dots,n_0$\;

Initialize $(a^{(0)},b^{(0)})$ on $\{0,\dots,n_1\}\times\{0,\dots,n_0\}$\;

\For{$r = 0,\dots,B+R-1$}{
    Draw $a^* \sim q_A(\cdot)$ and $b^* \sim q_B(\cdot)$ independently\;

    Compute
    $
      \alpha_{\mathrm{MH}}
      = \min\left\{1,~
      \dfrac{\pi_{\mathrm{CR}}(a^*,b^*)}{\pi_{\mathrm{CR}}(a^{(r)},b^{(r)})}
      \right\}
    $\;

    Draw $U \sim \mathrm{Unif}(0,1)$ and set $(a^{(r+1)}, b^{(r+1)}) \gets 
    \begin{cases}
    (a^*, b^*), & U \le \alpha_{\mathrm{MH}}, \\
    (a^{(r)}, b^{(r)}), & \text{otherwise}.
    \end{cases}$

    Compute $p^{(r+1)} = p_{a^{(r+1)}b^{(r+1)}}$ via (5) in the main text\;
}

Use $\{(a^{(r)},b^{(r)})\}_{r=B+1}^{B+R}$ as draws from $\gamma(a,b)$ and record the corresponding $\{p^{(r)}\}_{r=B+1}^{B+R}$.

\end{algorithm}

\subsection{Algorithm for Lower Bounding Additional Privacy Budget}

Algorithm \ref{algo:epsplus_lb} approximates the data-adaptive lower bound on the additional privacy budget $\epsilon_{\mathrm{plus}}$ derived in Section 4.1.2 of the main text.
We describe a bisection to find the smallest $\epsilon$ such that $\widehat{\Delta}(\epsilon)\ge \tau$, but any bracketing root-finding procedure can be used instead.

\begin{algorithm}[H]
\spacingset{1.7}
\setlength{\baselineskip}{0.75\baselineskip}
\caption{Data-adaptive Lower Bound on the Additional Privacy Budget $\epsilon_{\mathrm{plus}}$}
\label{algo:epsplus_lb}

\KwIn{abstention region $A$; confidence level $1-\xi\in(0,1)$;
significance level $\alpha$; approximation size $L$; tolerance $\mathrm{tol}$;
current posterior $\gamma$ and evidence $\Psi$.}

\KwOut{data-adaptive necessary lower bound $\epsilon_{\mathrm{lb}}$.}

Compute $r(\Psi)=\min\{\Psi-t_{\mathrm{low}},~t_{\mathrm{high}}-\Psi\}$ and set
$\tau=(1-\xi)~r(\Psi)/\left(2\Psi(1-\Psi)\right)$\;
Set $\mu_1(a,b\mid\tilde{\bm n})\propto \gamma(a,b)\mathbf{1}\{p_{ab}\le\alpha\}$ and
$\mu_0(a,b\mid\tilde{\bm n})\propto \gamma(a,b)\mathbf{1}\{p_{ab}>\alpha\}$\;

\For{$l=1$ \KwTo $L$}{
Draw $(a_l,b_l)\sim \mu_1(\cdot\mid\tilde{\bm n})$ and $(a'_l,b'_l)\sim \mu_0(\cdot\mid\tilde{\bm n})$\;
Set $d_l = |a_l-a'_l|+|b_l-b'_l|$\;
}

Define $\widehat{\Delta}(\epsilon)
= \dfrac{1}{L}\displaystyle\sum_{l=1}^L
\tanh\left(\dfrac{\epsilon d_l}{2}\right)$\;
Initialize $\epsilon_{\min}=0$ and choose $\epsilon_{\max}$ such that
$\widehat{\Delta}(\epsilon_{\max})\ge \tau$\;
\While{$\epsilon_{\max}-\epsilon_{\min}>\mathrm{tol}$}{
  Set $\epsilon_{\mathrm{mid}}=(\epsilon_{\min}+\epsilon_{\max})/2$\;
  Set $(\epsilon_{\min}, \epsilon_{\max}) \gets
  \begin{cases}
    (\epsilon_{\mathrm{mid}}, \epsilon_{\max}), & \widehat{\Delta}(\epsilon_{\mathrm{mid}}) < \tau,\\
    (\epsilon_{\min}, \epsilon_{\mathrm{mid}}), & \text{otherwise}.
  \end{cases}$\;
}

Set $\epsilon_{\mathrm{lb}}=\epsilon_{\max}$ and report $\epsilon_{\mathrm{lb}}$.

\end{algorithm}

\subsection{Algorithms for Frequentist-calibrated Decision Rules}

The following two algorithms implement the frequentist-calibrated decision rules with Type I error control described in Section 4.2 of the main text.
Algorithm \ref{algo:worst_case} implements the worst-case calibration and Algorithm \ref{algo:neyman} describes the data-adaptive calibration.

\begin{algorithm}[H]
\spacingset{1.7}
\setlength{\baselineskip}{0.75\baselineskip}
\caption{Worst-case Calibration of the Threshold}
\label{algo:worst_case}

\KwIn{group sizes $(n_1,n_0)$; significance level $\alpha$; Type I error level $\alpha_{\mathrm{Freq}}$; Monte Carlo size $R$; privatized counts $\tilde{\bm n}$.}
\KwOut{threshold $t^*_{\mathrm{LFC}}$ and decision rule $\delta_{\mathrm{LFC}}(\tilde{\bm n})$.}

\For{$K = 0,1,\dots,n$}{
    \For{$r = 1$ \KwTo $R$}{
        Draw $A_r \sim \mathrm{Hypergeometric}(n,K,n_1)$ and set $(a_{11}^{(r)}, b_{01}^{(r)}) = (A_r, K-A_r)$\;
        Apply the Geometric mechanism to $(a_{11}^{(r)}, b_{01}^{(r)})$ to obtain $(\tilde a_{11}^{(r)}, \tilde b_{01}^{(r)})$\;
        Compute $\psi^{(r)} = \Pr(p_{\mathrm{FRT}} \le \alpha \mid \tilde a_{11}^{(r)},\tilde b_{01}^{(r)})$\;
    }
    Compute $t_K$ as the empirical $(1-\alpha_{\mathrm{Freq}})$-quantile of $\{\psi^{(r)}\}_{r=1}^R$\;
}

Set $t^*_{\mathrm{LFC}} = \max_{K} t_K$ and report decision $\delta_{\mathrm{LFC}} = \mathbf{1}\left(\Pr(p_{\mathrm{FRT}} \le \alpha \mid \tilde{\bm n}) > t^*_{\mathrm{LFC}}\right)$.

\end{algorithm}

\begin{algorithm}[H]
\spacingset{1.7}
\setlength{\baselineskip}{0.75\baselineskip}
\caption{Data-adaptive Calibration of the Threshold}
\label{algo:neyman}

\KwIn{group sizes $(n_1,n_0)$; significance level $\alpha$; 
Type I error level $\alpha_{\mathrm{Freq}}$; Monte Carlo size $R$; 
confidence level $1-\zeta$ with $\zeta\in(0,\alpha_{\mathrm{Freq}})$; privatized counts $\tilde{\bm n}$.}

\KwOut{threshold $t^*_{\mathrm{Neyman}}(\tilde{\bm n})$ and decision $\delta_{\mathrm{Neyman}}(\tilde{\bm n})$.}

\For{$K = 0,1,\dots,n$}{
    \For{$r = 1$ \KwTo $R$}{
        Draw $A_r \sim \mathrm{Hypergeometric}(n,K,n_1)$ and set 
        $(a_{11}^{(r)}, b_{01}^{(r)}) = (A_r, K-A_r)$\;
        Apply the Geometric mechanism to $(a_{11}^{(r)}, b_{01}^{(r)})$ to obtain $(\tilde a_{11}^{(r)}, \tilde b_{01}^{(r)})$\;
        Compute $\psi^{(r)} = \Pr(p_{\mathrm{FRT}}\le\alpha \mid \tilde a_{11}^{(r)}, \tilde b_{01}^{(r)})$\;
    }
    Use the empirical distribution of 
    $\{(\tilde a_{11}^{(r)},\tilde b_{01}^{(r)})\}_{r=1}^R$ 
    to approximate $Q_K$\; 
    Form the smallest high-probability set $A_K$ with mass at least $1-\zeta$\;
    Compute $t'_K$ as the empirical $(1-\alpha')$-quantile of $\{\psi^{(r)}\}_{r=1}^R$,
    where $\alpha'=\alpha_{\mathrm{Freq}}-\zeta$\;
}

Define $C_{1-\zeta}(\tilde{\bm n})=\{K:\tilde{\bm n}\in A_K\}.$
Set $t^*_{\mathrm{Neyman}}(\tilde{\bm n})
=
\max_{K\in C_{1-\zeta}(\tilde{\bm n})} t'_K$ and report $\delta_{\mathrm{Neyman}}(\tilde{\bm n})
=
\mathbf{1}\left(
\Pr(p_{\mathrm{FRT}}\le\alpha \mid \tilde{\bm n})
>
t^*_{\mathrm{Neyman}}(\tilde{\bm n})
\right).$

\end{algorithm}

\section{Additional Simulation Studies}\label{sec:addsims}

\subsection{Additional Simulations for Estimating FRT \texorpdfstring{$p$-values}{p-values}}

We provide additional simulation results that complement Section 5.1 of the main text. 
They focus on the point estimation accuracy of several DP-FRT estimators. We use the same simulation design as in the main text.

\begin{figure}[t]
    \centering
    \includegraphics[width=0.9\linewidth]{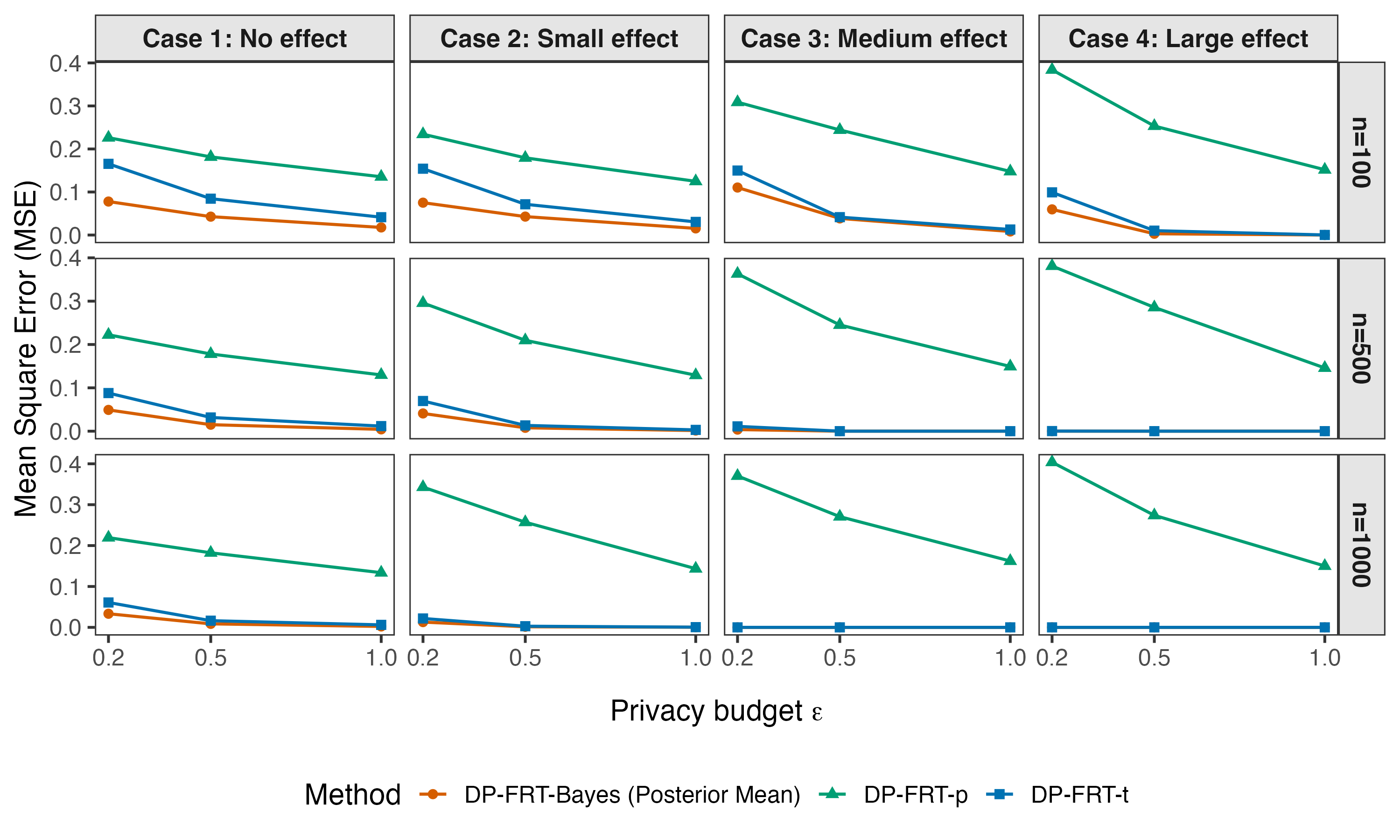}
    \caption{MSE of different DP-FRT point estimators of $p$-values in Cases 1--4 when $\epsilon \in \{0.2, 0.5, 1\}$ and $n \in \{100, 500, 1000\}$.
    }
    \label{fig:ALL_MSE}
\end{figure}

Figure \ref{fig:ALL_MSE} displays the MSE of three DP-FRT point estimators for the FRT $p$-value: DP-FRT-Bayes (posterior mean), DP-FRT-p, and DP-FRT-t.
We find that DP-FRT-Bayes consistently achieves the smallest MSE, particularly when $\epsilon$ is small or the sample size is limited.
The direct perturbation approaches DP-FRT-p and DP-FRT-t exhibit markedly larger MSE, with DP-FRT-p performing worst in most settings due to the high variability induced by directly perturbing the $p$-value.
DP-FRT-t improves upon DP-FRT-p but still remains less efficient, and it offers no uncertainty quantification.
These results demonstrate the clear advantage of DP-FRT-Bayes in stabilizing inference under DP when using the Geometric mechanism.

\begin{figure}[t]
    \centering
    \includegraphics[width=0.9\linewidth]{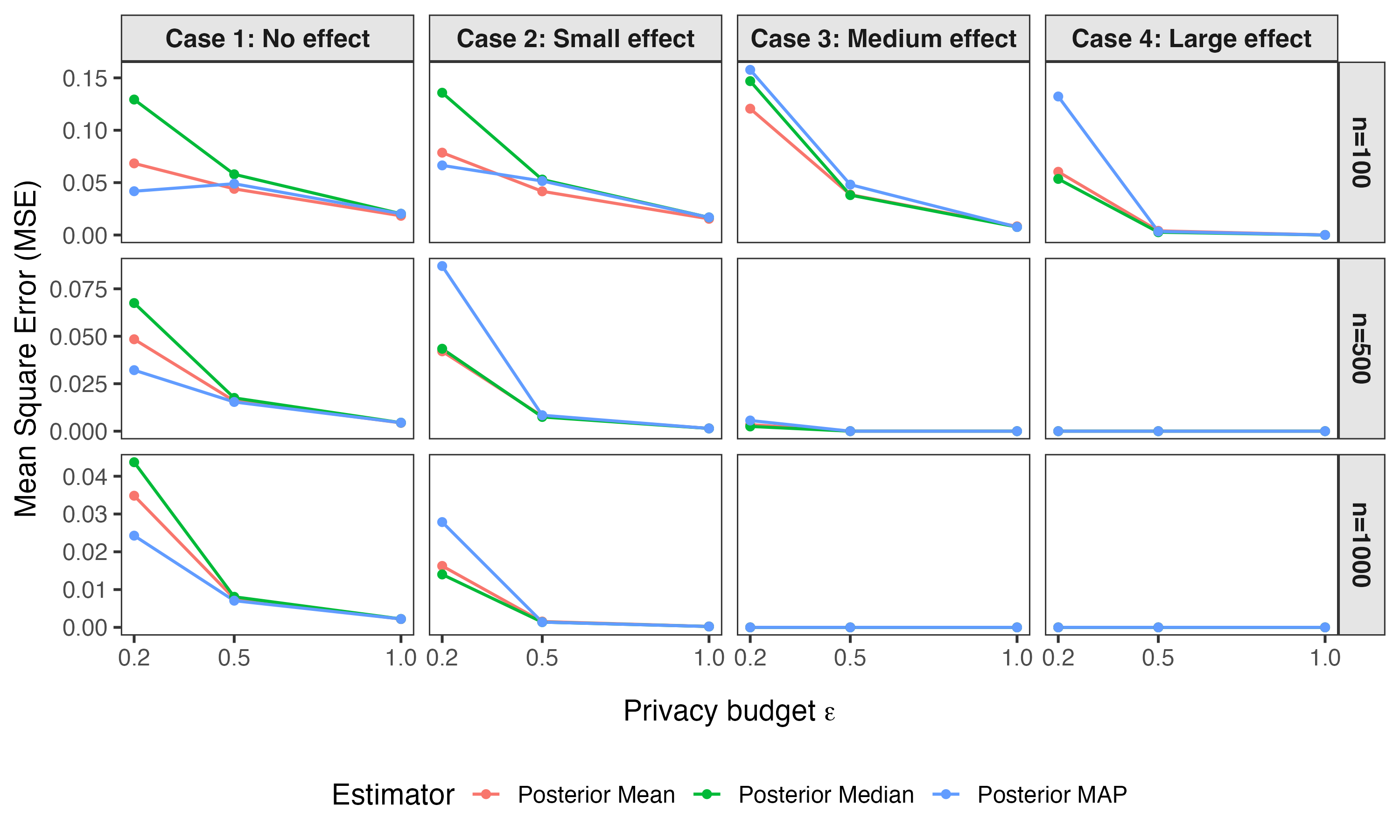}
    \caption{MSE of different DP-FRT-Bayes point estimators of $p$-values in Cases 1--4 when $\epsilon \in \{0.2, 0.5, 1\}$ and $n \in \{100, 500, 1000\}$. The y-axis scales differ across panels.}
    \label{fig:Bayes_MSE}
\end{figure}

Figure \ref{fig:Bayes_MSE} displays the MSEs of three point estimators for the FRT $p$-value when using DP-FRT-Bayes. 
As expected, the MSEs of all estimators decrease as the privacy budget or sample size increase.
The posterior mean and median have comparable MSEs across most settings. The posterior MAP estimator occasionally exhibits larger MSE when $\epsilon$ is small, reflecting its sensitivity to multimodal posterior distributions induced by heavy DP noise. We recommend using the posterior mean estimator for its overall stability and accuracy.

\begin{figure}[t]
  \centering
  \begin{subfigure}[b]{\linewidth}
    \centering
    \includegraphics[width=0.9\linewidth]{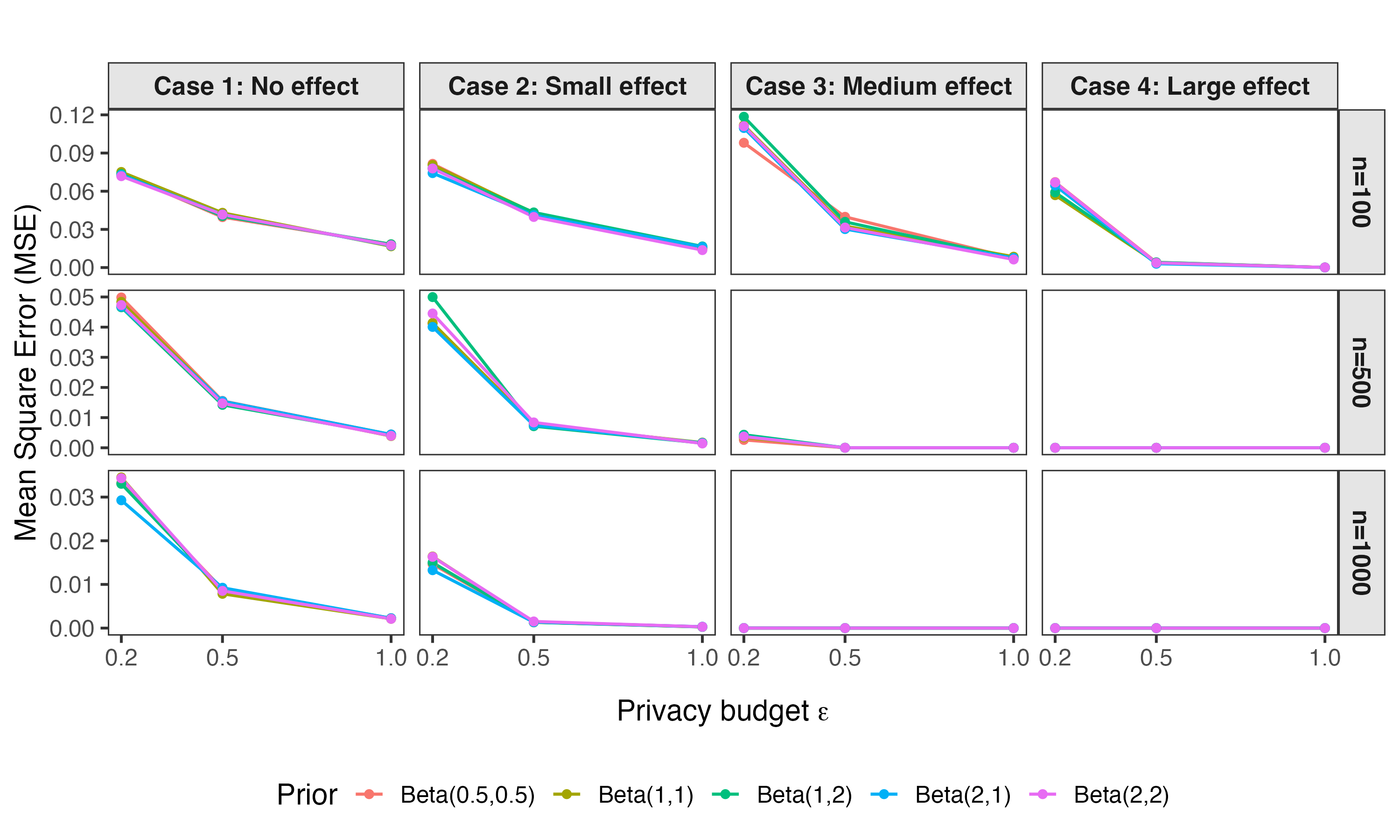}
    \caption{Beta-binomial Priors}
  \end{subfigure}
  \begin{subfigure}[b]{\linewidth}
    \centering
    \includegraphics[width=0.9\linewidth]{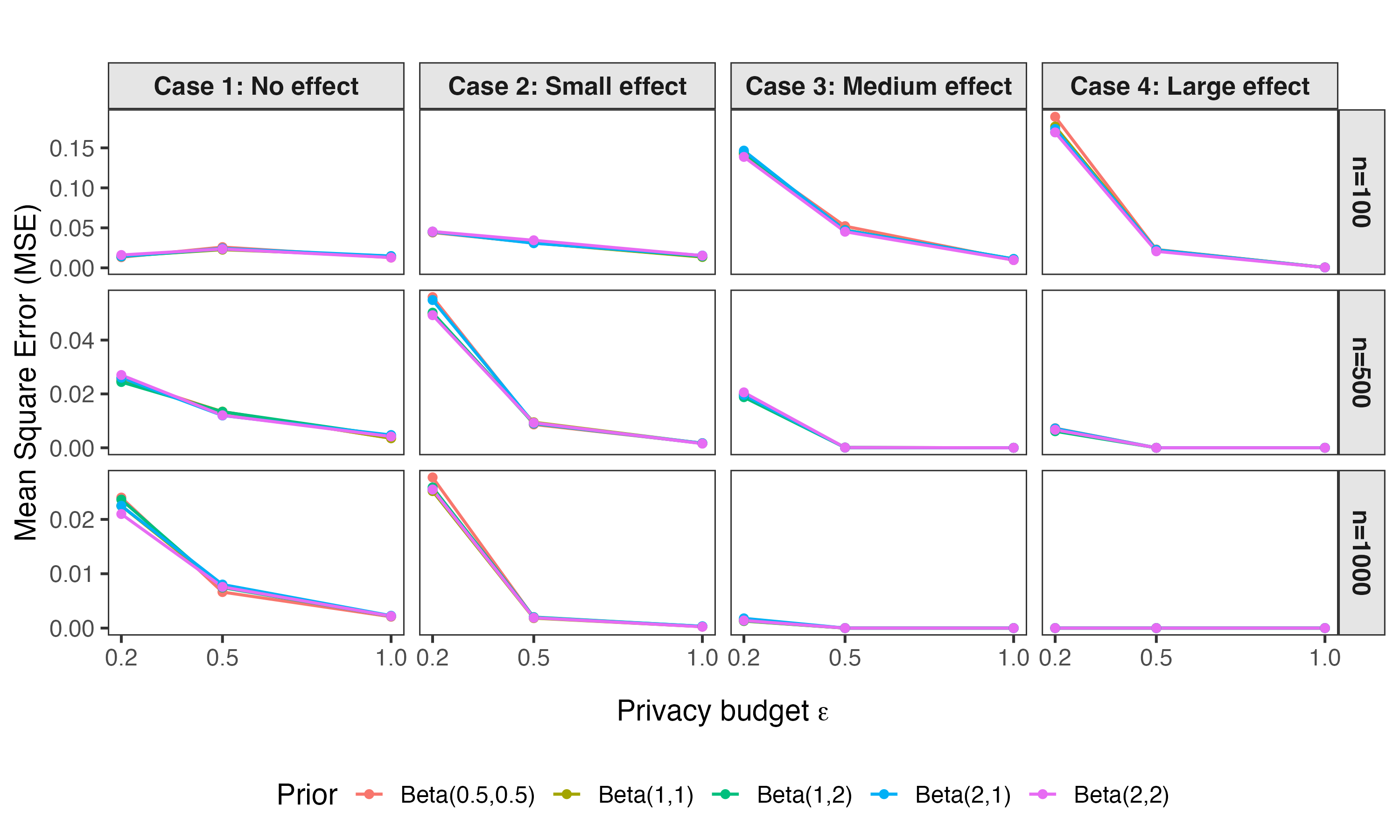}
    \caption{Common Success Rate Beta Priors}
  \end{subfigure}
  \caption{Prior sensitivity of the DP-FRT-Bayes posterior mean estimator for FRT $p$-value under (a) independent Beta-binomial priors and (b) common success rate Beta priors for $\epsilon \in \{0.2, 0.5, 1\}$ and $n \in \{100, 500, 1000\}$ in Cases 1--4. The y-axis scales differ across panels.}
  \label{fig:Bayes_prior}
\end{figure}

Figure \ref{fig:Bayes_prior} examines the prior sensitivity of the DP-FRT-Bayes posterior mean estimator for $p$-values under the independent Beta-binomial prior, which is (8) in the main text, and the common success rate Beta prior, which is (9) in the main text.
Note that choosing $\mathrm{Beta}(1,1)$ in the Beta-binomial priors yields the uniform prior in (7) in the main text.

\clearpage

\noindent Across all settings, the MSE curves for different priors are nearly indistinguishable, indicating that DP-FRT-Bayes is robust to the choice among these priors in these settings.

\subsection{Additional Simulations for Evaluating Decision Rules}

In this section, we present additional simulation results that complement Section 5.2 of the main text.
For the Bayesian sequential decision procedure, we focus on how a second-stage DP release with the additional budget reduces abstention.
For the frequentist tests, we focus on Type I error control under the sharp null and power behavior under alternatives with nonzero treatment effects.

\begin{table}[!t]
\spacingset{1.2}
\centering
\caption{First-stage abstention rates (\%), average $\epsilon_{\mathrm{lb}}$ with $\xi=0.05$, and escape rates (\%) after a second release at $\epsilon_{\rm plus} \in\{1,2,5,10\}\times\epsilon_{\mathrm{lb}}$ in Cases 5--8 with $\epsilon=0.2$.}
\begin{tabular*}{\linewidth}{@{\extracolsep{\fill}}l cccccc}
\toprule
\textbf{~~~~~Case} & Abstain & $\epsilon_{\mathrm{lb}}$ & Escape$_{1\times}$ & Escape$_{2\times}$ & Escape$_{5\times}$ & Escape$_{10\times}$ \\
\midrule
\multicolumn{7}{l}{\textbf{(a) $n=100$}} \\
~5: No effect     & 87.6 & 0.074 & 0.1 & 5.4 & 41.7 & 72.4 \\
~6: Small effect  & 90.6 & 0.080 & 0.0 & 4.2 & 30.2 & 61.0 \\
~7: Medium effect & 94.6 & 0.080 & 0.0 & 6.0 & 24.5 & 46.8 \\
~8: Large effect  & 93.2 & 0.087 & 0.0 & 4.5 & 23.9 & 43.8 \\
\midrule
\multicolumn{7}{l}{\textbf{(b) $n=500$}} \\
~5: No effect     & 50.5 & 0.057 & 1.2 & 13.7 & 46.5 & 75.4 \\
~6: Small effect  & 75.3 & 0.074 & 1.3 & 4.4  & 39.8 & 66.6 \\
~7: Medium effect & 75.6 & 0.070 & 1.9 & 11.1 & 34.4 & 61.4 \\
~8: Large effect  & 49.4 & 0.065 & 0.2 & 12.3 & 50.4 & 72.7 \\

\midrule
\multicolumn{7}{l}{\textbf{(c) $n=1000$}} \\
~5: No effect     & 28.5 & 0.047 & 0.0 & 14.4 & 54.0 & 76.5 \\
~6: Small effect  & 79.5 & 0.078 & 0.3 & 7.8  & 46.9 & 63.5 \\
~7: Medium effect & 45.8 & 0.056 & 1.5 & 10.9 & 50.9 & 69.2 \\
~8: Large effect  & 5.2  & 0.046 & 0.0 & 36.5 & 73.1 & 96.2 \\
\bottomrule
\end{tabular*}
\label{tab:TabS1_eps02}
\end{table}

\begin{table}[!t]
\spacingset{1.2}
\centering
\caption{First-stage abstention rates (\%), average $\epsilon_{\mathrm{lb}}$ with $\xi=0.05$, and escape rates (\%) after a second release using $\epsilon_{\rm plus} \in\{1,2,5,10\}\times\epsilon_{\mathrm{lb}}$ in Cases 5--8 with $\epsilon=0.5$.}
\begin{tabular*}{\linewidth}{@{\extracolsep{\fill}}l cccccc}
\toprule
\textbf{~~~~~Case} & Abstain & $\epsilon_{\mathrm{lb}}$ & Escape$_{1\times}$ & Escape$_{2\times}$ & Escape$_{5\times}$ & Escape$_{10\times}$ \\
\midrule

\multicolumn{7}{l}{\textbf{(a) $n=100$}} \\
~5: No effect     & 39.2 & 0.129 & 1.0 & 13.3 & 54.1 & 81.6 \\
~6: Small effect  & 56.3 & 0.138 & 0.9 & 12.6 & 51.2 & 71.4 \\
~7: Medium effect & 73.0 & 0.149 & 1.4 & 6.6  & 46.0 & 68.9 \\
~8: Large effect  & 81.2 & 0.187 & 0.4 & 3.8  & 42.5 & 69.8 \\

\midrule
\multicolumn{7}{l}{\textbf{(b) $n=500$}} \\
~5: No effect     & 17.5 & 0.154 & 1.1 & 9.7  & 52.0 & 81.7 \\
~6: Small effect  & 40.9 & 0.149 & 0.2 & 12.0 & 50.1 & 71.1 \\
~7: Medium effect & 35.5 & 0.154 & 0.0 & 10.7 & 46.8 & 64.5 \\
~8: Large effect  & 20.3 & 0.139 & 2.5 & 19.7 & 36.0 & 70.9 \\

\midrule
\multicolumn{7}{l}{\textbf{(c) $n=1000$}} \\
~5: No effect     & 10.8 & 0.142 & 0.0 & 8.3  & 49.1 & 80.6 \\
~6: Small effect  & 39.8 & 0.175 & 1.5 & 7.8  & 44.5 & 74.9 \\
~7: Medium effect & 12.5 & 0.170 & 0.0 & 11.2 & 54.4 & 80.8 \\
~8: Large effect  & 0.0   & --    & --  & --   & --   & --   \\

\bottomrule
\end{tabular*}
\label{tab:TabS1_eps05}
\end{table}

Table \ref{tab:TabS1_eps02} and Table \ref{tab:TabS1_eps05} summarize the behavior of the two-stage Bayesian sequential decision. 
The first-stage abstention rate reports how often the posterior evidence $\Psi$ falls inside the abstention region $A$, and $\epsilon_{\mathrm{lb}}$ is the average data-adaptive lower bound on the additional privacy budget required to have a substantial chance of leaving $A$. The remaining columns report the empirical escape probability after a second independent release, where the top-up budget is set to $\epsilon_{\mathrm{plus}} \in \{1,2,5,10\}\times \epsilon_{\mathrm{lb}}$.
Across both tables, the escape probability increases monotonically as $\epsilon_{\mathrm{plus}}$ is scaled up, illustrating how additional privacy budget translates into sharper posterior evidence and fewer abstentions. The reported $\epsilon_{\mathrm{lb}}$ varies with the underlying scenario and sample size, reflecting that the amount of extra privacy budget needed to exit $A$ is data-adaptive and instance-specific rather than a fixed constant. 

\begin{figure}[!t]
    \centering
    \includegraphics[width=0.9\linewidth]{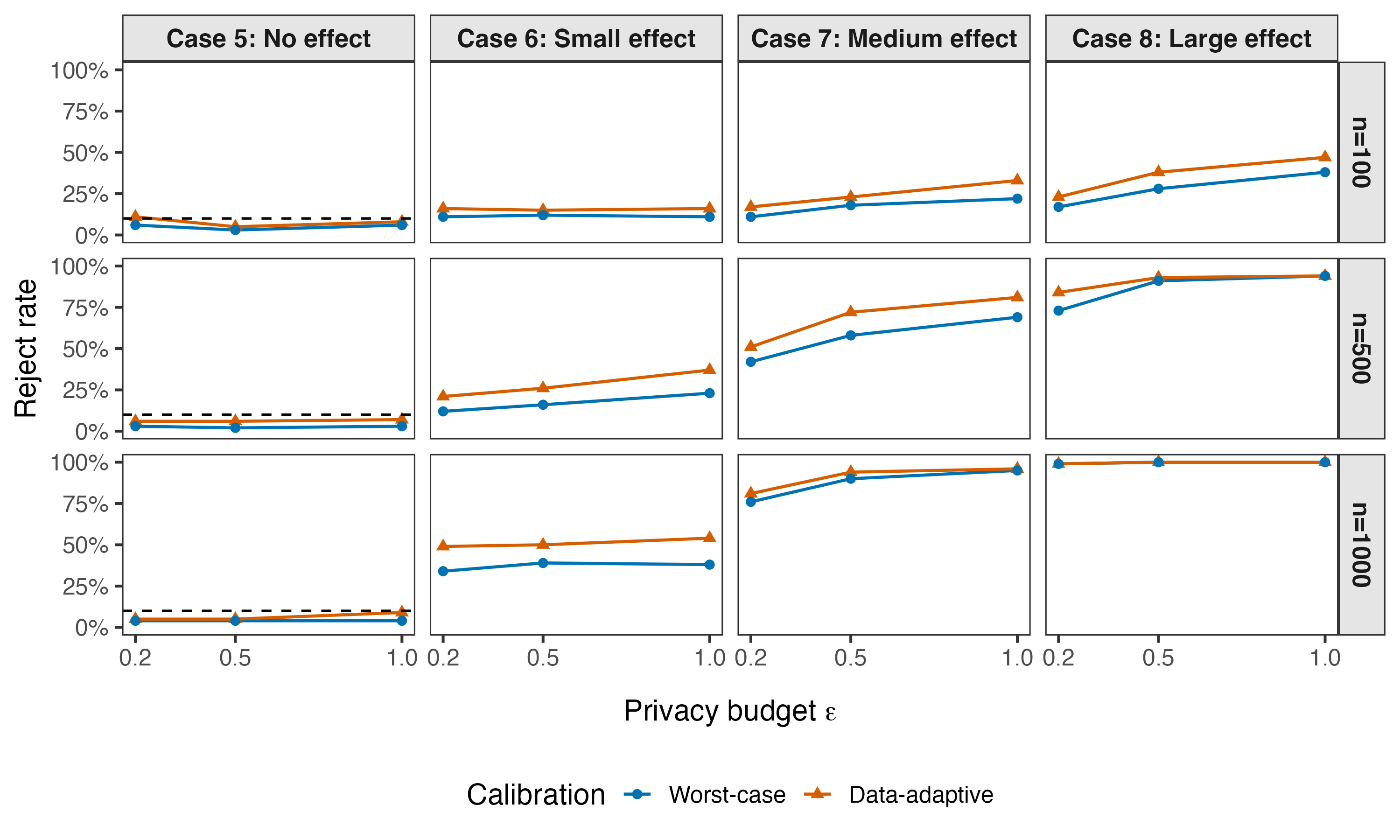}
    \caption{Rejection rates of the frequentist-calibrated rules for $\epsilon \in \{0.2, 0.5, 1\}$ in Cases 5--8 under $n \in \{100, 500, 1000\}$. Horizontal black dashed line indicates the nominal Type I error level $\alpha_{\rm Freq}=0.1$, while $\zeta = 0.01$ is used for the data-adaptive calibration method. Rates based on 100 simulation runs.}
    \label{fig:Freq}
\end{figure}

Figure \ref{fig:Freq} shows the rejection rates of the frequentist-calibrated decision rules across privacy budgets and sample sizes.
Both the worst-case calibration in Section 4.2.1 of the main text and the data-adaptive calibration in Section 4.2.2 of the main text control the Type I error at the nominal level under the sharp null (Case 5).
The rejection rates increase with larger effect sizes or sample sizes (Cases 6--8), and the data-adaptive calibration is slightly more powerful than the worst-case calibration, as illustrated in the main text.

\section{Theory Supporting Claims from the Main Text}
\label{sec:addtheorys}

Section \ref{sec:s31} proves that inferences from DP-FRT-Bayes are invariant to clipping the DP counts. Section \ref{sec:s_concentration} establishes the rate at which the DP-FRT-Bayes posterior concentrates around the confidential FRT $p$-value. Section \ref{sec:s_fidelity} bounds the probability that the binary Bayes decision differs from the confidential FRT decision. Section \ref{sec:s32} proves the statement that the chance of staying in the abstention region goes to zero as the additional privacy budget increases. Section \ref{sec:s_seqfidelity} extends the decision-fidelity bound to the sequential procedure. Section \ref{sec:psi_justification} presents two heuristic interpretations of $\Psi$. Section \ref{sec:alttest} presents a significance test based on an alternative test statistic and compares it with the test based on $\Psi$.

\subsection{Clipping Invariance of the Posterior} \label{sec:s31}

In Section 3.2 of the main text, we claim that posterior inferences in DP-FRT-Bayes are identical regardless of whether or not the DP counts are clipped.
This fact eliminates the need for post-processing to enforce feasible ranges on privatized counts. To be more specific, the Geometric mechanism can result in privatized counts $\tilde{n}_{11}$ and $\tilde{n}_{01}$ that fall outside the intervals $[0, n_1]$ and $[0, n_0]$, respectively. 
Lemma \ref{lem:clip} shows that it is not necessary to truncate these noisy counts to the feasible range of the true counts if one uses DP-FRT-Bayes for posterior inference. The posterior distribution remains unchanged when the released counts are clipped under the Geometric mechanism.

\begin{lemma}
\label{lem:clip}
Let $n\in\{0,\dots,T\}$ denote a true count, and let $\tilde n = n+\eta$ be generated from the Geometric mechanism. Let $m$ be a random variable used for inferences about the unknown $n$ with prior distribution $\{\pi_t\}_{t=0}^T$ on $\{0,\dots,T\}$. The likelihood function is $\tilde n \mid (m=t)\sim K(\cdot\mid t)$ with $K(\tilde n\mid t)=\kappa_\rho(|\tilde n-t|)$, where $\kappa_\rho$ is multiplicatively separable:
$\kappa_\rho(a+b)=\kappa_\rho(a)\kappa_\rho(b)/\kappa_\rho(0)$ for all integers $a,b\ge0$.
Define $\tilde n^{\mathrm{clip}}=\min(\max(\tilde n,0),T)$.
Then, for all $k\in\{0,\dots,T\}$,
\begin{equation}
\Pr(m=k\mid \tilde n)=\Pr(m=k\mid \tilde n^{\mathrm{clip}}).
\end{equation}
\end{lemma}

\begin{proof}
By Bayes' rule, for any $k\in\{0,\dots,T\}$ and any realized value $\tilde n=x$,
\begin{equation}
\Pr(m=k\mid \tilde n=x)=
\frac{\pi_k \kappa_\rho(|k-x|)}
{\sum_{t=0}^{T}\pi_t \kappa_\rho(|t-x|)}.
\end{equation}
If $x\in[0,T]$ there is nothing to prove. If $x<0$, then for all $k\in\{0,\dots,T\}$, $|k-x|=|k-0|+|0-x|$. Hence, by multiplicative separability, we have
\begin{equation}
\kappa_\rho(|k-x|)
=
\frac{\kappa_\rho(|k-0|)~\kappa_\rho(|0-x|)}{\kappa_\rho(0)}.
\end{equation}
Substituting this into the Bayes formula shows that the factor $\kappa_\rho(|0-x|)/\kappa_\rho(0)$ cancels between numerator and denominator, yielding
\begin{equation}
\Pr(m=k\mid \tilde n=x)=
\frac{\pi_k \kappa_\rho(|k-0|)}
{\sum_{t=0}^{T}\pi_t \kappa_\rho(|t-0|)}
=
\Pr(m=k\mid \tilde n^{\mathrm{clip}}=0).
\end{equation}
If $x>T$, then for all $k\in\{0,\dots,T\}$ we have $|k-x|=|T-k|+|x-T|$, and the same argument gives
\begin{equation}
\Pr(m=k\mid \tilde n=x)=
\frac{\pi_k \kappa_\rho(|T-k|)}
{\sum_{t=0}^{T}\pi_t \kappa_\rho(|T-t|)}
=
\Pr(m=k\mid \tilde n^{\mathrm{clip}}=T).
\end{equation}
Since these Bayes formulas do not depend on the particular value of $x$, $\Pr(m=k\mid \tilde n=x)$ is constant over $x\le0$ and over $x\ge T$; because $\{\tilde n^{\mathrm{clip}}=0\}=\{\tilde n\le0\}$ and $\{\tilde n^{\mathrm{clip}}=T\}=\{\tilde n\ge T\}$, conditioning on $\tilde n^{\mathrm{clip}}$ returns this common value. Thus, $\Pr(m=k\mid \tilde n)=\Pr(m=k\mid \tilde n^{\mathrm{clip}})$ in all cases.
\end{proof}

The argument applies coordinate-wise for $(n_{11},n_{01})$, because the kernel is multiplicatively separable and the joint kernel factorizes across coordinates.

\subsection{Concentration of the DP-FRT-Bayes Posterior}\label{sec:s_concentration}

In Section 3.2 of the main text, we state that the DP-FRT-Bayes posterior concentrates around the confidential FRT $p$-value as the privacy budget $\epsilon$ grows. Theorem \ref{thm:pvalue_concentration} formalizes this and gives the rate.

Let $S=\{0,\dots,n_1\}\times\{0,\dots,n_0\}$ be the support of $(m_{11},m_{01})$, and write $\|(a,b)-(a',b')\|_1=|a-a'|+|b-b'|$ for the $\ell_1$ lattice distance between tables. For any $G\subseteq S$ and $t>0$, define the distance-weighted prior odds
\begin{equation}
R_G(t)=\sum_{(a,b)\in G}\frac{\pi(a,b)}{\pi(n_{11},n_{01})}\exp\{-t\|(a,b)-(n_{11},n_{01})\|_1\},\qquad R_\emptyset(t)=0,
\label{eq:distance_weighted_prior_odds}
\end{equation}
where tables farther in $\ell_1$ distance from the confidential counts $(n_{11},n_{01})$ receive exponentially smaller weight. Define $\min\emptyset=+\infty$. For a tolerance $\nu\ge0$, let $G_\nu=\{(a,b)\in S: |p_{ab}-p_{n_{11},n_{01}}|>\nu\}$, whose smallest lattice distance $d_\nu=\min_{(a,b)\in G_\nu}\|(a,b)-(n_{11},n_{01})\|_1$ is the fewest single-unit changes to the confidential counts needed to move the FRT $p$-value by more than $\nu$.

\begin{theorem}
\label{thm:pvalue_concentration}
Assume the prior assigns positive mass to the confidential counts, i.e., $\pi(n_{11},n_{01})>0$, and fix $\nu\ge0$ with $d_\nu<+\infty$. Then, except on an event of Geometric-noise probability at most $1\wedge 4e^{-\epsilon d_\nu/8}$,
\begin{equation}
\Pr\left(|p_{\mathrm{FRT}}-p_{n_{11},n_{01}}|>\nu\mid\tilde{\bm n}\right)\le 1\wedge\left\{e^{\epsilon d_\nu/2}R_{G_\nu}(\epsilon)\right\}.
\label{eq:pvalue_concentration_R}
\end{equation}
\end{theorem}

The proofs of Theorem \ref{thm:pvalue_concentration}, Theorem \ref{thm:fidelity} and Corollary \ref{cor:seq_fidelity} rely on three auxiliary facts. Throughout, $\bm m=(m_{11},m_{01})$ denotes the true counts with support $S$ and prior $\pi$, and $\bm\eta=(\eta_{11},\eta_{01})$ is the Geometric noise in (4) of the main text, with $\eta_{11},\eta_{01}\stackrel{\mathrm{i.i.d.}}{\sim}\mathrm{Geom}(e^{-\epsilon})$.

\begin{lemma}
\label{lem:post_mass}
Let $G\subseteq S$, and assume $\pi(n_{11},n_{01})>0$. With $R_G(t)$ as in \eqref{eq:distance_weighted_prior_odds}, conditional on $\tilde{\bm n}$, we have
\begin{equation}
\Pr(\bm m\in G\mid\tilde{\bm n})
\le
\exp\{2\epsilon\|\bm\eta\|_1\}\, R_G(\epsilon).
\label{eq:post_mass_R}
\end{equation}
\end{lemma}

\begin{proof}
If $G=\emptyset$ the claim is trivial, so assume $G\ne\emptyset$. The common normalizing constant of the Geometric kernel cancels in posterior ratios, so the unnormalized posterior weight is
\begin{equation}
\tilde w(a,b)=\pi(a,b)\exp\{-\epsilon(|\tilde n_{11}-a|+|\tilde n_{01}-b|)\}.
\end{equation}
Retaining the contribution of $(n_{11},n_{01})$ lower-bounds the normalizer by $\pi(n_{11},n_{01})\exp\{-\epsilon\|\bm\eta\|_1\}$. For any $(a,b)\in S$, the triangle inequality gives
\begin{equation}
    |\tilde n_{11}-a|+|\tilde n_{01}-b|\ge\|(a,b)-(n_{11},n_{01})\|_1-\|\bm\eta\|_1,
\end{equation}
so that
\begin{equation}
\frac{\tilde w(a,b)}{\pi(n_{11},n_{01})e^{-\epsilon\|\bm\eta\|_1}}
\le
\frac{\pi(a,b)}{\pi(n_{11},n_{01})}\exp\{-\epsilon\|(a,b)-(n_{11},n_{01})\|_1+2\epsilon\|\bm\eta\|_1\}.
\end{equation}
Summing over $(a,b)\in G$ yields \eqref{eq:post_mass_R}.
\end{proof}

\begin{lemma}
\label{lem:geom_l1_tail}
Let $X_\epsilon=|\eta_{11}|+|\eta_{01}|$ and $\bar F_\epsilon(x)=\Pr(X_\epsilon\ge x)$, with $\bar F_\epsilon(x)=1$ for $x\le0$, and set $\mathcal M(\epsilon)=(1+e^{-\epsilon/2})^2/(1+e^{-\epsilon})$. Then $\mathbb E\{e^{\epsilon|\eta|/2}\}=\mathcal M(\epsilon)\le2$, and for all $x\in\mathbb R$,
\begin{equation}
\bar F_\epsilon(x)
\le
1\wedge \mathcal M(\epsilon)^2 e^{-\epsilon x/2}
\le
1\wedge 4e^{-\epsilon x/2},
\label{eq:tail_chernoff}
\end{equation}
while for $x>0$,
\begin{equation}
\bar F_\epsilon(x)
\le
1\wedge
\frac{4e^{-\epsilon x}\{(1-e^{-\epsilon})x+1\}}{(1+e^{-\epsilon})^2}.
\label{eq:tail_exact_rate}
\end{equation}
\end{lemma}

\begin{proof}
Set $\rho=e^{-\epsilon}$ and $c_\rho=(1-\rho)/(1+\rho)$, so $\Pr(|\eta|=0)=c_\rho$ and $\Pr(|\eta|=j)=2c_\rho\rho^j$ for $j\ge1$. Convolving the two coordinates gives $\Pr(X_\epsilon=s)=4c_\rho^2s\rho^s$ for $s\ge1$, hence for $x>0$ with $k=\lceil x\rceil$,
\begin{equation}
\bar F_\epsilon(x)=\frac{4\rho^{k}\{(1-\rho)k+\rho\}}{(1+\rho)^2}
=\frac{4e^{-\epsilon k}\{(1-e^{-\epsilon})k+e^{-\epsilon}\}}{(1+e^{-\epsilon})^2}.
\label{eq:geom_tail_def}
\end{equation}
Writing $u=e^{-\epsilon/2}$,
\begin{equation}
\mathbb E\{e^{\epsilon|\eta|/2}\}=c_\rho\,\frac{1+u}{1-u}=\frac{(1+u)^2}{1+u^2}=\mathcal M(\epsilon),
\end{equation}
and $\mathcal M(\epsilon)\le2$ follows from $(1-u)^2\ge0$. The Chernoff bound \eqref{eq:tail_chernoff} then follows from Markov's inequality with exponent $\epsilon/2$ and the independence of the two coordinates; for $x\le0$ its right side is at least $1$. For \eqref{eq:tail_exact_rate}, substitute $\rho^k\le\rho^x$ and $(1-\rho)k+\rho\le(1-\rho)(x+1)+\rho=(1-\rho)x+1$, valid since $k\le x+1$, into \eqref{eq:geom_tail_def}; this requires $x>0$, as $(1-\rho)x+1$ can be negative for sufficiently negative $x$.
\end{proof}

\begin{lemma}
\label{lem:flip}
Assume $\pi(n_{11},n_{01})>0$, and let $G\subseteq S$ with $(n_{11},n_{01})\notin G$ and $c\in(0,1]$. If a $\sigma(\tilde{\bm n})$-measurable event $B$ satisfies $B\subseteq\{\Pr(\bm m\in G\mid\tilde{\bm n})\ge c\}$, then
\begin{equation}
\Pr(B)\le\bar F_\epsilon(\tau),
\qquad
\tau=\frac{1}{2\epsilon}\log\frac{c}{R_G(\epsilon)}.
\end{equation}
Here $\bar F_\epsilon(\tau)=1$ when $\tau\le0$, and $\Pr(B)=0$ when $R_G(\epsilon)=0$.
\end{lemma}

\begin{proof}
If $R_G(\epsilon)=0$ then $\pi\equiv0$ on $G$ and $B$ is empty. Otherwise, on $B$, Lemma \ref{lem:post_mass} gives $c\le e^{2\epsilon\|\bm\eta\|_1}R_G(\epsilon)$, that is, $\|\bm\eta\|_1\ge\tau$. Hence $\Pr(B)\le\Pr(\|\bm\eta\|_1\ge\tau)=\bar F_\epsilon(\tau).$
\end{proof}

We now prove Theorem \ref{thm:pvalue_concentration}.

\begin{proof}
Let $G_\nu=\{(a,b)\in S:|p_{ab}-p_{n_{11},n_{01}}|>\nu\}$. By Lemma \ref{lem:post_mass},
\begin{equation}
\Pr(|p_{\mathrm{FRT}}-p_{n_{11},n_{01}}|>\nu\mid\tilde{\bm n})
=\Pr(\bm m\in G_\nu\mid\tilde{\bm n})
\le e^{2\epsilon\|\bm\eta\|_1}R_{G_\nu}(\epsilon).
\end{equation}
On the event $\{\|\bm\eta\|_1\le d_\nu/4\}$ the right side is at most $e^{\epsilon d_\nu/2}R_{G_\nu}(\epsilon)$. By the Chernoff bound \eqref{eq:tail_chernoff},
\begin{equation}
\Pr(\|\bm\eta\|_1>d_\nu/4)\le\bar F_\epsilon(d_\nu/4)\le 4e^{-\epsilon d_\nu/8},
\end{equation}
so combining the two bounds gives the stated result.
\end{proof}

\begin{remark}
Using the exact-rate tail \eqref{eq:tail_exact_rate} in place of the Chernoff bound \eqref{eq:tail_chernoff} further sharpens the exceptional probability to
\begin{equation}
\bar F_\epsilon(d_\nu/4)\le1\wedge\frac{4e^{-\epsilon d_\nu/4}\{(1-e^{-\epsilon})d_\nu/4+1\}}{(1+e^{-\epsilon})^2},
\end{equation}
so the exceptional probability decays at rate $\epsilon d_\nu/4$ rather than $\epsilon d_\nu/8$. We keep the more readable version in Theorem \ref{thm:pvalue_concentration}.
\end{remark}

\subsection{Fidelity of the Binary Bayes Decision}\label{sec:s_fidelity}

In Section 4.1 of the main text, we state that the binary Bayes rule closely reproduces the decision the FRT would reach on the confidential counts. Theorem \ref{thm:fidelity} bounds the probability that the two disagree.

Recall $S=\{0,\dots,n_1\}\times\{0,\dots,n_0\}$, and for $h\in\{0,1\}$ define $S_h=\{(a,b)\in S:\mathbf{1}(p_{ab}\le\alpha)=h\}$, so that $S_1$ is the rejection set. Let $\delta^{\mathrm{conf}}=\mathbf{1}\{p_{n_{11},n_{01}}\le\alpha\}=\mathbf{1}\{(n_{11},n_{01})\in S_1\}$ be the confidential FRT decision. Write $c^*=\lambda_0/(\lambda_0+\lambda_1)$ for the threshold in (14) of the main text and $c_{\mathrm B}=\min\{c^*,1-c^*\}=\min(\lambda_0,\lambda_1)/(\lambda_0+\lambda_1)$.

\begin{theorem}
\label{thm:fidelity}
Assume the prior assigns positive mass to the confidential counts, i.e., $\pi(n_{11},n_{01})>0$, and $S_{1-\delta^{\mathrm{conf}}}\neq\emptyset$. Then
\begin{equation}
\Pr\left\{\delta_{\mathrm{Bayes}}(\tilde{\bm n})\ne\delta^{\mathrm{conf}}\right\}\le 1\wedge 4\left\{\frac{R_{S_{1-\delta^{\mathrm{conf}}}}(\epsilon)}{c_{\mathrm B}}\right\}^{1/4}.
\label{eq:fidelity_quarter}
\end{equation}
\end{theorem}

\begin{proof}
Let $G=S_{1-\delta^{\mathrm{conf}}}$, so $(n_{11},n_{01})\notin G$, and write $R=R_G(\epsilon)$. Let $\Psi=\Pr(p_{\mathrm{FRT}}\le\alpha\mid\tilde{\bm n})=\Pr(\bm m\in S_1\mid\tilde{\bm n})$ be the posterior rejection evidence in (14). A disagreement places posterior mass at least $c_{\mathrm B}$ on $G$: if $\delta^{\mathrm{conf}}=1$ it requires $\Psi\le c^*$, so that $\Pr(\bm m\in G\mid\tilde{\bm n})=1-\Psi\ge1-c^*\ge c_{\mathrm B}$, while if $\delta^{\mathrm{conf}}=0$ it requires $\Psi>c^*\ge c_{\mathrm B}$. Hence the disagreement event lies in $\{\Pr(\bm m\in G\mid\tilde{\bm n})\ge c_{\mathrm B}\}$, and Lemma \ref{lem:flip} with $c=c_{\mathrm B}$ gives the certificate
\begin{equation}
\Pr\{\delta_{\mathrm{Bayes}}(\tilde{\bm n})\ne\delta^{\mathrm{conf}}\}
\le
\bar F_\epsilon(\tau_{\mathrm B}),
\qquad
\tau_{\mathrm B}=\frac{1}{2\epsilon}\log\frac{c_{\mathrm B}}{R}.
\label{eq:fidelity_certificate}
\end{equation}
Applying the Chernoff bound \eqref{eq:tail_chernoff} with $e^{-\epsilon\tau_{\mathrm B}/2}=(R/c_{\mathrm B})^{1/4}$ yields the fourth-root bound in the statement.
\end{proof}

\begin{remark}
Using the exact-rate tail \eqref{eq:tail_exact_rate} in place of the Chernoff bound \eqref{eq:tail_chernoff} further sharpens the fourth-root bound to a square-root bound. When $\tau_{\mathrm B}>0$, applying \eqref{eq:tail_exact_rate} to \eqref{eq:fidelity_certificate} with $e^{-\epsilon\tau_{\mathrm B}}=(R/c_{\mathrm B})^{1/2}$ gives
\begin{equation}
\Pr\{\delta_{\mathrm{Bayes}}(\tilde{\bm n})\ne\delta^{\mathrm{conf}}\}
\le
1\wedge
\frac{4\{(1-e^{-\epsilon})\tau_{\mathrm B}+1\}}{(1+e^{-\epsilon})^2}
\left(\frac{R}{c_{\mathrm B}}\right)^{1/2},
\end{equation}
which replaces the exponent $1/4$ by $1/2$. We keep the more readable fourth-root version in Theorem \ref{thm:fidelity}.
\end{remark}

\subsection{Escape from the Abstention Region}\label{sec:s32}

In Section 4.1.2 of the main text, we state that, conditional on $\Psi\in A$ where $A$ is the abstention region,
the probability of remaining in the abstention region after the top-up release
decays exponentially fast in the additional privacy budget $\epsilon_{\mathrm{plus}}$.
Intuitively, as $\epsilon_{\mathrm{plus}}\to\infty$, the second-stage release reveals the true cell counts with probability tending to one. Consequently, the refined
posterior evidence $\Psi^{+}$ concentrates on the non-private decision
$\mathbf 1(p_{\mathrm{FRT}}\le\alpha)$.
Lemma \ref{lem:lower_abstention} formalizes this exponential escape guarantee.

\begin{lemma}
\label{lem:lower_abstention}
Let $\Psi=\Pr(p_{\mathrm{FRT}}\le\alpha\mid\tilde{\bm{n}})$ and $\Psi^{+}=\Pr(p_{\mathrm{FRT}}\le\alpha\mid\tilde{\bm{n}},\tilde{\bm{n}}^{+})$. Let $A=(t_{\mathrm{low}},t_{\mathrm{high}})$ be the abstention region with 
$0<t_{\mathrm{low}}<t_{\mathrm{high}}<1$. 
Assume $\Pr(\Psi\in A)>0$. There exists a finite constant $c>0$ such that, for all $\epsilon_{\mathrm{plus}}>0$,
\begin{equation}
\Pr\left(\Psi^{+}\notin A \mid \Psi\in A\right)
 \ge 
1 - c e^{-\epsilon_{\mathrm{plus}}}.
\label{eq:hp_escape}
\end{equation}
\end{lemma}

\begin{proof}
The proof proceeds in three steps: (1) establishing posterior concentration on the true cell counts after the top-up release; (2) translating posterior concentration into escape from the abstention region; and, (3) converting the unconditional escape bound into the desired conditional probability bound.

In the first step, we analyze the posterior concentration around the true counts.
Define the event
$\mathcal{E} = \{\eta_{11}^{+}=0,~\eta_{01}^{+}=0\}.$
Since $\eta^{+}_{11},\eta^{+}_{01}\stackrel{\text{i.i.d.}}{\sim}\mathrm{Geom}(\rho_{+})$, we have
\begin{equation}
\Pr(\mathcal{E}^c)
=
1 - \left(\frac{1-\rho_{+}}{1+\rho_{+}}\right)^2 
=
\frac{4\rho_{+}}{(1+\rho_{+})^2}
 \le 
4\rho_{+}.
\end{equation}
On the event $\mathcal{E}$, we have $\tilde n_{11}^+=n_{11}$ and 
$\tilde n_{01}^+=n_{01}$, so
\begin{equation}
\kappa_{\rho_{+}}(\tilde n_{11}^+-a) \kappa_{\rho_{+}}(\tilde n_{01}^+-b)
=
\kappa_{\rho_{+}}(n_{11}-a) \kappa_{\rho_{+}}(n_{01}-b).
\end{equation}
For $(a,b)=(n_{11},n_{01})$ this factor equals 
$\kappa_{\rho_{+}}(0)\kappa_{\rho_{+}}(0)$, while for any $(a,b)\neq(n_{11},n_{01})$
we have $|a-n_{11}|+|b-n_{01}|\ge 1$. Thus, we have 
\begin{equation}
\label{eq:first_bound}
\kappa_{\rho_{+}}(n_{11}-a) \kappa_{\rho_{+}}(n_{01}-b)
 \le 
\kappa_{\rho_{+}}(0)^2 (\rho_{+})^{|a-n_{11}|+|b-n_{01}|}
 \le 
\kappa_{\rho_{+}}(0)^2 \rho_{+}.
\end{equation}

Next, we control the remaining factors uniformly in $\tilde{\bm n}$.
Let $S$ denote the finite support of $(m_{11},m_{01})$, which are the random variables representing the unknown true counts in the analyst's inference.
Because the prior is strictly positive on $S$, there exist constants
$0<\pi_{\min}\le\pi(a,b)\le\pi_{\max}<\infty$ for all $(a,b)\in S$.
For any integer $x$ and any $(a,b)\in S$,
\begin{equation}
\frac{\kappa_\rho(x-a)}{\kappa_\rho(x-n_{11})}
=
\rho^{ |x-a|-|x-n_{11}|},
\end{equation}
and by the triangle inequality, $\left||x-a|-|x-n_{11}|\right|\le |a-n_{11}|$.
Since $0<\rho<1$, this implies
\begin{equation}
\frac{\kappa_\rho(x-a)}{\kappa_\rho(x-n_{11})}
 \le 
\rho^{-|a-n_{11}|}.
\end{equation}
Applying this to both coordinates, we obtain, for any $(a,b)\in S$,
\begin{equation}
\frac{
\kappa_\rho(\tilde n_{11}-a) \kappa_\rho(\tilde n_{01}-b)
}{
\kappa_\rho(\tilde n_{11}-n_{11}) \kappa_\rho(\tilde n_{01}-n_{01})
}
 \le 
\rho^{-|a-n_{11}|-|b-n_{01}|},
\end{equation}
and hence
\begin{equation}
\frac{
\pi(a,b) \kappa_\rho(\tilde n_{11}-a) \kappa_\rho(\tilde n_{01}-b)
}{
\pi(n_{11},n_{01}) \kappa_\rho(\tilde n_{11}-n_{11}) \kappa_\rho(\tilde n_{01}-n_{01})
}
 \le 
\frac{\pi_{\max}}{\pi_{\min}} 
\rho^{-|a-n_{11}|-|b-n_{01}|}.
\end{equation}
Since $S$ is finite, the right-hand side admits a finite maximum over
$(a,b)\neq(n_{11},n_{01})$.
Thus there exists a finite constant $C_1>0$, depending only on the prior and
$\rho$, such that for all $\tilde{\bm n}$ and all $(a,b)\neq(n_{11},n_{01})$,
\begin{equation}
\label{eq:second_bound}
\frac{
\pi(a,b) \kappa_\rho(\tilde n_{11}-a) \kappa_\rho(\tilde n_{01}-b)
}{
\pi(n_{11},n_{01}) \kappa_\rho(\tilde n_{11}-n_{11}) \kappa_\rho(\tilde n_{01}-n_{01})
}
 \le C_1.
\end{equation}

Combining the bounds \eqref{eq:first_bound} and \eqref{eq:second_bound}, we obtain
that on $\mathcal{E}$,
\begin{equation}
\frac{\gamma^{+}(a,b)}{\gamma^{+}(n_{11},n_{01})}
 \le 
C_1 \rho_{+}
\qquad\text{for all }(a,b)\neq(n_{11},n_{01}).
\end{equation}
Summing over all $(a,b)\neq(n_{11},n_{01})$ in the finite support $S$ gives
\begin{equation}
1-\gamma^{+}(n_{11},n_{01})
=
\sum_{(a,b)\neq(n_{11},n_{01})}\gamma^{+}(a,b)
 \le 
C_2 \rho_{+}
\quad\text{on }\mathcal{E},
\end{equation}
for some finite $C_2>0$ independent of $\tilde{\bm n}$ and $\rho_{+}$.
Notice that on $\mathcal{E}^c$ we have the trivial bound $1-\gamma^{+}(n_{11},n_{01})\le 1$.
Therefore,
\begin{equation}
1-\gamma^{+}(n_{11},n_{01})
 \le 
\mathbf 1_{\mathcal{E}^c}
+
\mathbf 1_{\mathcal{E}} C_2 \rho_{+}.
\end{equation}
Taking expectations and using $\Pr(\mathcal{E}^c)\le 4\rho_{+}$ yields
\begin{equation}
\mathbb E\left(1-\gamma^{+}(n_{11},n_{01})\right)
 \le 
4\rho_{+} + C_2 \rho_{+}
 \le 
C \rho_{+}
 = 
C e^{-\epsilon_{\mathrm{plus}}},
\label{eq:true_mass_uncond}
\end{equation}
for some finite constant $C>0$ independent of $\epsilon_{\mathrm{plus}}$.

In the second step, we translate posterior concentration into abstention probability.
Let $p_{\mathrm{FRT}}=g(n_{11},n_{01})$ be the non-private FRT $p$-value.
Notice that the corresponding non-private decision $H=\mathbf 1(g(n_{11},n_{01})\le\alpha)$ is a
deterministic function of $(n_{11},n_{01})$.
Then, by the definition of $\Psi^{+}$, we have
\begin{equation}
\Psi^{+}
=
\sum_{a,b}\mathbf 1(g(a,b)\le\alpha) \gamma^{+}(a,b).
\end{equation}
If $H=1$, then $(n_{11},n_{01})$ is in the rejection region and
\begin{equation}
\Psi^{+}
=
\gamma^{+}(n_{11},n_{01})
+
\sum_{\substack{(a,b)\neq(n_{11},n_{01}) \\ g(a,b)\le \alpha}}\gamma^{+}(a,b)
 \ge 
\gamma^{+}(n_{11},n_{01}),
\end{equation}
hence
\begin{equation}
1-\Psi^{+}
 \le 
1-\gamma^{+}(n_{11},n_{01}).
\end{equation}
If $H=0$, then $(n_{11},n_{01})$ is in the acceptance region and
\begin{equation}
\Psi^{+}
=
\sum_{\substack{(a,b)\neq(n_{11},n_{01}) \\ g(a,b)\le \alpha}}\gamma^{+}(a,b)
 \le 
1-\gamma^{+}(n_{11},n_{01}).
\end{equation}

Denote $d_A = \min\{t_{\mathrm{low}},~1-t_{\mathrm{high}}\} > 0$.
If $1-\gamma^{+}(n_{11},n_{01})\le d_A$ and $H=0$, then $\Psi^{+}\le d_A\le t_{\mathrm{low}}$, and thus $\Psi^{+}\notin A$.
If $1-\gamma^{+}(n_{11},n_{01})\le d_A$ and $H=1$, then $1-\Psi^{+}\le d_A\le 1-t_{\mathrm{high}}$, so $\Psi^{+}\ge t_{\mathrm{high}}$ and hence $\Psi^{+}\notin A$.
Therefore, we have
\begin{equation}
\{\Psi^{+}\in A\}
 \subseteq 
\{1-\gamma^{+}(n_{11},n_{01})>d_A\}.
\end{equation}
Taking probabilities and applying Markov's inequality together with
\eqref{eq:true_mass_uncond} gives
\begin{equation}
\begin{aligned}
\Pr(\Psi^{+}\in A)
& \le 
\Pr \left(1-\gamma^{+}(n_{11},n_{01}) > d_A\right)\\
& \le 
\frac{\mathbb{E} \left( 1-\gamma^{+}(n_{11},n_{01}) \right)}{d_A}\\
& \le 
\frac{C}{d_A} e^{-\epsilon_{\mathrm{plus}}}
 = 
C_A e^{-\epsilon_{\mathrm{plus}}}.
\end{aligned}
\label{eq:psi_plus_A_uncond}
\end{equation}

In the final step, we derive the conditional escape probability that is of interest.
By Bayes' rule,
\begin{equation}
\Pr(\Psi^{+}\in A\mid\Psi\in A)
=
\frac{\Pr(\Psi^{+}\in A,\Psi\in A)}{\Pr(\Psi\in A)}
 \le 
\frac{\Pr(\Psi^{+}\in A)}{\Pr(\Psi\in A)}.
\end{equation}
Since $\Psi$ depends only on the first release, $\Pr(\Psi\in A)=p_A>0$ is a
constant independent of $\epsilon_{\mathrm{plus}}$. 
Combining this with \eqref{eq:psi_plus_A_uncond}, we obtain
\begin{equation}
\Pr(\Psi^{+}\in A\mid\Psi\in A)
 \le 
\frac{C_A}{p_A} e^{-\epsilon_{\mathrm{plus}}}.
\end{equation}
Setting $c=C_A/p_A$ yields
\begin{equation}
\Pr(\Psi^{+}\notin A\mid\Psi\in A)
= 1 - \Pr(\Psi^{+}\in A\mid\Psi\in A)
 \ge 
1 - c e^{-\epsilon_{\mathrm{plus}}},
\end{equation}
which completes the proof.
\end{proof}

\subsection{Fidelity of the Sequential Decision}\label{sec:s_seqfidelity}

In Section 4.1.2 of the main text, we state that the fidelity guarantee of Theorem \ref{thm:fidelity} extends to the refined ternary decision $\delta_{\mathrm{Bayes}}^{+}$ after a top-up release. Let $\delta_{\mathrm{Bayes}}^{+}$ denote the ternary rule (15) in the main text evaluated at the refined evidence $\Psi^{+}$, which either commits a decision in $\{0,1\}$ or abstains. Because a top-up is issued only when the first stage abstains, we bound the probability of a committed misclassification after abstention, namely, the event that the first stage abstains and the refined stage subsequently commits to a decision that differs from the confidential FRT decision $\delta^{\mathrm{conf}}$. Set $c_{\mathrm U}=\min\{t_{\mathrm{high}},1-t_{\mathrm{low}}\}$. Because the analyst may choose $\epsilon_{\mathrm{plus}}$ based on the first release through the lower bound in (20) of the main text, Corollary \ref{cor:seq_fidelity} allows $\epsilon_{\mathrm{plus}}$ to depend on $\tilde{\bm n}$.

\begin{corollary}
\label{cor:seq_fidelity}
Assume the prior assigns positive mass to the confidential counts, i.e., $\pi(n_{11},n_{01})>0$, and $S_{1-\delta^{\mathrm{conf}}}\neq\emptyset$, and suppose the top-up rule uses budget $\epsilon_{\mathrm{plus}}(\tilde{\bm n})\ge\epsilon_{\min}$, possibly depending on the first release, for a preset floor $\epsilon_{\min}>0$. Then the refined ternary rule satisfies
\begin{equation}
\Pr\left\{\delta_{\mathrm{Bayes}}^{+}\notin\{u,\delta^{\mathrm{conf}}\},\ \delta^*_{\mathrm{Bayes}}(\tilde{\bm n})=u\right\}\le 1\wedge 16\left\{\frac{R_{S_{1-\delta^{\mathrm{conf}}}}(\epsilon+\epsilon_{\min})}{c_{\mathrm U}}\right\}^{1/4}.
\label{eq:seq_fidelity_R}
\end{equation}
\end{corollary}

\begin{proof}
Write $G=S_{1-\delta^{\mathrm{conf}}}$, so $(n_{11},n_{01})\notin G$, and let
\begin{equation}
E=\{\delta_{\mathrm{Bayes}}^{+}\notin\{u,\delta^{\mathrm{conf}}\}\}\cap\{\delta^*_{\mathrm{Bayes}}(\tilde{\bm n})=u\}
\end{equation}
be the committed-misclassification event. On $E$ the first stage abstains, so a top-up is issued. Condition on $\tilde{\bm n}$, which fixes $\bm\eta$ and $\epsilon_{\mathrm{plus}}=\epsilon_{\mathrm{plus}}(\tilde{\bm n})$; $\bm\eta^{+}=(\eta_{11}^{+},\eta_{01}^{+})$ then has i.i.d.\ $\mathrm{Geom}(e^{-\epsilon_{\mathrm{plus}}})$ coordinates. As in Lemma \ref{lem:post_mass}, applied to the two-release kernel $\kappa_\rho\kappa_{\rho_+}$ in (18) of the main text,
\begin{equation}
\Pr(\bm m\in G\mid\tilde{\bm n},\tilde{\bm n}^{+})
\le
\exp\{2\epsilon\|\bm\eta\|_1+2\epsilon_{\mathrm{plus}}\|\bm\eta^{+}\|_1\}\,
R_G(\epsilon+\epsilon_{\mathrm{plus}}).
\end{equation}
On $E$ the refined posterior places mass at least $c_{\mathrm U}$ on $G$. If $\delta^{\mathrm{conf}}=0$, a committed wrong rejection forces $\Psi^{+}>t_{\mathrm{high}}$, so the mass on $G=S_1$ exceeds $t_{\mathrm{high}}\ge c_{\mathrm U}$; if $\delta^{\mathrm{conf}}=1$, a committed wrong non-rejection forces $\Psi^{+}<t_{\mathrm{low}}$, so the mass on $G=S_0$ exceeds $1-t_{\mathrm{low}}\ge c_{\mathrm U}$. Write $R_+=R_G(\epsilon+\epsilon_{\mathrm{plus}})$; if $R_+=0$ then $E$ has conditional probability zero, so assume $R_+>0$. Combining the two preceding bounds gives, on $E$,
\begin{equation}
\|\bm\eta^{+}\|_1\ge(2\epsilon_{\mathrm{plus}})^{-1}\{\log(c_{\mathrm U}/R_+)-2\epsilon\|\bm\eta\|_1\}.
\end{equation}
The Chernoff bound \eqref{eq:tail_chernoff} for $\bm\eta^{+}$ then gives
\begin{equation}
\Pr(E\mid\tilde{\bm n})
\le
1\wedge
\mathcal M(\epsilon_{\mathrm{plus}})^2
\left\{\frac{R_+}{c_{\mathrm U}}\right\}^{1/4}
 e^{\epsilon\|\bm\eta\|_1/2}.
\end{equation}
Taking expectations over $\bm\eta$ therefore gives
\begin{equation}
\Pr(E)\le\mathbb E\!\left[\mathcal M(\epsilon_{\mathrm{plus}})^2\left\{\frac{R_+}{c_{\mathrm U}}\right\}^{1/4}e^{\epsilon\|\bm\eta\|_1/2}\right].
\end{equation}
Now $R_+\le R_G(\epsilon+\epsilon_{\min})$ by monotonicity of $R_G$ and $\epsilon_{\mathrm{plus}}\ge\epsilon_{\min}$, while $\mathcal M(\epsilon_{\mathrm{plus}})\le2$ and $\mathbb E\{e^{\epsilon\|\bm\eta\|_1/2}\}=\mathcal M(\epsilon)^2\le4$ by Lemma \ref{lem:geom_l1_tail}. Together with the trivial bound $\Pr(E)\le1$, these give the stated result. When $\epsilon_{\mathrm{plus}}$ is fixed in advance, $\epsilon_{\min}=\epsilon_{\mathrm{plus}}$ and the constant sharpens to $\mathcal M(\epsilon)^2\mathcal M(\epsilon_{\mathrm{plus}})^2$.
\end{proof}

\subsection{Additional Interpretations of the Statistic in Frequentist-calibrated Rules}
\label{sec:psi_justification}

In Section 4.2 of the main text, we calibrate a frequentist threshold based on the test statistic $\Psi(\tilde{\bm n})=\Pr(p_{\mathrm{FRT}}\le\alpha\mid\tilde{\bm n})$. The same calibration pipeline applies to any function of the privatized counts. Here, we provide two interpretations of $\Psi(\tilde{\bm n})$ that further support why it can be a desirable choice for testing the non-private rejection event from the privatized data.

Fix the rejection level $\alpha$, the privacy parameter $\epsilon$, and the group sizes $(n_1,n_0)$. The analyst specifies a prior $\pi$ over the latent cell counts $(m_{11},m_{01})\in\{0,\dots,n_1\}\times\{0,\dots,n_0\}$. The privatized counts $\tilde{\bm n}=(\tilde n_{11},\tilde n_{01})$ are generated by the Geometric mechanism with conditional pmf $K(\tilde{\bm n}\mid a,b)=\kappa_\rho(\tilde n_{11}-a)\,\kappa_\rho(\tilde n_{01}-b)$. All probabilities and expectations are taken under the joint distribution induced by $\pi$ and $K$. 

We define the non-private rejection indicator $J = \mathbf{1}\{p_{m_{11},m_{01}}\le\alpha\}$, where $p_{ab}$ is the FRT $p$-value. Because $J$ is a deterministic function of $(m_{11},m_{01})$, we have $\Psi(\tilde{\bm n})=\mathbb E \left(J\mid\tilde{\bm n}\right)$. Let $\pi_1=\Pr(J=1)$ and $\pi_0=1-\pi_1$, and assume $0<\pi_1<1$. Averaging the mechanism kernel over each hypothesis class gives the conditional pmf of $\tilde{\bm n}$ given $J$. We have 
\begin{equation}
\label{eq:f1_f0}
f_h(\tilde{\bm n})
=
\Pr(\tilde{\bm n}\mid J=h)
=
\frac{1}{\pi_h}\sum_{\substack{(a,b):\\\mathbf{1}\{p_{ab}\le\alpha\}=h}}\pi(a,b)\,K(\tilde{\bm n}\mid a,b),
\qquad h\in\{0,1\},
\end{equation}
which reduces the composite hypotheses $\{(a,b):p_{ab}\le\alpha\}$ and $\{(a,b):p_{ab}>\alpha\}$ to a simple-vs-simple testing problem on $\tilde{\bm n}$.

We establish two optimality properties. The first (Lemma \ref{lem:np_psi}) shows that thresholding $\Psi(\tilde{\bm n})$ is a most powerful test in the Neyman--Pearson sense for the prior-weighted mixture problem defined by \eqref{eq:f1_f0}. The second (Lemma \ref{lem:dpi_psi}) shows that $\Psi(\tilde{\bm n})$ is a minimal sufficient statistic for $J$ given $\tilde{\bm n}$ in the information-theoretic sense, so that any alternative summary that does not determine $\Psi(\tilde{\bm n})$ strictly discards information about the non-private rejection event.

\begin{lemma}[Neyman--Pearson optimality]
\label{lem:np_psi}
Under the model in \eqref{eq:f1_f0},
\begin{equation}
\label{eq:psi_lr}
\Psi(\tilde{\bm n})
=
\frac{\pi_1 f_1(\tilde{\bm n})}{\pi_1 f_1(\tilde{\bm n}) + \pi_0 f_0(\tilde{\bm n})},
\end{equation}
so $\Psi(\tilde{\bm n})$ is a monotone increasing function of the likelihood ratio $f_1(\tilde{\bm n})/f_0(\tilde{\bm n})$. For any Type I error level, the test that rejects when $\Psi(\tilde{\bm n})$ exceeds a threshold $c$, with randomization on the boundary $\{\Psi(\tilde{\bm n})=c\}$ if needed to attain the size exactly, is most powerful for testing $f_0$ against $f_1$.
\end{lemma}

\begin{proof}
By Bayes' rule,
\begin{align*}
\Psi(\tilde{\bm n})
&=
\Pr(J=1\mid\tilde{\bm n})
=
\frac{\Pr(\tilde{\bm n}\mid J=1)\,\Pr(J=1)}{\Pr(\tilde{\bm n})} \\
&=
\frac{\pi_1 f_1(\tilde{\bm n})}{\pi_1 f_1(\tilde{\bm n}) + \pi_0 f_0(\tilde{\bm n})},
\end{align*}
which establishes \eqref{eq:psi_lr}. Since $\pi_0,\pi_1>0$, the map $\ell\mapsto\pi_1\ell/(\pi_1\ell+\pi_0)$ is strictly increasing in $\ell=f_1(\tilde{\bm n})/f_0(\tilde{\bm n})$, so $\Psi(\tilde{\bm n})$ is a monotone increasing function of the likelihood ratio. By the Neyman--Pearson lemma, thresholding the likelihood ratio (and hence $\Psi$), with randomization at the threshold when the target size is not attained exactly, yields a most powerful test of $f_0$ against $f_1$ at any fixed size; such randomization is generally needed since $\tilde{\bm n}$ is discrete.
\end{proof}

Lemma \ref{lem:np_psi} implies that no other summary of $\tilde{\bm n}$ can achieve higher power at a given Type I error level for the prior-weighted mixture problem. We next strengthen this conclusion from an information-theoretic perspective. Let $I(X;Y)$ denote the Shannon mutual information between random variables $X$ and $Y$, and let $I(X;Y\mid Z)$ denote the conditional mutual information given $Z$.

\begin{lemma}[Information-theoretic sufficiency]
\label{lem:dpi_psi}
$I(J;\Psi(\tilde{\bm n}))=I(J;\tilde{\bm n})$. Moreover, for any measurable $g(\tilde{\bm n})$,
$I(J;g(\tilde{\bm n}))\le I(J;\tilde{\bm n})$,
with equality if and only if $\Psi(\tilde{\bm n})$ is $\sigma(g(\tilde{\bm n}))$-measurable.
\end{lemma}

\begin{proof}
By definition, $\Pr(J=1\mid\tilde{\bm n})=\Psi(\tilde{\bm n})$. By the tower property,
\begin{equation}
\Pr(J=1\mid\Psi(\tilde{\bm n}))
=
\mathbb{E}\bigl[\Pr(J=1\mid\tilde{\bm n})\,\big|\,\Psi(\tilde{\bm n})\bigr]
=
\mathbb{E}\bigl[\Psi(\tilde{\bm n})\,\big|\,\Psi(\tilde{\bm n})\bigr]
=
\Psi(\tilde{\bm n}).
\end{equation}
Since $\Psi(\tilde{\bm n})$ is $\sigma(\tilde{\bm n})$-measurable, $\sigma(\tilde{\bm n},\Psi(\tilde{\bm n}))=\sigma(\tilde{\bm n})$, so
$\Pr(J=1\mid\tilde{\bm n},\Psi(\tilde{\bm n}))=\Pr(J=1\mid\tilde{\bm n})=\Psi(\tilde{\bm n})=\Pr(J=1\mid\Psi(\tilde{\bm n}))$.
Because $J$ is $\{0,1\}$-valued, its conditional distribution given any $\sigma$-algebra is determined by the conditional probability of $\{J=1\}$; hence $J\perp\!\!\!\perp\tilde{\bm n}\mid\Psi(\tilde{\bm n})$. By the chain rule for mutual information,
\begin{equation}
I(J;\tilde{\bm n})
=
I(J;\Psi(\tilde{\bm n}))
+
I(J;\tilde{\bm n}\mid\Psi(\tilde{\bm n})).
\end{equation}
The conditional independence gives $I(J;\tilde{\bm n}\mid\Psi(\tilde{\bm n}))=0$. Hence, $I(J;\Psi(\tilde{\bm n}))=I(J;\tilde{\bm n})$.

For the second claim, since $g(\tilde{\bm n})$ is a deterministic function of $\tilde{\bm n}$, the chain rule gives
\begin{equation}
I(J;\tilde{\bm n})
=
I(J;g(\tilde{\bm n}))
+
I(J;\tilde{\bm n}\mid g(\tilde{\bm n})).
\end{equation}
Thus, $I(J;g(\tilde{\bm n}))\le I(J;\tilde{\bm n})$, with equality if and only if $I(J;\tilde{\bm n}\mid g(\tilde{\bm n}))=0$, i.e., $J\perp\!\!\!\perp\tilde{\bm n}\mid g(\tilde{\bm n})$.

It remains to show that $J\perp\!\!\!\perp\tilde{\bm n}\mid g(\tilde{\bm n})$ holds if and only if $\Psi(\tilde{\bm n})$ is $\sigma(g(\tilde{\bm n}))$-measurable. Suppose $J\perp\!\!\!\perp\tilde{\bm n}\mid g(\tilde{\bm n})$. Then
$\Psi(\tilde{\bm n})=\Pr(J=1\mid\tilde{\bm n})=\Pr(J=1\mid g(\tilde{\bm n}))$ almost surely,
so $\Psi(\tilde{\bm n})$ is $\sigma(g(\tilde{\bm n}))$-measurable. Conversely, suppose $\Psi(\tilde{\bm n})=h(g(\tilde{\bm n}))$ almost surely for some measurable $h$. Then
\begin{equation}
\Pr(J=1\mid\tilde{\bm n},g(\tilde{\bm n}))
=
\Pr(J=1\mid\tilde{\bm n})
=
\Psi(\tilde{\bm n})
=
h(g(\tilde{\bm n})).
\end{equation}
Moreover, by the tower property, $\Pr(J=1\mid g(\tilde{\bm n}))=\mathbb{E}[\Psi(\tilde{\bm n})\mid g(\tilde{\bm n})]=\mathbb{E}[h(g(\tilde{\bm n}))\mid g(\tilde{\bm n})]=h(g(\tilde{\bm n}))$. Thus, $\Pr(J=1\mid\tilde{\bm n},g(\tilde{\bm n}))=\Pr(J=1\mid g(\tilde{\bm n}))$. Since $J$ is $\{0,1\}$-valued, this gives $J\perp\!\!\!\perp\tilde{\bm n}\mid g(\tilde{\bm n})$.
\end{proof}

As an illustration of an alternative statistic, consider the difference in privatized proportions, $T(\tilde{\bm n})=\tilde n_{11}/n_1-\tilde n_{01}/n_0$. Because the FRT $p$-value depends on the marginal total $n_{+1}=m_{11}+m_{01}$ through the hypergeometric reference distribution, different pairs $(\tilde n_{11},\tilde n_{01})$ with the same $T(\tilde{\bm n})$ generally produce different values of $\Psi$. Thus, $\Psi(\tilde{\bm n})$ is not measurable with respect to $T(\tilde{\bm n})$. Lemma \ref{lem:dpi_psi} then gives
$
I(J;T(\tilde{\bm n}))
<
I(J;\Psi(\tilde{\bm n}))
=
I(J;\tilde{\bm n}),
$
confirming that $T(\tilde{\bm n})$ strictly discards information about the non-private rejection event, whereas $\Psi(\tilde{\bm n})$ preserves all of it.

\subsection{Illustration of a Significance Test Based on an Alternative Statistic}\label{sec:alttest}

\begin{figure}[t]
    \centering
    \includegraphics[width=0.9\linewidth]{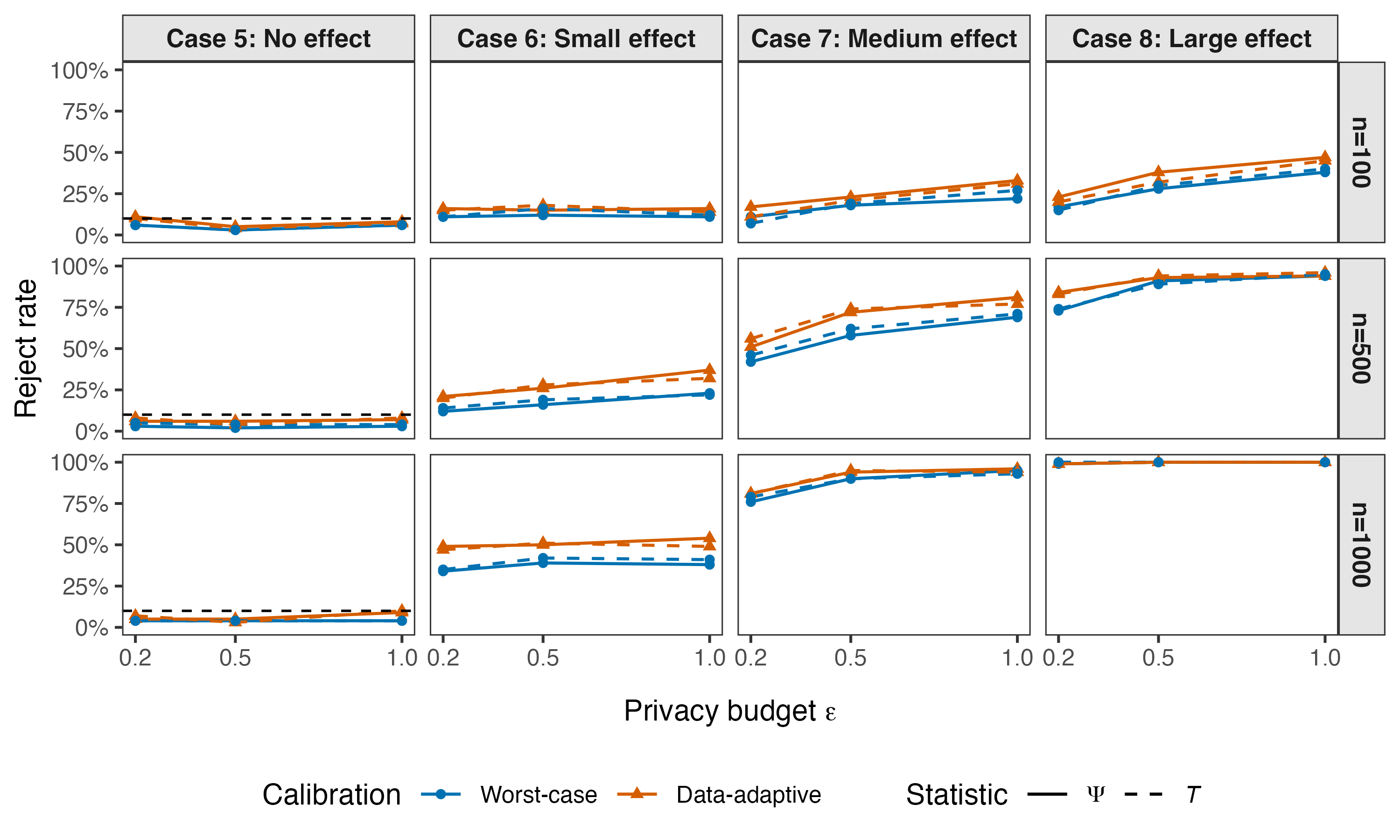}
    \caption{Rejection rates of the frequentist-calibrated rules based on $\Psi$ and on the one-sided statistic $T(\tilde{\bm n})$ for $\epsilon \in \{0.2, 0.5, 1\}$ in Cases 5--8 under $n \in \{100, 500, 1000\}$. Solid and dashed lines represent the tests based on $\Psi$ and $T(\tilde{\bm n})$, respectively, while blue and orange lines indicate worst-case and data-adaptive calibration, respectively. Horizontal black dashed line indicates the nominal Type I error level $\alpha_{\rm Freq}=0.1$, while $\zeta = 0.01$ is used for the data-adaptive calibration. Rates based on 100 simulation runs.}
    \label{fig:Freq_T}
\end{figure}

While appealing, the properties described in Section \ref{sec:psi_justification} do not by themselves guarantee tests based on $\Psi$ have a power advantage over tests based on other statistics. To illustrate this empirically, we compare the power of the $\Psi$-based test with that of a test based on $T(\tilde{\bm n})$, using the same simulation setup as Figure \ref{fig:Freq}. We calibrate both statistics with the worst-case and data-adaptive procedures in Section 4.2 of the main text and evaluate their rejection rates under the sharp null (Case 5) and the causal alternatives (Cases 6--8). Figure \ref{fig:Freq_T} summarizes the results. The tests for both statistics control the Type I error at the nominal level under the sharp null, and the data-adaptive calibration is more powerful than the worst-case calibration for both. Within each calibration, the $\Psi$-based and $T$-based tests attain nearly the same power, suggesting analysts could use them interchangeably to make decisions about Fisher's sharp null. 
As mentioned in the main text, $\Psi$ admits a direct interpretation as the posterior probability of rejection at level $\alpha$, allowing the Bayesian denoising framework to support decision-making via a Bayes rule or significance testing within a single, coherent pipeline. We note that many of the quantities required to compute $\Psi$ depend only on $n$, $n_1$, $n_0$, and $\epsilon$. These quantities can be precomputed and stored in lookup tables for reuse across thresholds, thereby reducing the computational overhead of using $\Psi$.

\section{Proofs for Theoretical Results in the Main Text}\label{sec:proof}

\subsection{Proof of Lemma 3.1}

We now prove Lemma 3.1, stated in Section 3.1 of the main text, which derives the $\ell_1$-sensitivity of the FRT $p$-value under the CRE with binary outcomes.

{\bf Lemma 3.1.} {\em Under the design of CRE with binary outcomes and the test statistic $\hat{\tau}$, the $\ell_1$-sensitivity of $p_{\rm FRT}$ is $\Delta_p = \max \left\{n_1/n, n_0/n\right\}$.}

\begin{proof}
Define $A(\bm Z)=\sum_{i=1}^n z_i Y_i^{\rm obs}$ and $a=A(\bm Z^{\rm obs})$. Since $n_{01}(\bm Z)=n_{+1}-A(\bm Z)$, we have $\hat{\tau}(\bm Z;\bm Y^{\rm obs})=[(1/n_1)+(1/n_0)]A(\bm Z)-n_{+1}/n_0$; hence, $\hat{\tau}$ is strictly increasing in $A(\bm Z)$. Therefore, $p_{\mathrm{FRT}}=|\mathcal Z|^{-1}\sum_{\bm Z\in\mathcal Z}\bm{1} (A(\bm Z)\ge a)$.

Let $D$ and $D'$ differ only at unit $j$, and set $s=Y'_j-Y_j\in\{-1,+1\}$. Then for $D'$, $A'(\bm Z)=A(\bm Z)+s z_j$ and $a'=a+s z^{\rm obs}_j$. If $z_j=z^{\rm obs}_j$ the indicator is unchanged; if $z_j\ne z^{\rm obs}_j$ it can change by at most one in absolute value. 
Averaging over $\mathcal Z$ gives
\begin{equation}
\left|p_{\mathrm{FRT}}(D)-p_{\mathrm{FRT}}(D')\right|
\le \frac{1}{|\mathcal Z|}\sum_{\bm Z\in\mathcal Z}\mathbf 1\left(z_j\neq z^{\rm obs}_j\right)
=\begin{cases}
n_1/n,& z^{\rm obs}_j=0,\\
n_0/n,& z^{\rm obs}_j=1.
\end{cases}
\end{equation}
Maximizing over $j$ yields $\Delta_p\le \max\{n_1/n,n_0/n\}$.
Tightness holds in two extremal constructions. If all $Y_i^{\rm obs}=0$ and $z^{\rm obs}_j=1$, then $p_{\mathrm{FRT}}(D)=1$, while after flipping $Y_j^{\rm obs}$ to 1, we obtain $p_{\mathrm{FRT}}(D')=\Pr(Z_j=1)=n_1/n$. So the gap is $n_0/n$. Symmetrically, if all $Y_i^{\rm obs}=1$ and $z^{\rm obs}_j=0$, the gap is $n_1/n$.
In particular, under the balanced design $n_1=n_0=n/2$, the sensitivity is $\Delta_p=1/2$.
\end{proof}

\subsection{Proof of Lemma 3.3}

We prove Lemma 3.3, stated in Section 3.1 of the main text, which computes the $\ell_1$-sensitivity of the test statistic $\hat{\tau}$ under the CRE with binary outcomes.

{\bf Lemma 3.3.} {\em Under the design of CRE with binary outcomes, the $\ell_1$-sensitivity of $\hat{\tau}$ is $\Delta_{\hat{\tau}} = \max\left\{1/n_1, 1/n_0\right\}$.}

\begin{proof}
If $Z_j^{\rm obs}=1$, flipping $Y_j^{\rm obs}$ changes $n_{11}$ by $\pm1$ and keeps $n_{01}$ unchanged, so $|\hat{\tau}(D)-\hat{\tau}(D')|=1/n_1$. If $Z_j^{\rm obs}=0$, it changes $n_{01}$ by $\pm1$ and keeps $n_{11}$ unchanged, so $|\hat{\tau}(D)-\hat{\tau}(D')|=1/n_0$. Maximizing over $j$ yields the claim. In a balanced design, $\Delta_{\hat{\tau}}=2/n$.
\end{proof}

\subsection{Proof of Lemma 4.1}

We prove Lemma 4.1, stated in Section 4.1.1 of the main text, which gives a necessary and sufficient condition for the abstention option to degenerate.

{\bf Lemma 4.1.} {\em The abstention option degenerates if and only if $\lambda_u \ge H/2$, where $H = (2\lambda_0\lambda_1)/(\lambda_0 + \lambda_1)$ is the harmonic mean of $\lambda_0$ and $\lambda_1$.}

\begin{proof}
Write $c^*=\lambda_0/(\lambda_0+\lambda_1)$. By (15) in the main text, the rule assigns $\delta=1$ when $\Pr(p_{\rm FRT}\le\alpha\mid\tilde{\bm n})>t_{\mathrm{high}}$ and $\delta=0$ when $\Pr(p_{\rm FRT}\le\alpha\mid\tilde{\bm n})<t_{\mathrm{low}}$, where
\begin{equation}
t_{\mathrm{low}}=\min\left\{c^*,~\frac{\lambda_u}{\lambda_1}\right\},
\qquad
t_{\mathrm{high}}=\max\left\{c^*,~1-\frac{\lambda_u}{\lambda_0}\right\}.
\end{equation}
The abstention region is $A=(t_{\mathrm{low}},t_{\mathrm{high}})$, so the abstention option degenerates if and only if $A=\emptyset$, that is, $t_{\mathrm{high}}\le t_{\mathrm{low}}$.

By construction $t_{\mathrm{high}}\ge c^*\ge t_{\mathrm{low}}$; hence $t_{\mathrm{high}}\le t_{\mathrm{low}}$ holds if and only if $t_{\mathrm{high}}=t_{\mathrm{low}}=c^*$, which requires both $1-\lambda_u/\lambda_0\le c^*$ and $\lambda_u/\lambda_1\ge c^*$. Since $1-c^*=\lambda_1/(\lambda_0+\lambda_1)$, the first condition rearranges to $\lambda_u/\lambda_0\ge\lambda_1/(\lambda_0+\lambda_1)$ and the second to $\lambda_u/\lambda_1\ge\lambda_0/(\lambda_0+\lambda_1)$; both are equivalent to
\begin{equation}
\lambda_u\ge\frac{\lambda_0\lambda_1}{\lambda_0+\lambda_1}=\frac{H}{2}.
\end{equation}
Conversely, if $\lambda_u<H/2$, then $\lambda_u/\lambda_1<c^*<1-\lambda_u/\lambda_0$, so $t_{\mathrm{low}}=\lambda_u/\lambda_1<t_{\mathrm{high}}=1-\lambda_u/\lambda_0$ and $A\neq\emptyset$. Therefore the abstention option degenerates if and only if $\lambda_u\ge H/2$, in which case (15) reduces to the binary rule (14) in the main text. Under the strict inequalities in (15), when $\lambda_u\ge H/2$ the rule still returns $u$ only on the knife-edge event $\{\Pr(p_{\rm FRT}\le\alpha\mid\tilde{\bm n})=c^*\}$, where both binary decisions attain posterior risk $H/2\le\lambda_u$; we break this tie toward $\delta=0$, so that (15) coincides with (14) there as well.
\end{proof}

\subsection{Proof of Theorem 4.2}

We prove Theorem 4.2, stated in Section 4.1.2 of the main text, which provides an upper bound on the probability of exiting the abstention region after the second release when spending an additional $\epsilon_{\mathrm{plus}}$ of privacy budget.

{\bf Theorem 4.2.} {\em Let $A=(t_{\mathrm{low}},t_{\mathrm{high}})$ be the abstention region with $0<t_{\mathrm{low}}<t_{\mathrm{high}}<1$, and define $r(\Psi) = \min\{\Psi-t_{\mathrm{low}},~t_{\mathrm{high}}-\Psi\}$ whenever $\Psi\in A$. Then, for every $\epsilon_{\mathrm{plus}}>0$,
\begin{equation*}
\Pr\left(\Psi^{+}\notin A \mid  \Psi\in A\right)
 \le
2~
\mathbb E \left(
\frac{\Psi(1-\Psi)}{r(\Psi)}~
\Delta(\tilde{\bm n}; \epsilon_{\rm plus})
 ~\Bigg|~ \Psi\in A
\right).
\end{equation*}}

\begin{proof}
The proof proceeds in three steps: (1) a channel-based representation of posterior refinement; (2) bounding the channel's total variation contraction via DP; and (3) converting mean movement into an upper bound on the probability of exiting the abstention region.
For simplicity, denote $H=\mathbf 1(p_{\rm FRT} \le \alpha)$. All expectations and probabilities in this proof are taken with respect to
the joint law of $(H,\tilde{\bm n},\tilde{\bm n}^{+})$.

In the first step, we formulate the binary-input channel given the first release.
Fix a realization $\tilde{\bm n}$ and condition on this event.
Under this conditional law, we have $\Psi(\tilde{\bm n}) = \Pr(H=1\mid \tilde{\bm n})\in[0,1]$.
For $h\in\{0,1\}$, define the conditional output distributions as
\begin{equation}
Q_h(\cdot)
=
\mathcal L\left(\tilde{\bm n}^{+} \mid H=h,\tilde{\bm n}\right).
\end{equation}
Thus, conditionally on $\tilde{\bm n}$, $(H,\tilde{\bm n}^{+})$ is a
binary-input channel with input prior $\Pr(H=1)=\Psi(\tilde{\bm n})$ and output kernel
$\{Q_0,Q_1\}$.

Let $Y$ denote a generic random variable with distribution
$\mathcal L(\tilde{\bm n}^{+}\mid \tilde{\bm n})$. Then, by Bayes' rule,
the refined posterior can be written as $\Psi^{+}(Y)
= \Pr(H=1\mid \tilde{\bm n},Y)$.
Conditionally on $\tilde{\bm n}$, one can compute
\begin{equation}
\label{eq:ABS_TV_conditional_tight}
\mathbb E\left(|\Psi^{+}-\Psi| \mid \tilde{\bm n}\right)
= 2 \Psi(1-\Psi) 
\mathrm{TV}(Q_1,Q_0),
\end{equation}
where $\mathrm{TV}(\cdot,\cdot)$ denotes total variation distance.

In the second step, we bound $\mathrm{TV}(Q_1,Q_0)$ via DP in a data-adaptive manner.
Let $K(\cdot\mid a,b)$ denote the conditional law of the top-up release
$\tilde{\bm n}^{+}$ given $(m_{11},m_{01})=(a,b)$, i.e., the Geometric mechanism kernel.
Conditionally on $\tilde{\bm n}$ and $H=h$, the posterior of the counts is
supported on $S_h$, and $Q_h$ is the corresponding mixture,
\begin{equation}
Q_h(\cdot)
=
\sum_{(a,b)\in S_h}
K(\cdot \mid a,b) 
\Pr\left(m_{11}=a,m_{01}=b \mid H=h,\tilde{\bm n}\right),
\qquad h\in\{0,1\}.
\end{equation}
Consider two arbitrary mixtures $Q_1=\sum_i\alpha_i P_i$ and
$Q_0=\sum_j\beta_j R_j$ with weights summing to 1. By convexity of total variation, we have $\mathrm{TV}(Q_1,Q_0)
\le
\sum_{i,j}\alpha_i\beta_j\mathrm{TV}(P_i,R_j).$
Applied here, writing $\mu_h(a,b)=\Pr(m_{11}=a,m_{01}=b \mid H=h,\tilde{\bm n})$ for $(a,b)\in S_h$, we obtain
\begin{equation}
\label{eq:TV_mixture_avg_tight}
\mathrm{TV}(Q_1,Q_0)
\le
\sum_{(a,b)\in S_1}\sum_{(a',b')\in S_0}
\mu_1(a,b)\mu_0(a',b')~
\mathrm{TV}\left(K(\cdot\mid a,b),K(\cdot\mid a',b')\right).
\end{equation}
The Geometric mechanism on the counts has $\ell_1$-sensitivity equal to one and
satisfies $\epsilon_{\mathrm{plus}}$-DP with respect to the adjacency
$d((a,b), (a',b'))=1$. By \cite{GI2024}, we know that for an $\epsilon$-DP mechanism
$K$, any adjacent inputs $z,z'$ obey
\begin{equation}
\mathrm{TV}\left(K(\cdot\mid z),K(\cdot\mid z')\right)
 \le 
s(\epsilon)
=\tanh(\epsilon/2).
\end{equation}
Equip $S$ with the $\ell_1$-metric $d\left((a,b),(a',b')\right)=|a-a'|+|b-b'|$. Then, by group privacy, for any two
$(a,b),(a',b')\in S$, the mechanism satisfies $(d((a,b),(a',b'))~\epsilon_{\mathrm{plus}})$-DP when viewed as a function of the input lattice point. Hence,
\begin{equation}
\mathrm{TV}\left(K(\cdot\mid a,b),K(\cdot\mid a',b')\right)
 \le 
s\left(d\left((a,b),(a',b')\right)\epsilon_{\mathrm{plus}}\right).
\end{equation}
Combining with \eqref{eq:TV_mixture_avg_tight}, for each fixed $\tilde{\bm n}$, we obtain
\begin{equation}
\label{eq:TV_final_bound}
\mathrm{TV}(Q_1,Q_0)
\le
\sum_{(a,b)\in S_1}\sum_{(a',b')\in S_0}
\mu_1(a,b)\mu_0(a',b')~
s\left(\epsilon_{\mathrm{plus}}~d((a,b),(a',b'))\right)
= \Delta(\tilde{\bm n}).
\end{equation}
Substituting \eqref{eq:TV_final_bound} into
\eqref{eq:ABS_TV_conditional_tight}, we obtain, for each fixed $\tilde{\bm n}$,
\begin{equation}
\label{eq:ABS_final_conditional_tight}
\mathbb E\left(|\Psi^{+}-\Psi|\mid \tilde{\bm n}\right)
 \le 
2 \Psi(1-\Psi)~\Delta(\tilde{\bm n}).
\end{equation}

In the final step, we establish how the mean movement translates into the abstention probability.
Fix $\tilde{\bm n}$ such that $\Psi\in A$. The minimal distance from $\Psi$ to the boundaries of $A$ is $r(\Psi) = \min\{\Psi-t_{\mathrm{low}},~t_{\mathrm{high}}-\Psi\}>0$.
If $\Psi\in A$ and $\Psi^{+}\notin A$, then necessarily
$|\Psi^{+}-\Psi|\ge r(\Psi)$, so
\begin{equation}
\{\Psi\in A,~\Psi^{+}\notin A\}
\subseteq
\left\{\Psi\in A,~|\Psi^{+}-\Psi|\ge r(\Psi)\right\}.
\end{equation}
Conditionally on this fixed $\tilde{\bm n}$ with $\Psi\in A$, we obtain
\begin{equation}
\Pr\left(\Psi^{+}\notin A\mid \tilde{\bm n}\right)
 \le 
\Pr\left(|\Psi^{+}-\Psi|\ge r(\Psi)\mid \tilde{\bm n}\right).
\end{equation}
Applying Markov's inequality to the nonnegative random variable
$|\Psi^{+}-\Psi|$ yields
\begin{equation}
\Pr\left(|\Psi^{+}-\Psi|\ge r(\Psi)\mid \tilde{\bm n}\right)
\le
\frac{\mathbb E\left(|\Psi^{+}-\Psi|\mid \tilde{\bm n}\right)}
     {r(\Psi)}.
\end{equation}
Combining with \eqref{eq:ABS_final_conditional_tight} gives, for every
$\tilde{\bm n}$ such that $\Psi\in A$,
\begin{equation}
\Pr\left(\Psi^{+}\notin A\mid \tilde{\bm n}\right)
 \le 
2~\Delta(\tilde{\bm n})~
\frac{\Psi(1-\Psi)}{r(\Psi)}.
\end{equation}
Finally,
\begin{equation}
\Pr\left(\Psi^{+}\notin A\mid \Psi\in A\right)
=
\mathbb E\left(
\Pr\left(\Psi^{+}\notin A\mid \tilde{\bm n}\right)
 \mid
\Psi\in A
\right).
\end{equation}
Using the bound from the previous step and the fact that
$\Psi$, $r(\Psi)$, and $\Delta(\tilde{\bm n})$ are functions of $\tilde{\bm n}$, we obtain
\begin{equation}
\Pr\left(\Psi^{+}\notin A \mid \Psi\in A\right)
 \le 
2~
\mathbb E \left(
\frac{\Psi(1-\Psi)}{r(\Psi)}~\Delta(\tilde{\bm n})
 ~\Bigg|~ \Psi\in A
\right),
\end{equation}
which completes the proof.
\end{proof}

\subsection{Proof of Corollary 4.3}

We now prove Corollary 4.3, stated in Section 4.1.2 of the main text, which gives a data-adaptive necessary lower bound on the additional privacy budget $\epsilon_{\mathrm{plus}}$.

{\bf Corollary 4.3.} {\em Let $A=(t_{\mathrm{low}},t_{\mathrm{high}})$ be the abstention region with $0<t_{\mathrm{low}}<t_{\mathrm{high}}<1$. Fix a confidence level $1-\xi\in(0,1)$. For $\Psi\in A$, if the refined posterior $\Psi^+$ satisfies $\Pr\left(\Psi^+\notin A \mid \tilde{\bm n}\right) \ge 1-\xi$, then
\begin{equation*}
\epsilon_{\mathrm{plus}} \ge \inf\left\{\epsilon>0:
\Delta(\tilde{\bm n};\epsilon)\ \ge\
\frac{(1-\xi)~r(\Psi)}{2 \Psi(1-\Psi)}
\right\},
\end{equation*}
where $r(\Psi) = \min\{\Psi-t_{\mathrm{low}},~t_{\mathrm{high}}-\Psi\}$, and $\Delta(\tilde{\bm n}; \epsilon)$ is defined as in Theorem 4.2.}

\begin{proof}
From the proof of Theorem 4.2, for any fixed $\tilde{\bm n}$ with $\Psi\in A$,
\begin{equation}
\Pr(\Psi^{+}\notin A\mid \tilde{\bm n})
 \le 
2~\Delta(\tilde{\bm n};\epsilon_{\mathrm{plus}})~
\frac{\Psi(1-\Psi)}{r(\Psi)}.
\end{equation}
Assume that for some $\epsilon_{\mathrm{plus}}>0$ the desired bound $\Pr(\Psi^{+}\notin A\mid \tilde{\bm n}) \ge 1-\xi$ holds. Then,
\begin{equation}
1-\xi
 \le 
\Pr\left(\Psi^{+}\notin A\mid \tilde{\bm n}\right)
 \le 
2~\Delta(\tilde{\bm n};\epsilon_{\mathrm{plus}})~
\frac{\Psi(1-\Psi)}{r(\Psi)},
\end{equation}
which implies
\begin{equation}
\label{eq:min_condition_frontier_compact}
\Delta(\tilde{\bm n};\epsilon_{\mathrm{plus}}) \ge
\frac{(1-\xi)~r(\Psi)}{2\Psi(1-\Psi)}.
\end{equation}
Since $s(x)=\tanh(x/2)$ is increasing on $[0,\infty)$ and satisfies $0\le s(x)<1$,
for any fixed $\tilde{\bm n}$ and any two values $0\le \epsilon_1\le \epsilon_2$ we have $s \left(\epsilon_1~d((a,b),(a',b'))\right) \le s \left(\epsilon_2~d((a,b),(a',b'))\right)$
for all $(a,b)\in S_1$ and $(a',b')\in S_0$.
Multiplying by the nonnegative weights $\mu_1(a,b)\mu_0(a',b')$ and summing over
$S_1\times S_0$ yields $\Delta(\tilde{\bm n};\epsilon_1) \le \Delta(\tilde{\bm n};\epsilon_2)$, so $\Delta(\tilde{\bm n};\epsilon)$ is nondecreasing in $\epsilon$.

Consequently, for any threshold $\tau\in(0,1)$ the superlevel set
$\{\epsilon>0:\Delta(\tilde{\bm n};\epsilon)\ge \tau\}$ is of the form $[\epsilon^\star,\infty)$
(possibly empty), where $\epsilon^\star=\inf\{\epsilon>0:\Delta(\tilde{\bm n};\epsilon)\ge \tau\}.$
Taking $\tau=(1-\xi)~r(\Psi)\left(2\Psi(1-\Psi)\right)^{-1}$ completes the proof.
\end{proof}

\subsection{Proof of Theorem 4.4}

We prove Theorem 4.4, stated in Section 4.2.1 of the main text, which guarantees Type I error control for the worst-case calibration procedure.

{\bf Theorem 4.4.} {\em Under the sharp null $H_0^{\rm F}$, we have $\Pr \left(\delta_{\mathrm{LFC}}(\tilde{\bm n})=1\right)\le \alpha_{\mathrm{Freq}}$, where the probability averages over randomization and the privacy mechanism.}

\begin{proof}
Fix $K$ and let $\Pr_K$ denote probability under $Q_K$. Since $t_K$ is the right-continuous $(1-\alpha_{\mathrm{Freq}})$-quantile of $F_\Psi^{(K)}$, we have $F_\Psi^{(K)}(t_K)\ge 1-\alpha_{\mathrm{Freq}}$, hence
\begin{equation}
\Pr_K\left(\Pr(p_{\rm FRT} \le \alpha \mid  \tilde{\bm n})>t_K\right)\le \alpha_{\mathrm{Freq}}.
\end{equation}
Since $t_{\mathrm{LFC}}^*\ge t_K$, it follows that 
\begin{equation}
\Pr_K\left(\Pr(p_{\rm FRT} \le \alpha \mid  \tilde{\bm n})>t_{\mathrm{LFC}}^*\right)\le \alpha_{\mathrm{Freq}}.
\end{equation}
Under $H_0^{\rm F}$ the potential outcomes are fixed, so $n_{+1}$ is a fixed constant and $\tilde{\bm n}\sim Q_{n_{+1}}$ exactly. Since the bound holds for every $K$, applying it at $K=n_{+1}$ shows that the Type I error is no greater than $\alpha_{\mathrm{Freq}}$.
\end{proof}

\subsection{Proof of Theorem 4.6}

We provide the proof of Theorem 4.6, stated in Section 4.2.2 of the main text, establishing Type I error control for the data-adaptive procedure.

{\bf Theorem 4.6.} {\em Fix $\zeta \in (0,\alpha_{\mathrm{Freq}})$ and set $\alpha'=\alpha_{\mathrm{Freq}}-\zeta$. Under the sharp null $H_0^{\rm F}$, we have $\Pr \left(\delta_{\mathrm{Neyman}}(\tilde{\bm n})=1\right)\le \alpha_{\mathrm{Freq}}$.}

\begin{proof}
Fix $K$ and denote $\Psi = \Pr(p_{\rm FRT} \le \alpha \mid  \tilde{\bm n})$. We first decompose
\begin{equation}
\begin{aligned}
\Pr_K\left(\Psi > t^*_{\mathrm{Neyman}}(\tilde{\bm n})\right)
&= \Pr_K\left(
\Psi > t^*_{\mathrm{Neyman}}(\tilde{\bm n}),~\tilde{\bm n} \in A_K
\right) 
+ \Pr_K\left(\Psi > t^*_{\mathrm{Neyman}}(\tilde{\bm n}),~\tilde{\bm n} \notin A_K\right) \\
&\leq \Pr_K\left(
\Psi > t^*_{\mathrm{Neyman}}(\tilde{\bm n}),~\tilde{\bm n} \in A_K
\right) 
+ \Pr_K\left(\tilde{\bm n} \notin A_K\right).
\end{aligned}
\end{equation}

On the event $\{\tilde{\bm n}\in A_K\}$ we have $K\in C_{1-\zeta}(\tilde{\bm n})$, so by the definition of $t^{*}_{\mathrm{Neyman}}(\tilde{\bm n})$, $t^{*}_{\mathrm{Neyman}}(\tilde{\bm n})\ge t_K'$.
Hence, $\left\{\Psi > t^*_{\mathrm{Neyman}}(\tilde{\bm n})\right\} \subseteq \left\{
\Psi > t_K' \right\}$ on $\{\tilde{\bm n} \in A_K\}$.
Therefore,
\begin{equation}
\Pr_K\left(
\Psi > 
t^*_{\mathrm{Neyman}}(\tilde{\bm n}),~\tilde{\bm n} \in A_K
\right) \le 
\Pr_K\left(
\Psi > t_K'
\right)
\le \alpha'.
\end{equation}

By Lemma 4.5 in the main text, we have $\Pr_K\left(\tilde{\bm n} \notin A_K\right) \le \zeta$. Combining the two bounds gives
\begin{equation}
\Pr_K\left(
\Psi > 
t^*_{\mathrm{Neyman}}(\tilde{\bm n})
\right) \le \alpha' + \zeta = \alpha_{\mathrm{Freq}}.
\end{equation}

Since this holds for every $K$, including the case where $K$ equals the true total successes $n_{+1}$, the Type I error is bounded by $\alpha_{\mathrm{Freq}}$.
\end{proof}

\end{document}